\documentclass[rmp,aps,twocolumn,nofootinbib,eqsecnum]{revtex4}
\pdfoutput=1
\usepackage{graphics}
\usepackage{epsfig}
\usepackage{amsmath}
\usepackage{supertabular}
\usepackage{amssymb}
\usepackage{slashed}
\usepackage{color}

\newcommand{\nn}{\nonumber \\}
\newcommand{\no}{\nonumber}

\newcommand{\ve}{\varepsilon}
\newcommand{\mrm}{\mathrm}
\newcommand{\cO}{{\cal O}}
\newcommand{\cL}{{\cal L}}

\newcommand{\beq}{\begin{equation}}
\newcommand{\eeq}{\end{equation}}
\newcommand{\beqa}{\begin{eqnarray}}
\newcommand{\eeqa}{\end{eqnarray}}
\newcommand{\beqan}{\begin{eqnarray*}}
\newcommand{\eeqan}{\end{eqnarray*}}
\newcommand{\ba}{\begin{array}}
\newcommand{\ea}{\end{array}}

\newcommand\lsim{\mathrel{\rlap{\lower4pt\hbox{\hskip1pt$\sim$}}
    \raise1pt\hbox{$<$}}}
\newcommand\gsim{\mathrel{\rlap{\lower4pt\hbox{\hskip1pt$\sim$}}
    \raise1pt\hbox{$>$}}}

\newcommand{\real}{{\rm Re}}
\newcommand{\imag}{{\rm Im}}
\newcommand{\eps}{\epsilon}
\newcommand{\epsp}{\epsilon^\prime}

\begin{document}
\title{Kaon Decays in the Standard Model}

\author{Vincenzo Cirigliano}
\email{cirigliano@lanl.gov}
\affiliation{Theoretical Division, Los Alamos National Laboratory, Los
 Alamos, NM 87545, USA}
\author{Gerhard Ecker}
\email{gerhard.ecker@univie.ac.at}
\affiliation{University of Vienna, Faculty of Physics, Boltzmanngasse 5,
A-1090 Wien, Austria}
\author{Helmut Neufeld}
\email{helmut.neufeld@univie.ac.at}
\affiliation{University of Vienna, Faculty of Physics, Boltzmanngasse 5,
A-1090 Wien, Austria}
\author{Antonio Pich}
\email{Antonio.Pich@ific.uv.es}
\affiliation{Departament de F\'{\i}sica Te\`orica, IFIC,
Universitat de Val\`encia -- CSIC, \\ Apartat de Correus 22085, E-46071
Val\`encia, Spain}
\author{Jorge Portol\'es}
\email{Jorge.Portoles@ific.uv.es}
\affiliation{Departament de F\'{\i}sica Te\`orica, IFIC,
Universitat de Val\`encia -- CSIC, \\ Apartat de Correus 22085, E-46071
Val\`encia, Spain}

\begin{abstract}
A comprehensive overview of kaon decays is presented. The
  Standard Model predictions are discussed
in detail, covering both the underlying short-distance electroweak
dynamics and the important
interplay of QCD at long distances. Chiral perturbation theory
provides a universal framework for treating leptonic, semileptonic and
nonleptonic decays including rare and radiative modes.
All allowed decay modes with branching ratios of at least $10^{-11}$
are analyzed. Some decays with even smaller rates are also
included. Decays that are strictly forbidden in the Standard Model
are not considered in this review.
The present
experimental status and the prospects for future improvements are
reviewed.
\end{abstract}

\maketitle
\tableofcontents

\section{Introduction}
\label{sec:intro}
Kaon decays have played a key role in the shaping of the Standard
Model (SM) \cite{Glashow:1961tr,Salam:1968rm,Weinberg:1967tq}, from
the discovery of kaons \cite{Rochester:1947mi} until today. Prominent
examples are the introduction of strangeness
\cite{Pais:1952zz,GellMann:1953zza}, parity violation ($K \to
2 \pi, 3 \pi$ puzzle) \cite{Dalitz:1954cq,Lee:1956qn}, quark mixing
\cite{Cabibbo:1963yz,Kobayashi:1973fv}, the discovery of CP violation
\cite{Christenson:1964fg},  suppression of flavor-changing neutral
currents (FCNC) and the GIM mechanism \cite{Glashow:1970gm}. Moreover,
kaon decays continue to have an important impact on flavor dynamics in
constraining physics beyond the SM.

The aim of this review is a comprehensive survey of kaon decays
allowed in the SM with branching ratios of at least $10^{-11}$. Some
decays are included with even smaller decay rates. We do not cover
decays that are strictly forbidden in the SM such as lepton-number
or lepton-flavor violating decays.

Kaon decays involve an intricate interplay between weak,
electromagnetic and strong interactions. A major theoretical challenge
has to do with the intrinsically non-perturbative nature of the strong
interactions in kaon physics. The last 25 years have seen the
development of a systematic approach to low-energy hadron physics in
the framework of chiral
perturbation theory (we refer to Sec.~\ref{subsec:long-distance} for
references). The approach provides a systematic expansion of decay
amplitudes in terms of momenta and meson masses. The momenta of
particles in the final states are sufficiently small so that the kaon
mass sets the scale for the quality of the expansion. The relevant
dimensionless ratio is $M_K^2/(4\pi F_\pi)^2 \simeq 0.18$.

Looking in more detail into the predictions of chiral
perturbation theory (CHPT) for kaon
decays, one observes a rather wide range, from cases where great
precision can be achieved [especially in (semi)leptonic decays] to
processes where hadronic uncertainties remain the major obstacle for
comparison with experiment (mainly in nonleptonic decays). However, an
important feature of CHPT in general is that it parametrizes the
intrinsic hadronic uncertainties by a number of parameters, the
so-called low-energy constants (LECs). The quality of theoretical
predictions hinges to a large extent on the available information on
those LECs. Over the years, a lot of progress has been achieved in the
theoretical understanding and the phenomenological knowledge of
LECs. In recent years, lattice simulations have
made important contributions
to this field. For the time being, lattice QCD gives access mainly to
strong LECs (see Sec.~\ref{sec:lecs}). This is one of
the reasons why much better precision can be achieved in semileptonic
than in nonleptonic decays.

Kaon decays form a fascinating chapter of particle physics in themselves
but they also give access to fundamental parameters of the SM such as
the Cabibbo-Kobayashi-Maskawa (CKM) matrix elements $V_{us}$ and
$V_{td}$. Although we stay strictly within the SM for this review, kaon
decays have of course the potential to catch glimpses of New Physics,
especially in those cases where the SM makes precise
predictions. The constraints on New Physics are most effective in
combination with other information from the low-energy/high-intensity
frontier that are out of scope for this review. However,
the impact of such constraints depends crucially on the quality of SM
predictions. A state-of-the-art survey of kaon decays in the SM is
therefore expected to be very useful also for
physics beyond the SM.

We review the status of $K$ decays at a time when both theory and
experiment progress at a somewhat slower pace than ten years
ago. Nevertheless, there is still considerable activity in the
field. The experimental program for kaon decays concentrates on
specific channels of particular interest for physics beyond the SM but
results for other channels can be expected as by-products.
The main players in the near
future on the experimental side will be NA62 at CERN \cite{Collazuol:2009zz},
$\mathrm{K}^0$TO \cite{Nanjo:2009xx} and TREK
\cite{Paton:2006xxx,Kohl:2010zz} at
J-PARC, KLOE-2 at DA$\Phi$NE \cite{AmelinoCamelia:2010me}, KLOD
\cite{Bolotov:2009zz} and
OKA \cite{Kurshetsov:2009zz} at IHEP Protvino,
and the proposed Project-X at Fermilab \cite{Project-X:2010zz}.  

Kaon decays have been treated in several reviews and lecture notes
during the past 20 years
\cite{Bryman:1988ut,Battiston:1990gw,Winstein:1992sx,Littenberg:1993qv,
Ritchie:1993ua,D'Ambrosio:1994ae,Buras:1996cw,Buchholz:1997da,Barker:2000gd,
Buchalla:2001ux,Buras:2004uu,Buchalla:2008jp}.

The review is organized along the following lines. We start with a
brief summary of the theoretical framework, including both the
short-distance aspects and the low-energy realization in terms
of CHPT. For the latter, we also recapitulate the present knowledge of
the coupling constants in the chiral Lagrangians (LECs). As already
emphasized and in contrast to most previous reviews of the field, we then
discuss essentially all kaon decays allowed in the SM that have either
already been measured or that may become accessible experimentally in
the not too distant future. We divide the actual review
of different channels into three parts: leptonic and semileptonic
decays, dominant nonleptonic decays ($K \to 2 \pi, 3 \pi$) and,
finally, rare and radiative decays.

In Sec.~\ref{sec:frame}, we recall the effective Lagrangians in the SM,
for both semileptonic and nonleptonic decays, after integrating out the
$W$ and $Z$ bosons and the heavy quarks ($t,b,c$). For nonleptonic transitions, the
leading QCD corrections are summed up with the help of the operator
product expansion (OPE) and the renormalization group. The second part of
this section contains a brief introduction to CHPT, concentrating on
the various chiral Lagrangians relevant for $K$ decays. Estimates of
LECs are presented in Sec.~\ref{sec:lecs}. In addition to
phenomenological determinations, the large-$N_C$ limit of QCD provides
a useful framework for theoretical estimates. Strong LECs are
dominated by the exchange of meson resonances.
The resulting  numerical estimates are in general in good
agreement with upcoming lattice determinations. Through the
hadronization of short-distance operators
appearing in the effective Lagrangian for nonleptonic transitions at
the quark level, electroweak LECs can be expressed in terms of strong
and electromagnetic couplings.

In Sec.~\ref{sec:semilep}, we review the current status of leptonic
and semileptonic modes (including radiative channels). $K_{\ell 2}$ and
$K_{\ell 3}$ decay rates can be predicted  with great accuracy, providing
non-trivial tests of the SM and allowing for the extraction of
$V_{us}$. $K_{\ell 4}$ decays are used as sensitive probes of chiral
dynamics in $\pi\pi$ scattering whereas $K_{e5}$ decays are included
for completeness only.

The dominant nonleptonic decays $K \to 2 \pi, 3\pi$
are treated in Sec.~\ref{sec:nonlep}. The two-pion modes are used to
extract the leading-order (LO) nonleptonic LECs $G_8$, $G_{27}$ on the
basis of a next-to-leading-order (NLO) calculation.
To determine the s-wave $\pi\pi$ phase shift
difference $\delta_0(M_K)- \delta_2(M_K)$, inclusion of
isospin-violating corrections is mandatory. This applies also to the
SM prediction of the CP-violating ratio  $\epsilon'/\epsilon$ although
the theoretical precision still does not match the experimental accuracy.
For the three-pion modes, NLO corrections significantly improve
the agreement between theory and experiment. In addition to the
CP-violating decay $K_S \to 3 \pi^0$, special attention has been
given to CP-violating asymmetries in the linear Dalitz plot
parameter for the three-pion decays of charged kaons. Somewhat
unexpectedly, $K \to 3 \pi$ decays with at least two $\pi^0$ in the
final state allow for a precise extraction of
s-wave $\pi\pi$ scattering lengths by analyzing the cusp near
threshold.

Rare and radiative decays are considered in
Sec.~\ref{sec:rare}.  We first summarize the status of the rare
decays $K_L \to  \pi^0 \nu \bar{\nu}$ and $K^\pm  \to  \pi^\pm  \nu
\bar{\nu}$ that can be predicted with a  precision surpassing any other
FCNC process involving quarks. The modes $K \to \pi\pi \nu \bar{\nu}$
can also be predicted with good accuracy but experimental limits are
still far above the rates expected in the SM. Most of the remaining
radiative modes $K \to \gamma^{(*)} \gamma^{(*)}$, $K \to \ell^+
\ell^-$, $K \to \pi \ell^+ \ell^-$, $K \to \pi \gamma \gamma^{(*)}$
and $K \to \pi \pi \gamma^{(*)}$ are dominated by long-distance
dynamics to be analyzed in the CHPT framework. All channels have
been calculated to NLO but in many cases estimates of the dominant
next-to-next-to-leading-order (NNLO) effects are also
available. Although of minor phenomenological
interest at present, the decays
$K^0 \to 3 \gamma$,
$K_L \to \gamma \gamma \ell^+ \ell^-$,
$K_L \to \gamma \nu \bar{\nu}$,
$K_S \to \ell^+_1 \ell^-_1 \ell^+_2 \ell^-_2$,
$K^0 \to \pi^0 \pi^0 \gamma$,
$K^0 \to \pi^0 \pi^0 \gamma \gamma$
and $K \to 3 \pi \gamma$
are included for completeness. Conclusions and an outlook are
presented in Sec.~\ref{sec:conc}.
Some one-loop functions are collected in the appendix.

\section{Theoretical framework}
\label{sec:frame}

\subsection{Short-distance description}
\label{subsec:short-distance}

The SM predicts strangeness-changing
transitions with $\Delta S=1$ via
$W$ exchange between two weak charged currents.
At the kaon mass scale, the heavy $W$ boson can be integrated out
and the interaction is described in terms of effective four-fermion
operators.

Semileptonic transitions are mediated by the effective Lagrangian
\begin{eqnarray}
{\cal L}_{\rm eff} &=&  - \frac{G_F}{\sqrt{2}} \ S_{\rm EW}^{1/2} \
\left[\bar{\ell} \gamma_\mu  (1 - \gamma_5) \nu_\ell\right]  \ \left[\bar{u}_i
\gamma^\mu (1 - \gamma_5)  \, V_{ij}  \,  d_j\right]  \nn
&&{}  +~ \ {\rm h.c.} \label{eq:Lsemi}
\end{eqnarray}
where $V_{ij}$ denotes the element $ij$ of the
CKM matrix \cite{Cabibbo:1963yz,Kobayashi:1973fv}
and $G_F= 1.1663788\, (7)\times 10^{-5}\,\mathrm{GeV^{-2}}$ \cite{Webber:2010zf}
is the Fermi constant as extracted from muon decay.
The universal short-distance factor
\begin{eqnarray}
S_{\rm EW} &=& 1 + \frac{2 \alpha}{\pi} \left(1-\frac{\alpha_s}{4\pi}
\right)\ln \frac{M_Z}{M_\rho} + \cO \left(\frac{\alpha \alpha_s}{\pi^2}
\right) \nn
&=&  1.0223 \pm 0.0005
\label{SEW}
\end{eqnarray}
encodes electroweak corrections not included in $G_F$ \cite{Sirlin:1977sv,Sirlin:1981ie}
and small QCD effects \cite{Marciano:1993sh}.

$W$ exchange between two quark currents generates the $\Delta S=1$ four-quark operator
\beq\label{eq:Q2}
Q_2 \, = \,
\left[\bar s\gamma^\mu (1 - \gamma_5) u\right] \ \left[\bar u \gamma_\mu (1 - \gamma_5) d\right] ,
\eeq
mediating nonleptonic $K$ decays. Gluonic corrections bring further $\Delta S=1$ operators,
which mix under renormalization
\cite{Gaillard:1974nj,Altarelli:1974exa,Vainshtein:1975sv,Shifman:1975tn}:
\beqa\label{eq:Q1-6}
Q_1  & = &  \left[\bar s^\alpha \gamma^\mu (1 - \gamma_5) u^\beta\right] \left[\bar u^\beta \gamma_\mu (1 - \gamma_5) d^\alpha\right], \nn
Q_3 & = &  \left[\bar s \gamma^\mu (1 - \gamma_5) d\right]  \sum_{q=u,d,s} \left[\bar q \gamma_\mu (1 - \gamma_5) q \right], \nn
Q_4 & = &  \left[\bar s^\alpha \gamma^\mu (1 - \gamma_5) d^\beta\right]
\sum_{q=u,d,s} \left[\bar q^\beta \gamma_\mu (1 - \gamma_5) q^\alpha\right], \nn
Q_5 & = & \left[ \bar s \gamma^\mu (1 - \gamma_5) d\right]
\sum_{q=u,d,s} \, \left[\bar q \gamma_\mu (1 + \gamma_5) q\right],
\quad\\
Q_6 & = & \left[\bar s^\alpha \gamma^\mu (1 - \gamma_5) d^\beta\right]
 \sum_{q=u,d,s} \left[\bar q^\beta \gamma_\mu (1 + \gamma_5) q^\alpha\right],
\nonumber
\eeqa
where $\alpha, \beta$ denote color indices and color-singlet currents are understood
whenever color labels are not explicit ($\bar q\,\Gamma q\equiv
  \bar q^\alpha\Gamma q^\alpha$).

Owing to the presence of very different mass scales
($M_\pi < M_K \ll M_W$), the QCD corrections are amplified by large logarithms.
The short-distance logarithmic corrections can be summed up using the OPE
and the renormalization
group, all the way down from $M_W$ to scales $\mu < m_c$
\cite{Gilman:1979bc,Gilman:1979ud}.
One gets in this way an effective Lagrangian, defined in the
three-flavor theory \cite{Buras:1998raa},
\beq\label{eq:Leff}
\cL_{\mathrm{eff}}^{\Delta S=1}= - \frac{G_F}{\sqrt{2}}\,
 V_{ud}^{\phantom{*}}V^*_{us}\,  \sum_{i=1}^{13}
 C_i(\mu) \, Q_i (\mu),
\eeq
which is a sum of local four-fermion operators $Q_i$,
constructed with the light degrees of freedom ($m<\mu$),
modulated
by Wilson coefficients $C_i(\mu)$, which are functions of the
heavy masses ($M_Z, M_W, m_t, m_b, m_c >\mu$)
and CKM parameters.
The unitarity of the CKM matrix,
\beq\label{eq:lambda_q}
\lambda_u + \lambda_c + \lambda_t = 0 ,
\qquad\qquad \lambda_q\equiv V_{qd}^{\phantom{*}}V^*_{qs},
\eeq
allows to write the Wilson coefficients in the form
\beq\label{eq:tau}
C_i(\mu) \, =\, z_i(\mu) + \tau\, y_i(\mu),
\eeq
where $\tau = -\lambda_t/\lambda_u$.
The CP-violating decay amplitudes are proportional to the components
$y_i(\mu)$.

The Wilson coefficients are known at the NLO
\cite{Buras:1996dq,Buras:1993dy,Buras:1992tc,Ciuchini:1995cd,Ciuchini:1993vr,Ciuchini:1992tj}.
This includes all corrections of $\mathcal{O}(\alpha_s^n t^n)$ and
$\mathcal{O}(\alpha_s^{n+1} t^n)$, where
$t = \ln{(M_1/M_2)}$ refers to the logarithm of any ratio of
heavy mass scales $M_1,M_2\geq\mu$.
Moreover, the full $m_t/M_W$ dependence (at lowest order
in $\alpha_s$) has been taken into account.

The combination $Q_- = Q_2 - Q_1$
and the `penguin' operators $Q_i, \; i=3,4,5,6,$ induce pure $\Delta
I= 1 / 2$
transitions and  transform like $(8_L,\, 1_R)$ under chiral $\rm SU(3)_L \times
SU(3)_R$ transformations in flavor space,
while $Q^{(27)} = 2 Q_2 + 3 Q_1 - Q_3$ transforms like a $(27_L,\,
1_R)$ operator that induces both
$\Delta I= 1 / 2$ and $\Delta I= 3 / 2$ transitions.

The inclusion of virtual electromagnetic interactions brings in
the additional electromagnetic penguin
operators \cite{Bijnens:1983ye,Lusignoli:1988fz,Buras:1987wc,Sharpe:1987cx}

\beqa\label{eq:Q7-10}
Q_7 & = & \frac{3}{2} \, \left[\bar s \gamma^\mu (1 - \gamma_5) d\right] \sum_{q=u,d,s}
 e_q \, \left[\bar q \gamma_\mu (1 + \gamma_5) q\right],
\nn
Q_8 & = & \frac{3}{2} \,\left[ \bar s^\alpha \gamma^\mu (1 - \gamma_5) d^\beta\right]
\sum_{q=u,d,s} e_q \, \left[\bar q^\beta \gamma_\mu (1 + \gamma_5) q^\alpha\right],
\nn
Q_9 & = & \frac{3}{2} \, \left[\bar s \gamma^\mu (1 - \gamma_5) d\right]
\sum_{q=u,d,s}  e_q \, \left[\bar q \gamma_\mu (1 - \gamma_5) q\right],
\\
Q_{10} & = & \frac{3}{2} \, \left[\bar s^\alpha \gamma^\mu (1 - \gamma_5) d^\beta\right]
\sum_{q=u,d,s} e_q \, \left[\bar q^\beta \gamma_\mu (1 - \gamma_5) q^\alpha\right], \nonumber
\eeqa
where $e_q$ denotes the corresponding quark charges in units of $e=\sqrt{4\pi\alpha}$.
Their Wilson coefficients get also higher-order electroweak
contributions from  $Z$-penguin and $W$-box
diagrams.
Under the action of the chiral group $\rm SU(3)_L \times SU(3)_R$ the
operators $Q_7$ and $Q_8$ transform like combinations of $(8_L,\,
1_R)$ and $(8_L,\, 8_R)$ operators, while $Q_9$ and $Q_{10}$
transform like combinations of $(8_L,\, 1_R)$ and $(27_L,\,
1_R)$.

\begin{flushleft}
\begin{figure}[t]
\leavevmode
\includegraphics[width=8 cm]{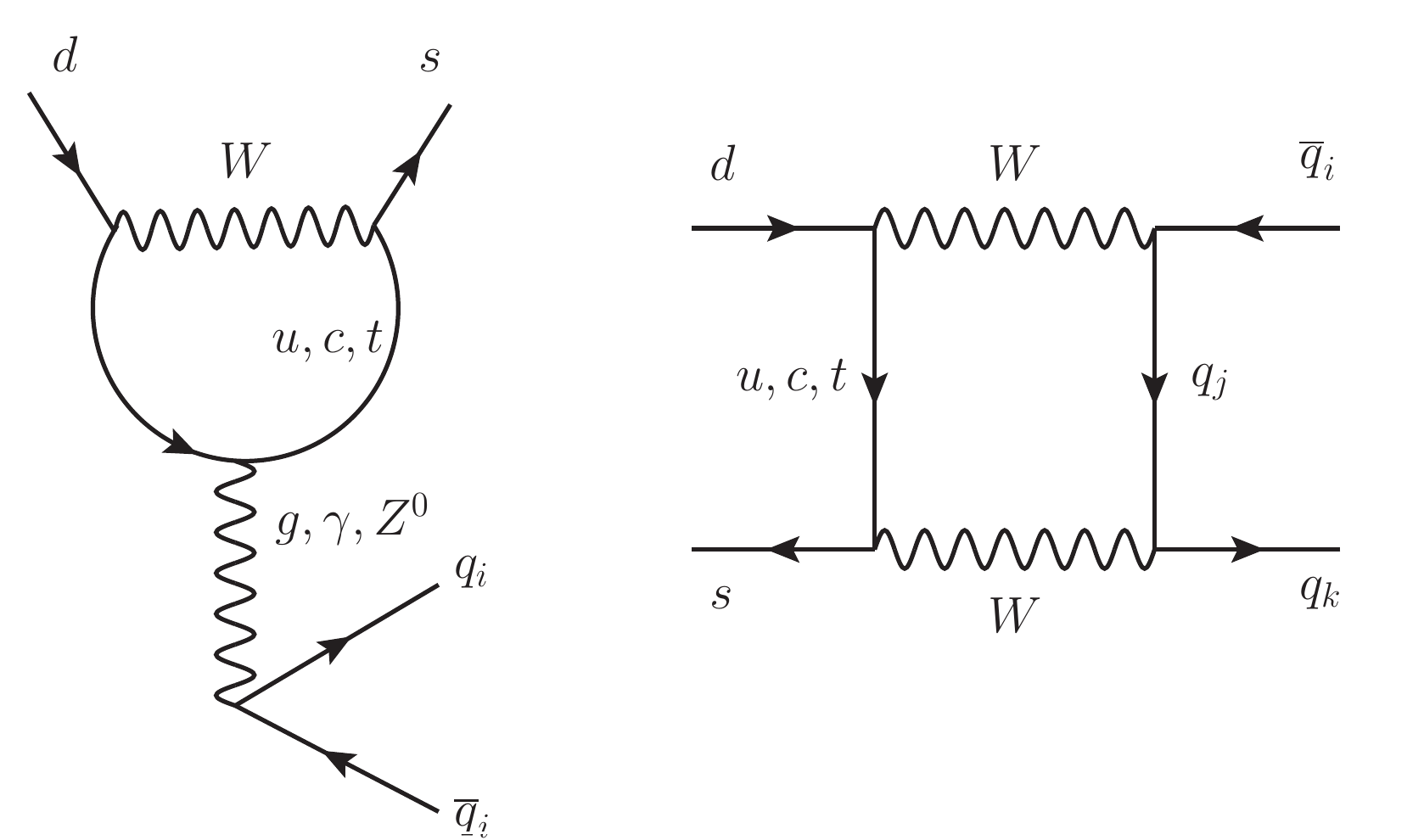}
\caption{\label{fig:PenguinBox}
Penguin and box topologies.}
\end{figure}
\end{flushleft}

Three more operators
need to be considered in processes with leptons in the final state
\cite{Gilman:1979ud,Buchalla:1995vs}:
\beqa\label{eq:Q11-13}
Q_{11} \equiv  Q_{7V} &= &  \left[\bar s \gamma^\mu (1 - \gamma_5) d \right]
\sum_{\ell=e,\mu} \left[\bar{\ell} \gamma_\mu  \ell \right], \nn
Q_{12} \equiv Q_{7A} & = & \left[\bar s \gamma^\mu (1 - \gamma_5) d \right]
\sum_{\ell=e,\mu}  \left[\bar{\ell} \gamma_\mu \gamma_5 \ell \right], \\
Q_{13} \equiv Q(\bar\nu\nu) & = & \left[\bar s \gamma^\mu (1 - \gamma_5) d \right]
\left[\bar \nu \gamma_\mu (1 - \gamma_5) \nu \right]. \quad\;\nonumber
\eeqa
The mixing between the $K^0$ and its antiparticle is induced through
box diagrams with two $W$ exchanges \cite{Gaillard:1974hs},
which in the three-flavor theory generate the effective Lagrangian
\cite{Gilman:1982ap}
\beq\label{eq:L_DS=2}
\cL_{\mathrm{eff}}^{\Delta S=2}= - \frac{G_F^2 M_W^2}{(4\pi)^2}\,
 C_{\Delta S=2}(\mu) \, Q_{\Delta S=2} (\mu),
\eeq
with
\beq\label{eq:Q_DS=2}
Q_{\Delta S=2} \, =\, \left[\bar s \gamma^\mu (1 - \gamma_5) d \right]\,
\left[\bar s \gamma_\mu (1 - \gamma_5) d \right] .
\eeq
There are in addition long-distance contributions
from two $\cL_{\mathrm{eff}}^{\Delta S=1}$ insertions.

The box and penguin topologies, shown in Fig.~\ref{fig:PenguinBox},
 introduce contributions proportional to
$\lambda_q$ ($q=u,c,t$) from virtual up-type quark exchanges. The hierarchy of
CKM factors is better visualized through the \textcite{Wolfenstein:1983yz} parametrization:
$\lambda_u \sim \lambda\equiv |V_{us}|$, $\lambda_c \sim -\lambda$ and $\lambda_t\sim -A^2\lambda^5$,
where $A = |V_{cb}|/\lambda^2$. The sum of the three contributions cancels for equal up-type quark masses,
owing to the unitarity relation (\ref{eq:lambda_q}). The quark-mass dependence
of the associated loop functions breaks the GIM cancellation and strongly enhances the relevance
of the top contribution in short-distance dominated processes. This is
especially true for CP-violating effects, which are
proportional to  \cite{Buras:1994ec}
\beq
\mathrm{Im}\,(\lambda_t) = - \mathrm{Im}\,(\lambda_c) \,\approx\, \eta\lambda^5 A^2 .
\eeq
The corresponding CP-conserving factors are given to a very good approximation by
\beqa\label{eq:ReLambda}
\mathrm{Re}(\lambda_c) & = & -\lambda \left(1-\frac{\lambda^2}{2}\right) , \nn
\mathrm{Re}(\lambda_t) & = & -A^2\lambda^5 \left(1-\frac{\lambda^2}{2}\right) (1-\bar\rho).
\eeqa
The value of $\lambda$ can be extracted from $K_{\ell 3}$ decays and is given in Eq.~(\ref{eq:Vus}).
The remaining CKM parameters are obtained from global SM fits to flavor-changing data
\cite{Bona:2006ah,Charles:2004jd}.
We will adopt the values \cite{CKMfit:2010xx}
\beqa\label{eq:CKMvalues}
A &=& 0.812{}^{+0.013}_{-0.027} ,\qquad\quad
\bar\rho \, =\,  0.144\pm 0.025 ,\nn
\bar\eta &\equiv& \left( 1-\frac{\lambda^2}{2}\right) \eta\, =\, 0.342\pm 0.016 .
\eeqa

In order to fully calculate the kaon decay amplitudes, we also
need to know the matrix elements of the operators $Q_i$ between the initial and
final states, which involve non-perturbative dynamics at low energies.
These hadronic matrix elements should cancel the renormalization-scale
dependence of the Wilson coefficients $C_i(\mu)$.
Methods to tackle this challenging task include lattice gauge theory
\cite{Colangelo:2010et,Kim:2010sd,Boucaud:2004aa,Laiho:2010ir,Boyle:2011kn,Goode:2011kb,Liu:2010fb},
the $1/N_C$ expansion \cite{Bardeen:1986uz,Bardeen:1986vz,Bardeen:1987vg,
Hambye:1998sma,Hambye:1999ic,Hambye:1999yy},
QCD sum rules \cite{Guberina:1985md,Pich:1985ab,Pich:1985st,Pich:1987sc,Pich:1995qp,Prades:1991sa,Jamin:1994sv},
functional bosonization techniques
\cite{Pich:1990mw,Bruno:1992za,Hambye:2003cy,Peris:2000sw,Knecht:1998nn,Knecht:2001bc,Friot:2004ba}
and dynamical models \cite{Bijnens:1994cx,Bijnens:1995br,Bijnens:1998ee,Bijnens:1999zn,Bijnens:2000im,Bijnens:2001ps,Bijnens:2006mr,
Bertolini:1997ir,Antonelli:1996qd,Antonelli:1995gw,Antonelli:1995nv,Bertolini:1994qk}.

CHPT provides a solid effective field theory framework to
analyze systematically the long-distance dynamics, which we discuss next. Nevertheless, non-perturbative
techniques remain necessary for matching the effective low-energy theory with the underlying QCD Lagrangian.

\subsection{Chiral perturbation theory}
\label{subsec:long-distance}
Most kaon decays are governed by physics at long distances.
Below the resonance region one can use symmetry considerations to define another
effective field theory in terms of the QCD Goldstone bosons.
CHPT describes \cite{Weinberg:1978kz,Gasser:1984gg}
the pseudoscalar-octet dynamics through a perturbative expansion
in powers of momenta and quark masses over the chiral symmetry-breaking scale
$\Lambda_\chi\sim 1\; {\rm GeV}$.
Chiral symmetry fixes the allowed operators, providing a comprehensive framework for
both semileptonic and nonleptonic kaon decays including
radiative corrections. For general reviews of CHPT, see
\textcite{Ecker:1994gg}, \textcite{Pich:1995bw}, \textcite{deRafael:1995zv},
\textcite{Pich:1998xt}, \textcite{Scherer:2002tk}, \textcite{Gasser:2003cg}
and \textcite{Bijnens:2006zp}. The standard textbook on the SM at
low energies is \textcite{Donoghue:1992dd}.

The effective mesonic chiral
Lagrangian in use today is listed schematically in Table~\ref{tab:Leff}.
The strong chiral Lagrangian relevant for (semi)leptonic decays to
NNLO is given by
\begin{eqnarray}
\cL_\mrm{strong}  &=& \underbrace{\frac{F_0^2}{4} \langle D_\mu U D^\mu
  U^\dagger +  \chi U^\dagger + \chi^\dagger  U \rangle}_{{\cal
  L}_{p^2}(2)}  \nn
&&{} +~  \underbrace{\sum_i L_i \,O^{p^4}_i}_{{\cal L}_{p^4}(10)} ~+~
\underbrace{\cL_\mrm{WZW}}_{{\cal L}_{p^4}^\mrm{odd}(0)}
  \label{eq:Lstrong}  \\
&&{} +~ \underbrace{\sum_i C_i  \,O^{p^6}_i}_{{\cal L}_{p^6}(90)} ~+~
\underbrace{\sum_i C_i^W  \,O^{p^6,\mrm{odd}}_i}_{{\cal L}_{p^6}
^\mrm{odd}(23)} . \nonumber
\end{eqnarray}
$F_0$ is the pion decay constant in the limit of chiral $\rm SU(3)$, the
$\rm SU(3)$ matrix field $U$ contains the pseudoscalar fields, $D_\mu$
is the covariant derivative in the presence of external vector and
axial-vector fields, the
scalar field $\chi$ accounts for explicit chiral symmetry breaking
through the quark masses $m_u, m_d, m_s$, and $\langle \ldots \rangle$
stands for the three-dimensional flavor trace.
Explicit forms of the
higher-order Lagrangians can be found in \textcite{Gasser:1984gg},
\textcite{Wess:1971yu}, \textcite{Witten:1983tw},
\textcite{Bijnens:1999sh,Bijnens:2001bb} and
\textcite{Ebertshauser:2001nj}.
%
\renewcommand{\arraystretch}{1.2}
\begin{center}
\begin{table}[floatfix]
\caption{\label{tab:Leff}
The $\Delta S=1$ chiral Lagrangian describing
  (semi)leptonic and nonleptonic kaon decays. Leptons are
  incorporated for radiative corrections in (semi)leptonic
  decays. The numbers in brackets denote the number of LECs. \\}
{\begin{tabular}{|l|c|}
\hline
 &  \\[-.4cm]
\hspace{1.5cm} ${\cal L}_\mrm{chiral\; order}$
~($\#$ of LECs)  &  \hspace*{.02cm}  loop order \hspace*{.02cm} \\[4pt]
\hline
 &  \\[-.3cm]
\hspace{.03cm} ${\cal L}_{p^2}(2)\, +\,
{\cal L}_{p^4}^\mrm{odd}(0)\, +\,
{\cal L}_{G_8 p^2}^{\Delta S=1}(1)\, +\,
{\cal  L}_{G_{27}p^2}^{\Delta S=1}(1)$  \hspace{.03cm}
& $L=0$
\\[3pt]
\hspace{.4cm}
$\mbox{} +\, {\cal L}_{G_8e^2p^0}^\mrm{emweak}(1) \, +\,
{\cal L}_{e^2p^0}^\mrm{em}(1)\, +\,
{\cal L}_{\mrm{kin}}^\mrm{lepton}(0)$
&
\\[10pt]
\hspace{.4cm}
$\mbox{} +\, {\cal L}_{p^4}(10)\, +\,
{\cal L}_{p^6}^\mrm{odd}(23)\, +\,
{\cal L}_{G_8p^4}^{\Delta S=1}(22)$
\hspace{.1cm}
& $L \le 1$
\\[3pt]
\hspace{.4cm}
$\mbox{} +\, {\cal L}_{G_{27}p^4}^{\Delta S=1}(28)\, +\,
{\cal L}_{G_8e^2p^2}^\mrm{emweak}(14)$ &
\\[3pt]
\hspace{.4cm}
$\mbox{} +\, {\cal L}_{e^2p^2}^\mrm{em}(13)\, +\,
{\cal L}_{e^2p^2}^\mrm{lepton}(5)$
& \\[10pt]
\hspace{.4cm} $\mbox{} +\, {\cal L}_{p^6}(90)$  & $L \le 2$ \\[5pt]
\hline
\end{tabular}}
\end{table}
\end{center}

The nonleptonic weak Lagrangian to NLO is given by
\begin{eqnarray}
\cL_\mrm{weak} &=&  \underbrace{G_8  F_0^4  \,\langle\lambda D^\mu U^\dagger
 D_\mu U \rangle}_{{\cal L}_{G_8 p^2}^{\Delta S=1}(1)} \nn
&&{} +~ \underbrace{G_{27} F_0^4 \left( L_{\mu 23} L^\mu_{11} +
\frac{2}{3} L_{\mu 21} L^\mu_{13}\right)}_{{\cal L}_{G_{27} p^2}^{\Delta
 S=1}(1)} \label{eq:Lweak} \\
&&{} +~    \underbrace{G_8 F_0^2 \sum_i  N_i O^8_i}_{{\cal
 L}_{G_8p^4}^{\Delta S=1}(22)}   +\,
\underbrace{ G_{27} F_0^2 \sum_i D_i  O^{27}_i}_{{\cal
 L}_{G_{27}p^4}^{\Delta S=1}(28)}  +  \,\mrm{h.c.}  \no
\end{eqnarray}
The matrix $L_{\mu}=i U^\dagger D_\mu U$  represents the octet of
$V-A$ currents to lowest order in derivatives;
$\lambda = (\lambda_6 - i \lambda_7)/2$ projects onto the
$\bar s\to \bar d$ transition.
The terms proportional to $G_8$ and $G_{27}$ transform  under chiral
transformations like
$(8_L,1_R)$ and $(27_L,1_R)$, respectively,
providing the most general effective realization of the corresponding
short-distance operators in Eq.~(\ref{eq:Leff}).
The Lagrangians can be found in
\textcite{Cronin:1967jq}, \textcite{Kambor:1989tz} and \textcite{Ecker:1992de}.

To include electromagnetic corrections for both (semi)leptonic and
nonleptonic decays, we also need the chiral Lagrangians
\beq
\cL_\mrm{lepton} \, =\, {\cal L}_{\mrm{kin}}^\mrm{lepton}(0) ~+~
\underbrace{e^2 F_0^2 \sum_i X_i O^\mrm{lepton}_i}_{{\cal
    L}_{e^2p^2}^\mrm{lepton}(5)},
\label{eq:Llepton}
\eeq
\begin{eqnarray}
\cL_\mrm{elm} &=& \underbrace{e^2 Z  F_0^4  \langle Q  U^\dagger Q
  U\rangle}_{{\cal L}_{e^2p^0}^\mrm{em}(1)} ~+~
\underbrace{e^2 G_8  g_\mrm{ewk} F_0^6 \,\langle\lambda U^\dagger Q
U\rangle}_{{\cal L}_{G_8e^2p^0}^\mrm{emweak}(1)} \nn
&&{} +~   \underbrace{e^2 F_0^2  \sum_i K _i O^{e^2 p^2}_i}_{{\cal
  L}_{e^2p^2}^\mrm{em}(13)} \nn
&&{}  +~
\underbrace{e^2  G_8 F_0^4\, \sum_i\; Z_i\, O^{EW}_i}_{{\cal
  L}_{G_8e^2p^2}^\mrm{emweak}(14)}  ~+~  \mrm{h.c.}, \label{eq:Lelweak}
\end{eqnarray}
where $Q$ is the diagonal matrix of quark charges.
The corresponding Lagrangians can be found in \textcite{Ecker:1988te},
\textcite{Bijnens:1983ye}, \textcite{Grinstein:1985ut},
\textcite{Urech:1994hd}, \textcite{Ecker:2000zr} and
\textcite{Knecht:1999ag}.
The terms proportional to $g_\mrm{ewk}$ and $Z_i$ provide the
low-energy realization
of the electromagnetic penguin operators (\ref{eq:Q7-10}),
while those with couplings $X_i$ account for the operators in
Eq.~(\ref{eq:Q11-13}) with explicit lepton fields.

The chiral realization of the $\Delta S=2$ effective Lagrangian
contains a single $\cO(p^2)$ operator:
\beq\label{eq:Leff_DS=2}
\cL_{\mathrm{eff}}^{\Delta S=2}= \frac{G_F^2 M_W^2}{(4\pi)^2}\,
 g_{_{\Delta S=2}} \, F_0^4\, \,\langle\lambda U^\dagger D^\mu U \rangle
 \,\langle\lambda U^\dagger D_\mu U \rangle .
\eeq
The $\cO(p^4)$ operators were discussed by \textcite{Kambor:1989tz}
and \textcite{EspositoFarese:1990yq}.

The CHPT framework determines the $K$ decay amplitudes
in terms of the LECs multiplying
the relevant operators in the chiral Lagrangian. These LECs
encode all information about the short-distance dynamics, while
the chiral operators yield the most general form of the
low-energy amplitudes compatible with chiral symmetry.
Chiral loops generate non-polynomial contributions, with
logarithms and threshold factors as required by unitarity.

Practically all kaon decays discussed in this review have been
calculated at least
up to $\mathcal{O}(p^4)$. The strong and nonleptonic parts of the corresponding
NLO amplitudes can be given in compact form in terms of the generating
functional. The matrix elements and form factors with at most
six external particles (an external photon counts as two particles),
at most one $W$ (semileptonic decays) and with at most three propagators
in the one-loop amplitudes were presented in closed form by
\textcite{Unterdorfer:2005au}. This includes most of the processes
covered in this review. Except
for electromagnetic and anomalous contributions, all amplitudes to
$\mathcal{O}(p^4)$ can be obtained from a Mathematica program written by Ren\'e
Unterdorfer. The code along with several examples can be downloaded
from http://homepage.univie.ac.at/Gerhard.Ecker/CPT-amp.html.

\section{Estimates of low-energy constants}
\label{sec:lecs}

A first-principle calculation of LECs requires to perform the
matching between CHPT and the underlying SM. This is a very difficult
task. In many cases one resorts to
phenomenology to determine the values of the LECs. For instance, most of the
$\cO(p^4)$ couplings of the strong chiral Lagrangian, $L_i$, are rather
well known \cite{Ecker:2007dj,Bijnens:2011tb} from low-energy data
($\pi\pi$ scattering, $\pi$ and $K$ decay constants
and masses, $\pi$ electromagnetic radius,
$\pi\to e\nu\gamma$, $\tau$ decay).
The electromagnetic LEC $Z$ arising at $\cO(e^2p^0)$ can be expressed
in terms of the squared-mass difference of the pions:
\begin{equation}\label{eq:Z}
M_{\pi^\pm}^2 - M_{\pi^0}^2 = 8 \pi \alpha Z F_0^2.
\end{equation}

The matching procedure is of course simpler whenever the hadronic
dynamics gets reduced to quark currents. For instance, in the CHPT language,
the short-distance enhancement of semileptonic decays
is encoded in the factor \cite{Knecht:1999ag,DescotesGenon:2005pw}
\begin{equation}
1 - \frac{e^2}{2} (X_6^r - 4 K_{12}^r) \equiv
1 - \frac{e^2}{2} X_6^{\rm phys},
\label{univfac}
\end{equation}
which is related to $S_{\rm EW}$ in the Lagrangian (\ref{eq:Lsemi}) by
\begin{equation}
e^2 X_6^{\rm phys} (M_\rho) =
 S_{\rm EW} - 1 +
e^2 \tilde{X}_6^{\rm phys} (M_\rho),
\label{matching}
\end{equation}
where
$\tilde{X}_6^{\rm phys} (M_\rho)$ denotes a residual long-distance
contribution \cite{DescotesGenon:2005pw}.

\subsection{The large-$\mathbf{N_C}$ limit of QCD}
\label{susubsec:largeNc}

The limit of an infinite number of quark colors
is a very useful starting point to understand many
features of QCD \cite{'tHooft:1974hx,'tHooft:1973jz,Witten:1979kh}.
Assuming confinement,
the strong dynamics at $N_C\to\infty$ is given
by tree diagrams with infinite sums of hadron exchanges,
which correspond to the tree approximation for some local
effective Lagrangian. Hadronic loops generate corrections
suppressed by factors of $1/N_C$.
Resonance chiral theory \cite{Ecker:1988te,Ecker:1989yg,Cirigliano:2006hb} provides
an appropriate framework to incorporate the massive mesonic states \cite{Pich:2002xy}.
Integrating out the resonance fields, one recovers the usual CHPT
Lagrangian with explicit values for the LECs, parametrized in terms of resonance
masses and couplings.
The resonance chiral theory generates Green functions that interpolate between
QCD and CHPT. Analyzing these Green functions, both for large and small
momenta, one gets QCD constraints on the resonance couplings and,
therefore, information on the LECs.

Truncating the infinite tower of meson resonances to the lowest states
with $0^{-+}$,
$0^{++}$, $1^{--}$ and $1^{++}$ quantum numbers, one gets
a very successful prediction of the $\cO(p^4 N_C)$ strong CHPT couplings
in terms of only three
parameters: $M_V$, $M_S$ and $F_0$. This provides a theoretical
understanding of the role of resonance saturation in low-energy
phenomenology, which was recently extended to
$\cO(p^6)$ \cite{Cirigliano:2006hb}.
Of particular interest for our present purposes are the resulting predictions for
the $\cO(p^6)$ couplings
$C_{12}$ and $C_{34}$, which govern the amount of SU(3) breaking in
the $K_{\ell 3}$ form factor at zero momentum transfer
\cite{Cirigliano:2006hb,Cirigliano:2005xn}:
\beq\label{Lecs-model-2}
C_{12} = - \frac{F_0^2}{8 M_S^4},
\quad
C_{34} =  \frac{3 \, F_0^2}{16 M_S^4} + \frac{F_0^2}{16}
\left(\frac{1}{M_S^2} - \frac{1}{M_P^2} \right)^2.
\eeq

The large-$N_C$ limit turns out to be very useful to analyze the nonleptonic
weak Lagrangian, because the T-product of two color-singlet quark currents
factorizes:
\beq\label{eq:factj}
\langle J \cdot J \rangle \, = \,
\langle J \rangle\;\langle J \rangle\;
\left\{ 1 \, + \,\cO\left(1\over N_C \right)\right\}.
\eeq
Since quark currents have well-known CHPT realizations,
the hadronization of the short-distance operators $Q_i$ can then be done in
a straightforward way. As a result, in this large-$N_C$ framework
the electroweak chiral couplings can be related to strong and
electromagnetic LECs of order $p^2$,  $p^4$, $p^6$ and $e^2p^2$,
respectively.  The lowest-order
electroweak LECs take the following values at large $N_C$
\cite{Pallante:2001he,Cirigliano:2003gt}:
\beqa\label{eq:G8matching}
\lefteqn{g_8^\infty\, =\, -{2\over 5}\,C_1(\mu)+{3\over 5}\,C_2(\mu)+C_4(\mu)
- 16\, L_5 B(\mu)\, C_6(\mu),}\hskip 3.6cm &&
\nn
\lefteqn{g_{27}^\infty\, =\, {3\over 5}\,[C_1(\mu)+C_2(\mu)],}\hskip 3.6cm  &&
\nn
\lefteqn{( e^2 g_8\, g_{\rm ewk})^\infty \, =\, -\, 3\, B(\mu) \, C_8(\mu)}\hskip 3.6cm  &&
\nn &&\hskip -1.58cm\mbox{}
-\frac{16}{3}\, C_6(\mu)\, e^2 (K_9-2 K_{10}) \hskip 1.8cm
\eeqa
and
\beq
g_{_{\Delta S=2}}^\infty\, =\,  C_{\Delta S=2}(\mu) ,
\eeq
where the dimensionless couplings $g_8, g_{27}$ are defined as
\begin{equation}
G_{8,27} = -{G_F \over \sqrt{2}}  V_{ud}^{\phantom{*}} V_{us}^*\;
  g_{8,27}. \label{eq:defgn}
\end{equation}

The operators $Q_i$ ($i\not=6,8$) factorize into products of
left- and right-handed vector currents,
which are renormalization-invariant quantities.
Thus, the large-$N_C$ factorization of these operators
does not generate any scale dependence.
The only anomalous dimensions that survive when $N_C\to\infty$
are the ones corresponding to $Q_6$ and $Q_8$ \cite{Bardeen:1986uz,Buras:1987wc}.
These operators  factorize into color-singlet
scalar and pseudoscalar currents, which are $\mu$ dependent.
The CHPT evaluation of the scalar and pseudoscalar currents provides,
of course, the right $\mu$ dependence, since only physical observables
can be realized in the low-energy theory. What one actually
finds is the chiral realization of the renormalization-invariant
products $m_q \,\bar{q} (1,\gamma_5) q$.
This generates in Eq.~(\ref{eq:G8matching}) the factors
\beqa\label{eq:B0_comp}
B(\mu) & = &
\left[\frac{M_K^2}{(m_s+m_d)(\mu)\, F_\pi}\right]^2
\times
\left\{ 1 + \frac{8 M_\pi^2}{F_\pi^2}\, L_5
\right.\\ &&\hskip -1cm\left.\mbox{}
- \frac{16 M_K^2}{F_\pi^2}\, (2 L_8-L_5)
 + 8 \,\frac{2 M_K^2+M_\pi^2}{F_\pi^2}\, (3 L_4-4 L_6)\right\},\nonumber
\eeqa
which exactly cancel the $\mu$ dependence of
$C_{6,8}(\mu)$ at large $N_C$.
There remains a dependence at NLO.

The large-$N_C$ expressions imply the numerical values
\beqa\label{eq:G8Nc}
g_8^\infty & = & \left( 1.13\pm 0.05_\mu \pm 0.08_{L_5} \pm 0.05_{m_s}\right)\nn
&&\mbox{}+\tau\,\left( 0.64\pm 0.15_\mu \pm 0.20_{L_5} {{}^{+\, 0.25}_{-\, 0.16}}_{m_s}\right),\nn
g_{27}^\infty & = & 0.46\pm 0.01_\mu,\nn
\left(g_8 g_{\mathrm{ewk}}\right)^\infty & = &
\left( -1.60\pm 0.86_\mu \pm 0.25_{K_i} {{}^{+\, 0.57}_{-\, 0.35}}_{m_s}\right)\nn
&&\mbox{}-\tau\,\left( 25.0\pm 4.5_\mu \pm 1.0_{K_i} {{}^{+\, 9.1}_{-\, 5.6}}_{m_s}\right),
\qquad\quad
\eeqa
where $\tau$ is the ratio of CKM matrix elements in Eq.~(\ref{eq:tau})
and the main sources of uncertainty are indicated.

While the CP-even part of $g_8^\infty$ gets contributions from all $(8_L,1_R)$ short-distance operators,
its CP-odd component is completely dominated by the strong penguin contribution,
proportional to $\tau y_6(\mu)$.
The CP-odd component of $g_{\mathrm{ewk}}$ is dominated by the electroweak penguin contribution,
proportional to $\tau y_8(\mu)$, while the CP-even part receives contributions of similar size from both
strong ($Q_6$) and electroweak ($Q_8$) penguin operators.
Explicit predictions for the higher-order electroweak LECs can be found in \textcite{Cirigliano:2003gt}.

It is important to stress that the large-$N_C$ limit is only applied to the matching between
the three-flavor quark theory and CHPT. The evolution from the electroweak scale down to $\mu < m_c$
has to be done without any unnecessary expansion
in powers of $1/N_C$; otherwise, one would miss large corrections
of the form $\ln (M/m) \big/ N_C$, with $M\gg m$ two widely
separated scales.
Similarly, the long-distance rescattering of the final pions in the $K$ decay
generates large logarithmic corrections through chiral loops,
which are of higher order in both the momentum
and $1/N_C$ expansions \cite{Pallante:2001he}.
These next-to-leading contributions, which give rise to
the large s-wave $\pi\pi$ strong phases, are rigorously incorporated through
the CHPT framework.

\subsection{Lattice determinations}
\label{subsubsec:Lattice}
Lattice results for strong LO and NLO LECs, both for $\rm SU(2)$ and for
$\rm SU(3)$, have recently been summarized by
the Flavianet Lattice Averaging Group --FLAG-- \cite{Colangelo:2010et}.
We refer to this comprehensive review for a detailed discussion.

\section{Leptonic and semileptonic decays}
\label{sec:semilep}

Purely leptonic and semileptonic  modes are among the theoretically
cleanest  $K$ decays.
Using  CHPT and lattice QCD,  $K_{\ell 2}$ and
$K_{\ell 3}$  decay rates
can be predicted with high accuracy and provide non-trivial tests  of
the SM and its extensions.
On the other hand, $K_{\ell 4}$ decays can be used as probes of chiral
dynamics in
$\pi \pi$ scattering.
In this section we review the current status of
$ K \to \ell \nu$ ($K_{\ell 2}$),  $K \to \pi \ell \nu$ ($K_{\ell 3}$),
and $K \to \pi \pi \ell \nu$ ($K_{\ell 4}$)  decays, as well as their
radiative counterparts.

\subsection{$K_{\ell 2}$ (and $\pi_{\ell 2}$)
decays}
\label{subsubsec:Pl2}

\subsubsection{Electromagnetic corrections}

The discussion of electromagnetic contributions to $K_{\ell 2}$ decays
serves as the simplest example of the treatment of
electromagnetism in (semi)leptonic processes.
We start with the parametrization of the
inclusive $P \to \ell \nu_\ell
(\gamma)$ decay rate proposed by
\textcite{Cirigliano:2007ga,Cirigliano:2007xi}  (here $P=K^\pm ,\pi^\pm$),
\begin{eqnarray}
\lefteqn{\Gamma_{P_{\ell 2 (\gamma)}} =
\Gamma^{(0)}_{P_{\ell 2}}\, S_{\rm EW}\,
\Bigg\{
1 + \frac{\alpha}{\pi}  \,  F (m_\ell/M_P)
\Bigg\} }\quad && \\
 &\times &{} \Bigg\{  1 - \frac{\alpha}{\pi}
\bigg[
\frac{3}{2}  \ln \frac{M_\rho}{M_P} +  c_1^{(P)}
 -  \frac{M_P^2}{M_\rho^2} \,  \tilde{c}_{2}^{(P)}  \,
\ln \frac{M_\rho^2}{m_\ell^2}
 \nn
&&{} +~
\frac{m_\ell^2}{M_\rho^2}\,
\bigg( c_2^{(P)}  \, \ln \frac{M_\rho^2}{m_\ell^2}
+  c_3^{(P)}
+ c_4^{(P)}\! (m_\ell/M_P) \bigg) \bigg]
\Bigg\}, \nonumber
\label{GPl2full}
\end{eqnarray}
\noindent
which is a modified version of the expression given in
\textcite{Marciano:1993sh}.
The decay rate in the absence of radiative corrections is given by
\begin{equation}
\Gamma^{(0)}_{P_{\ell 2}}
=
\frac{G_F^2 |V_P|^2  F_P^2 }{4
\pi} \,
M_P  \, m_\ell^2  \, \left(1 - \frac{m_\ell^2}{M_P^2} \right)^2,
\label{GPL20}
\end{equation}
where $V_\pi = V_{ud}$, $V_K = V_{us}$
and $F_{\pi^\pm}, F_{K^\pm}$ denote the pseudoscalar decay constants in pure
QCD including strong isospin breaking.\footnote{
For a discussion of subtleties
involved in this separation, see \textcite{Gasser:2010wz}.
}
The first term in curly brackets is the  universal
long-distance correction for a point-like meson. The explicit form of the
one-loop function $F(x)$ can be found in \textcite{Marciano:1993sh}. The
structure-dependent coefficients $c_1^{(P)}$ are independent of the lepton
mass $m_{\ell}$ and start at $\cO(e^2 p^2)$ in CHPT.
The other coefficients appear only at higher orders in
the chiral expansion. The one-loop result (order $e^2 p^2$) for
$c_1^{(P)}$ was given by \textcite{Knecht:1999ag},
\begin{eqnarray}
c_1^{(\pi)} \! \!
&=& \! \!
- \tilde E^r(M_\rho) +
\frac{Z}{4} \bigg(3 +2 \ln \frac{M_\pi^2}{M_\rho^2} + \ln
\frac{M_K^2}{M_\rho^2} \bigg),
\nn
c_1^{(K)} \! \!
&=& \! \!
- \tilde E^r(M_\rho)  +
\frac{Z}{4} \Big(3 +2 \ln \frac{M_K^2}{M_\rho^2} + \ln
\frac{M_\pi^2}{M_\rho^2} \Big), \quad
\label{c1}
\end{eqnarray}
where $Z$ is the $\cO(e^2p^0)$ electromagnetic coupling given in
Eqs.~(\ref{eq:Lelweak}) and (\ref{eq:Z}).
$\tilde E^r(M_\rho)$ is a certain linear combination of
LECs appearing in the Lagrangians $\cL_{e^2 p^2}^{\rm em}$
and $\cL_{e^2 p^2}^{\rm lepton}$:
\beqa
\tilde E^r &=& \frac{1}{2}
+ 4 \pi^2 \left(
\frac{8}{3} K_1^r
+ \frac{8}{3} K_2^r
+ \frac{20}{9} K_5^r
+ \frac{20}{9} K_6^r
\right.\nn &&\hskip 1.3cm\left.\mbox{}
-\frac{4}{3} X_1^r
- 4 X_2^r
+ 4 X_3^r
- \tilde{X}_6^{\rm phys}\right). \qquad\mbox{}
\label{Er}
\eeqa

The results in Eq.~(\ref{c1}) are a nice example demonstrating the power
of effective field theory methods. We see how the electromagnetic
corrections to $\pi_{\ell 2}$ and $K_{\ell 2}$ of $\cO(e^2p^2)$
are related. In particular,
$c_1^{(\pi)}$ and $c_1^{(K)}$ contain the same combination of LECs
$\tilde E^r$. Taking the  ratio
$\Gamma_{K_{\ell 2 (\gamma)}} / \Gamma_{\pi_{\ell 2 (\gamma)}}$,
the coupling constant $\tilde E^r$ cancels and the remaining expression
\begin{equation}
c_1^{(K)} - c_1^{(\pi)} = \frac{Z}{4} \ln \frac{M_K^2}{M_\pi^2}
\end{equation}
is uniquely determined in terms of measurable quantities
\cite{Knecht:1999ag}.

Finally, we note that using the matching
calculation of  \textcite{DescotesGenon:2005pw} for the LEC
combination $\tilde E^r (M_\rho)$,
one obtains
\begin{equation}
c_1^{(\pi)} =   -2.4 \pm 0.5, \qquad
c_1^{(K)} =   -1.9  \pm 0.5,
\end{equation}
where the errors given here are based on naive power counting of unknown
contributions arising at $\cO(e^2p^4)$.

\subsubsection{Extraction of $V_{us}/V_{ud}$}

As suggested by  \textcite{Marciano:2004uf}, a determination of
$|V_{us} / V_{ud}|$ can be obtained by
combining the experimental values for the decay rates
$K \to \mu \nu (\gamma)$ and
$\pi  \to \mu \nu (\gamma)$
with the lattice determination of $F_K / F_\pi$,
which is currently  performed  in the isospin limit  $(m_u= m_d$)
of QCD.
The relation to be used to extract $|V_{us}/V_{ud}|$ reads
\begin{eqnarray}
 \frac{\Gamma_{K_{\ell 2 (\gamma)}}}
{\Gamma_{\pi_{\ell 2 (\gamma)}}}  &=&
\frac{|V_{us}|^2} {|V_{ud}|^2}
 \frac{F_K^2 }{F_\pi^2} \,
\frac{
M_{K^\pm} ( 1 - m_\ell^2/M_{K^\pm}^2)^2
}{
M_{\pi^\pm}  ( 1 - m_\ell^2/M_{\pi^\pm}^2)^2
}
\nonumber \\
&& {} \times   \Big(
1 + \delta_{\rm EM} + \delta_{\rm SU(2)}
\Big),
\end{eqnarray}
where $F_\pi$ and  $F_K$ denote the decay constants in the isospin limit,
$\delta_{\rm EM}$  is the long-distance electromagnetic correction
(the short-distance part cancels in the ratio) and
the strong isospin-breaking correction $\delta_{\rm SU(2)}$
is defined by
\begin{equation}
 \frac{F_{K^\pm}^2 }{F_{\pi^\pm}^2} =
\frac{F_K^2 }{F_\pi^2} \, \left( 1 + \delta_{\rm SU(2)}\right) .
\end{equation}
The electromagnetic correction is given by
\cite{Knecht:1999ag,Cirigliano:2011tm}
\begin{eqnarray} \label{deltaEM}
\delta_{\mathrm {EM}} &=&
\frac{\alpha}{\pi}
\left(
F(m_\ell/M_K) - F(m_\ell/M_\pi)
+ \frac{3-Z}{4} \ln \frac{M_K^2}{M_\pi^2}
\right)
\nonumber
\\
&=& -0.0069 \pm 0.0017.
\end{eqnarray}
The $25\%$  uncertainty of the numerical value is an estimate of
corrections arising to higher order in the chiral expansion.
The correction parameter $\delta_{\rm SU(2)}$ reads
\cite{Cirigliano:2011tm}
\begin{eqnarray} \label{deltaSU2v2}
\delta_{\mathrm {SU(2)}} &=&
\sqrt{3} \, \varepsilon^{(2)}
\left[ -\frac{4}{3} \left( F_K / F_\pi -1 \right) \right.
\\
&& {} \left.\mbox{} +\frac{1}{3(4\pi)^2F_0^2}
\left(
M^2_{K} -M_\pi^2 - M_\pi^2 \ln \frac{M^2_K}{M_\pi^2}
\right) \right], \nonumber
\end{eqnarray}
where
\begin{equation} \label{defeps}
\varepsilon^{(2)} = \frac{\sqrt{3}}{4 R}, \quad
R = \frac{m_s - \widehat{m}}{m_d - m_u}, \quad
\widehat{m}=\frac{m_u + m_d}{2}.
\end{equation}
With the FLAG \cite{Colangelo:2010et} averages of
lattice calculations with $N_f = 2 + 1$ dynamical fermions,\footnote{
Based on the results from~\textcite{Aubin:2004fs,Bernard:2007ps,Bazavov:2009fk,Bazavov:2009bb,
Follana:2007uv,Aubin:2008ie,Lellouch:2009fg,Mawhinney:2009jy,Beane:2006kx,Blossier:2009bx,
Gockeler:2006ns,Durr:2010hr,Aoki:2008sm,Noaki:2009sk,Allton:2008pn}.}
\begin{equation}
R = 36.6 \pm 3.8,
\qquad
F_K / F_\pi = 1.193 \pm 0.006,
\label{eq:flag}
\end{equation}
\textcite{Cirigliano:2011tm} obtained
\begin{equation} \label{deltaSU2num}
\delta_{\mathrm {SU(2)}} =  - 0.0044 \pm 0.0005 \pm
  0.0011_{\mathrm {higher \, \, orders}},
\end{equation}
where the uncertainty due to higher-order
corrections in the chiral expansion was estimated
to be at a level of $25 \%$.
We note that the strong isospin-breaking correction
$\delta_{\mathrm {SU(2)}}$   is of the same
order of magnitude as the electromagnetic
correction $\delta_{\mathrm {EM}}$
 in Eq.~(\ref{deltaEM}) and should not
be neglected in the extraction of the ratio  $|V_{us}/V_{ud}|$.

Combined with the measured 
values for the
leptonic widths of the pion
\cite{Nakamura:2010zzb}
\begin{equation}
\Gamma_{\pi_{\mu 2 (\gamma)}} = 38.408 \pm 0.007 \ (\mu s)^{-1}
\end{equation}
and of the kaon \cite{Antonelli:2009ws}
\begin{equation}
\Gamma_{K_{\mu 2 (\gamma)}} = 51.25 \pm 0.16 \ (\mu s)^{-1},
\end{equation}
\textcite{Cirigliano:2011tm} obtained
\begin{eqnarray}
\frac{|V_{us}| F_K}{|V_{ud}| F_\pi}  & = &  0.23922\, (25)  \times
\Bigg( \frac{\Gamma_{K_{\ell 2 (\gamma)}}}
{\Gamma_{\pi_{\ell 2 (\gamma)}}}
\Bigg)^{1/2} \nonumber \\
&=&   0.2763\pm 0.0005.
\end{eqnarray}
Finally, taking as reference value for $F_K/F_\pi$  the FLAG
average in Eq.~(\ref{eq:flag}),
the ratio of the CKM matrix elements is given by \cite{Cirigliano:2011tm}
\beq
\frac{|V_{us}|}{|V_{ud}|} =  0.2316 \pm 0.0012.
\label{eq:VusKl2}
\eeq

\subsubsection{The ratio $R_{e/\mu}^{(K,\pi)}$}

In a first systematic calculation to $\cO(e^2 p^4)$, the
coefficients
$c_2^{(P)}, \,
c_3^{(P)}, \,
c_4^{(P)}, \,
\tilde{c}_2^{(P)}$
were determined by
\textcite{Cirigliano:2007ga,Cirigliano:2007xi}.
This allowed the determination of the ratios
$R_{e / \mu}^{(P)} =
\Gamma_{P_{e 2 (\gamma)}} /
\Gamma_{ P_{\mu 2 (\gamma)}}$   ($P=\pi, K$)
with an unprecedented theoretical accuracy.
In the SM, the ratios
$R^{(P)}_{e/\mu}$
are helicity suppressed as a consequence of the $V-A$ structure of the
charged currents, constituting sensitive probes of New Physics.
The two-loop effective theory results were complemented by a large-$N_C$
calculation of an associated counterterm and a
summation of leading logarithms $\alpha^n \ln^n (m_\mu/m_e)$
\cite{Marciano:1993sh}
giving \cite{Cirigliano:2007ga,Cirigliano:2007xi}
\beqa
R^{(\pi)}_{e/\mu} &=& (1.2352 \pm 0.0001) \times 10^{-4}, \nn
R^{(K)}_{e/\mu} &=& (2.477 \pm 0.001) \times 10^{-5}.
\eeqa
In the case of $R_{e/\mu}^{(K)}$ the uncertainty arising from matching
was increased
by a factor of four  to account for higher-order chiral corrections  of
$\cO(e^2 p^6)$.
The central value of  $R^{(\pi)}_{e/\mu}$
is in agreement with the results of
previous calculations
\cite{Marciano:1993sh,Finkemeier:1995gi}, pushing the theoretical
uncertainty below $10^{-4}$. The discrepancy with a
previous determination of
$R^{(K)}_{e/\mu}$ can be traced back to inconsistencies in the analysis of
\textcite{Finkemeier:1995gi}.

The above theoretical results are compatible with current experimental
measurements~\cite{Britton:1992pg,Britton:1993cj,Czapek:1993kc,:2009rv,:2011uv}
averaging to
\beqa
R^{(\pi)}_{e/\mu} \Big|_{\rm exp} &=& (1.230 \pm 0.004) \times 10^{-4}, \nn
R^{(K)}_{e/\mu} \Big|_{\rm exp}  &=& (2.488 \pm 0.012) \times 10^{-5}.
\eeqa
They  provide a clean basis to detect or
constrain non-standard physics in these
channels  by  comparison with upcoming measurements, which will push the
 fractional uncertainty
 from $0.3 \%$ to $0.05 \%$  in $ R^{(\pi)}_{e/\mu}$
\cite{Pocanic:2009gd,Sher:2009zz}
and  from $0.5 \%$ to $0.4 \%$  in $R^{(K)}_{e/\mu}$ \cite{Collazuol:2009zz}.

\subsection{$K_{\ell 2 \gamma}$}
\label{subsect:kl2g}

The radiative leptonic decays $K_{\ell 2 \gamma}$,
\begin{equation}
K^+(p) \to  \ell^+(p_\ell) \nu_\ell (p_\nu) \gamma (q),   \qquad \qquad \ell = e, \mu
\label{eq:kl2g}
\end{equation}
allow one to probe the low-energy structure of QCD and its anomalous couplings.
The matrix element reads
\begin{equation}
T = -  i e G_F V_{us}^{*}  \, \epsilon_{\mu}^{*}  \, \left( F_K L^\mu
-  H^{\mu \nu} l_{\nu} \right)   \label{eq:kl2gamp}
\end{equation}
where $\epsilon^\mu$ denotes the photon polarization vector and the
other quantities are
given by  ($W = p - q$)
\begin{eqnarray}
L^\mu  & = &  m_\ell\, \bar{u} (p_\nu) (1 + \gamma_5) \left(
\frac{2 p^\mu}{2 p \cdot q} - \frac{2 p_\ell^\mu  + \slashed{q} \gamma^\mu}{2 p_\ell \cdot q}
\right)  v (p_\ell),
\nonumber \\
l^\mu &=& \bar{u} (p_\nu) \gamma^\mu (1 - \gamma_5) v (p_\ell),
 \\
H^{\mu \nu} &=&
- \frac{1}{\sqrt{2} M_K} \Big[ i V (W^2)\, \epsilon^{\mu \nu \alpha \beta} q_\alpha p_\beta
\nonumber \\
&&\hskip 1.5cm\mbox{} -
A (W^2)\, (q\cdot W g^{\mu \nu} - W^{\mu} q^\nu) \, \Big].\nonumber
\end{eqnarray}
$V(W^2)$ and  $A(W^2)$ denote the
vector and axial-vector\footnote{Note that we are
using here  the normalization of $V$ and $A$ adopted in
\textcite{Nakamura:2010zzb}.}
Lorentz-invariant form factors characterizing
the general decomposition of the correlator of  weak and electromagnetic
currents,
\begin{eqnarray}
V_\mu^{\rm w} &=& \bar{s} \gamma_\mu u, \qquad A_\mu^{\rm w} = \bar{s} \gamma_\mu \gamma_5 u,
\nonumber \\
V_\mu^{\rm em} &=& (2 \bar{u} u - \bar{d} d - \bar{s} s)/3,
\label{eq:wemcurrents}
\end{eqnarray}
between the vacuum and the one-kaon state
at $q^2 = 0$, i.e., for real photons \cite{Bijnens:1992en}:
\begin{eqnarray}
\label{eq:correlator}
\Pi_{\mu \nu}(q,p) &=& \\
 && \hspace*{-1.5cm}  \int \! d^4x  \, e^{i q x}
\langle 0 | T \left( V_\mu^{\rm em}(x) ( V_\nu^{\rm w}(0) -
A_\nu^{\rm w}(0)) \right) | K^+ (p) \rangle .\no
\end{eqnarray}
The explicit relation of $A(W^2)$ and $V(W^2)$ to the above correlator
can be found in the appendix of \textcite{Bijnens:1992en}.
The amplitude (\ref{eq:kl2gamp}) is the sum of two terms: the so-called
\lq\lq inner-bremsstrahlung"  (IB)
term  (proportional to $L^\mu$), uniquely determined in terms of the
non-radiative amplitude,  and the \lq\lq structure-dependent"
(SD) term proportional
to $H^{\mu \nu}$.  Correspondingly, the spin-averaged
differential decay distribution can be decomposed
into three terms (IB, SD,  and INT, the latter
denoting the interference of IB and SD amplitudes):
\begin{equation}
\label{eq:dgdxy}
\begin{split}
& \frac{d^2 \Gamma}{dx dy} = A_{\mathrm{IB}}  \, f_{\mathrm{IB}} (x,y)
\\
& +
\frac{A_{\mathrm{SD}}}{2}  \left[ \left(V+A\right)^2  f_{\mathrm{SD^+}} (x,y) +
\left(V-A\right)^2  f_{\mathrm{SD^-}} (x,y) \right]
\\
&- \frac{A_{\mathrm{INT}}}{\sqrt{2}}  \left[ \left(V+A\right)  f_{\mathrm{INT^+}} (x,y) +
\left(V-A\right)  f_{\mathrm{INT^-}} (x,y) \right].
\end{split}
\end{equation}
Here
\begin{eqnarray}
A_{\mathrm{IB}} &=& 4 \frac{m_\ell^2}{M_K^2}
\left(\frac{F_K}{M_K}\right)^2  \, A_{\mathrm{SD}},   \nn
A_{\mathrm{SD}} &=& \frac{G_F^2 |V_{us}|^2  \alpha}{32 \pi^2}  M_K^5,  \nn
A_{\mathrm{INT}} &=&  4 \frac{m_\ell^2}{M_K^2}
\left(\frac{F_K}{M_K}\right)  \, A_{\mathrm{SD}},
\end{eqnarray}
the independent kinematical variables are
\begin{equation}
x = \frac{2 p \cdot q}{M_K^2},  \qquad
y = \frac{2 p \cdot p_\ell}{M_K^2},
\end{equation}
and the functions $f_{\mathrm{IB}}$, $f_{\mathrm{SD^\pm}}$, and
$f_{\mathrm{INT^\pm}}$ can be found in  \textcite{Bijnens:1992en}.
Note that the inner-bremsstrahlung and interference terms in the
rate
(\ref{eq:dgdxy})  are proportional to the
helicity suppression factor $(m_\ell/M_K)^2$,
so that $K_{e2\gamma}$ is dominated by the structure-dependent
contribution.
On the other hand, $K_{\mu 2 \gamma}$,
while sensitive to interference and structure-dependent terms,
is dominated by internal bremsstrahlung.

Internal bremsstrahlung is fixed in terms of a single hadronic
input, namely
the kaon decay constant  $F_K$.
The form factors $V(W^2)$ and  $A(W^2)$ can be calculated in CHPT,
with  $V(W^2)$ arising from the anomalous sector.
The first non-trivial contributions arise at
$\cO(p^4)$~\cite{Bijnens:1992en}:
\begin{eqnarray}
A_{p^4}  &=& \frac{4 \sqrt{2} M_K}{F_0}  (L_9^r + L_{10}^r)  = 0.042, \nn
V_{p^4}  &=& \frac{\sqrt{2} M_K}{8 \pi^2 F_0} = 0.096,
\end{eqnarray}
where the numerical values were obtained by using the central values
$F_0 = F_\pi$, $L_9^r (\mu = M_\rho) = 6.9 \times 10^{-3}$, and
$L_{10}^r (\mu = M_\rho) = - 5.5 \times 10^{-3}$.

The contributions of $\cO(p^6)$ to the vector form factor
involve  one-loop
graphs in the anomalous sector as well as $\mathcal{O}(p^6)$ counterterms.
They were worked out first by \textcite{Ametller:1993hg} and
later confirmed  by \textcite{Geng:2003mt}. To this order, $V(W^2)$
acquires a non-trivial dependence on
$W^2 = M_K^2 (1 - x)$, which can be approximated by the
linear parametrization
\begin{equation}
V(W^2) = V_0\,  \Big[1 + \tilde\lambda (1-x) \Big].
\label{eq:Vlinear}
\end{equation}
\textcite{Geng:2003mt} use estimates of the LECs in the anomalous sector
from vector meson dominance and the chiral quark model.
Their results can be summarized as:
\begin{eqnarray}
(V_0)_{p^4 + p^6} & = & 0.078  \pm 0.005,
\\
\tilde\lambda &=& 0.3 \pm 0.1 .
\end{eqnarray}
Our estimate of the uncertainty was
obtained by looking at the spread in results  obtained
within different models.

The contributions of $\cO(p^6)$ to $A(W^2)$ were worked
out first by
\textcite{Geng:2003mt}. They involve  two-loop graphs with insertion
of the lowest-order CHPT Lagrangian ${\cal L}_2$,  one-loop graphs
with one insertion of ${\cal L}_4$, and tree-level graphs from ${\cal L}_6$.
\textcite{Geng:2003mt} find that $A(W^2)$ is essentially flat in $W^2$
and receives a correction of about $20\%$ in the normalization, with an
uncertainty induced by the LECs at the percent level [the subscript $0$
indicates that we quote the $W$-independent part of $A(W^2)$]:
\begin{equation}
(A_0)_{p^4 + p^6}  =  0.034.
\end{equation}
The $\mathcal{O}(p^6)$ corrections  are of the expected size for both $V$ and
$A$ form factors.

Experimentally, the  $K_{e 2\gamma}$ mode provides the best constraint
on the combination
$V_0 + A_0$~\cite{Nakamura:2010zzb}. The data are dominated by the
$1484 \pm 63$ events collected by
KLOE~\cite{:2009rv}  in the range $10 ~{\rm MeV} < E_\gamma^{\rm cms}
< 250 ~{\rm MeV}$.
A fit to the measured spectrum leads to the results
\begin{eqnarray}
V_0 + A_0 & = & 0.125 \pm 0.007_{\rm stat} \pm 0.001_{\rm syst},
\nn
\tilde\lambda &=& 0.38 \pm 0.20_{\rm stat} \pm 0.02_{\rm syst},
\end{eqnarray}
which are in excellent agreement with the $\cO(p^6)$
CHPT predictions [from the work of
\textcite{Geng:2003mt} we extract
$(V_0 + A_0)_{p^4 + p^6} = 0.112 \pm 0.005$, where
the uncertainty captures the model dependence in the
LECs entering the $p^6$ calculation].
However, the $K_{\mu 2\gamma}$
extraction of the combination $V_0 + A_0$
differs from the  $K_{e 2\gamma}$ value \cite{Adler:2000vk}:
\begin{equation}
V_0 + A_0  = 0.165 \pm 0.007_{\rm stat} \pm 0.011_{\rm syst} ,
\end{equation}
assuming $V + A = V_0 + A_0$, i.e., no dependence on $W^2$.

Until very recently, the combination $V_0-A_0$ was only loosely constrained by
$K_{\ell 2 \gamma}$ measurements, the best determination,
$V_0-A_0 = 0.077 \pm 0.028$
(consistent with CHPT), coming from  $K \to \mu \nu e^+ e^-$~\cite{Poblaguev:2002ug}.
The situation has changed with the recent results from ISTRA+
reported in \textcite{Tchikilev:2010wy} and \textcite{Duk:2010bs}.
They studied the $K_{\mu 2 \gamma}$ mode in a kinematic region
in which the interference terms can be extracted, thus providing
direct
sensitivity to the combination $V_0-A_0$.
Different analyses assuming constant form factors lead to
\cite{Tchikilev:2010wy}
\begin{equation}
V_0 - A_0 = 0.126 \pm 0.027_{\rm stat} \pm 0.047_{\rm syst},
\end{equation}
and \cite{Duk:2010bs}
\begin{equation}
V_0 - A_0 = 0.21 \pm 0.04_{\rm stat} \pm 0.04_{\rm syst},
\end{equation}
respectively, about $1.5$ and $3$ sigmas above  the CHPT prediction
$V_0 - A_0 = 0.044 \pm 0.010$, where we take as uncertainty the difference
between the central values at $\mathcal{O}(p^6)$ and $\mathcal{O}(p^4)$.
At this stage it is clearly premature to claim a serious tension
between data and theory.

\subsection{$K_{\ell 3}$}
\label{subsubsec:Kl3}

The photon-inclusive decay rate for all four $K\to\pi\ell\nu$ modes
($K = K^\pm, K^0, \overline{K^0}; \, \ell = e, \mu$) can be written as
\beqa \label{Kl3}
\Gamma_{K_{\ell 3 (\gamma)}}  &=&
 \frac{G_{\rm F}^2 \, |V_{us}|^2 \, M_K^5 \, C_K^2 }{128 \, \pi^3} \,
S_{\rm EW}  \,  |f^{K^0  \pi^-}_{+}\!(0) |^2
\nn &&\mbox{}\times \,
 I^{(0)}_{K \ell} (\lambda_i) \,
 \left( 1 +
\delta_{\rm EM}^{K\ell} +
\delta_{\rm SU(2)}^{K \pi}
\right).
\eeqa
The
Clebsch-Gordan coefficient $C_K$
differs for neutral and charged kaons
($C_K = 1$ for $K^0_{\ell 3}$ and $C_K =
1/\sqrt{2}$ for $K^+_{\ell 3}$), while $I^{(0)}_{K \ell}$ is a phase space
integral depending on slope and curvature of the form
factors $f_\pm^{K\pi} (t)$,
defined by the QCD matrix elements
\beqa\label{eq:hadronic element}
\lefteqn{\langle \pi(p_\pi) | \bar{s}\gamma_{\mu}u | K(p_K)\rangle =}
&&\\ && \hskip 1.2cm
(p_\pi+p_K)_\mu\  f^{K \pi}_+ (t) + (p_K-p_\pi)_\mu\  f_-^{K \pi } (t),
\nonumber
\eeqa
where $t=(p_K-p_\pi)^2= (p_\ell+p_\nu)^2$.
As usual, the vector form factor of the $K^0$ decay at zero
momentum transfer has been pulled out in Eq.~(\ref{Kl3}).
The strong isospin-breaking correction is defined as
\begin{equation}
\delta_{\rm SU(2)}^{K \pi} =
\big( f_+^{K \pi}(0) / f_+^{K^0 \pi^-}(0) \big)^2 -1.
\end{equation}
The long-distance electromagnetic corrections
\begin{equation}
\delta_{\rm EM}^{K \ell} =
\delta_{\rm EM}^{K \ell}(\mathcal{D}_3) +
\delta_{\rm EM}^{K \ell}(\mathcal{D}_{4-3})
\end{equation}
receive contributions from three-particle and four-particle final states.
In the following sections we review  the theoretical quantities appearing
in the expression of the decay rate and then combine this information
with experimental input to extract the CKM element $V_{us}$.

\subsubsection{Electromagnetic effects in $K_{\ell 3 (\gamma)}$ decays}

The  calculation of the electromagnetic contributions to $\mathcal{O}(e^2 p^2)$ in
$K_{\ell 3}$ decays using the methods of CHPT with
virtual photons and leptons \cite{Knecht:1999ag}
was presented in \textcite{Cirigliano:2001mk}. Based
on this
analysis, full numerical results on the $K_{e 3}$ decay modes were given in
\textcite{Cirigliano:2004pv}, adopting a specific prescription for treating
real photon emission and a specific factorization scheme for soft photons.
This approach resulted in the partial inclusion of higher-order terms in
the chiral expansion.
In a recent publication \cite{Cirigliano:2008wn}, the
numerical
analysis of electromagnetic corrections was extended to a complete study
of $K_{\mu 3}$ decays. At the same time, previous results for the
$K_{e3}$ modes were updated using the new estimates
of electromagnetic LECs
\cite{Ananthanarayan:2004qk,DescotesGenon:2005pw}
that affect the structure-dependent electromagnetic
contributions. Rather than using the soft-photon factorization procedure of
\textcite{Cirigliano:2001mk}, the analysis was performed at fixed chiral
order $e^2 p^2$.

Table \ref{tab1} summarizes the numerical results for the long-distance
radiative corrections. Two characteristic features can be understood by
qualitative arguments. Firstly, the electromagnetic corrections for the
neutral $K$ decays are expected to be positive and sizable on account of
the final-state Coulomb interaction between $\ell^+$ and $\pi^-$
producing a correction factor of $\pi \alpha / v^{\rm rel}_{\ell^+ \pi^-}
\sim 2 \%$ over most of the Dalitz plot. While the exact correction and
the relative size of $K^0_{\mu 3}$ and $K^0_{e 3}$ depend on other effects
such as the emission of real photons, the qualitative expectation based on
the Coulomb interaction is confirmed by the detailed calculation.
Secondly,  the large hierarchy
$\delta^{K \mu}_{\rm EM} ({\cal D}_{4-3}) \ll \delta^{K e}_{\rm EM}
({\cal D}_{4-3})$
admits a simple interpretation in terms of bremsstrahlung
off the charged lepton
in the final state.  The probability of emitting soft photons is a
function of the lepton velocity $v_{\ell}$
which becomes logarithmically singular as $v_{\ell} \to 1$, thus enhancing
the electron emission.
For typical values of $v_{\ell}$ in ${\cal D}_{4-3}$,  the
semiclassical emission probability implies
$\delta^{K e}_{\rm EM} ({\cal D}_{4-3}) / \delta^{K \mu}_{\rm EM}
({\cal D}_{4-3}) \sim 20 \to 40$. \\

\begin{table}[floatfix]
\begin{center}

\caption{\label{tab1} Electromagnetic corrections to
$K_{\ell 3}$ decay rates \cite{Cirigliano:2008wn}.
$\mathcal{D}_3$ refers to
three-particle phase space, $\mathcal{D}_4$ to four-body kinematics. \\}

{\begin{tabular}{|c|c|c|c|}
\hline &&&\\[-.4cm]
  &   $\;\delta_{\rm EM}^{K \ell} ({\cal D}_3) (\%)\;$  &
   $\;\delta_{\rm EM}^{K \ell} ({\cal D}_{4-3}) (\%)\; $   &
$\delta_{\rm EM}^{K \ell}  (\%)$  \\[4pt]
\hline &&&\\[-.3cm]
$K^0_{e3} $ & 0.50  &  0.49 &  0.99 $\pm $ 0.22 \\
$K^\pm_{e3}$  & $-0.35$ &  0.45  & 0.10 $\pm $ 0.25  \\
$K^0_{\mu 3}$  & 1.38  &   0.02  & 1.40 $\pm $ 0.22  \\
$\; K^\pm_{\mu 3}\;$  & 0.007 & 0.009 & 0.016 $\pm $ 0.25  \\[5pt]
\hline
\end{tabular}}

\end{center}
\end{table}

The theoretical uncertainties assigned to the $\delta^{K
\ell}_{\rm EM}$ in Table \ref{tab1} arise from two sources: the input
parameters
(LECs and form factor parameters) used in the calculation and
unknown higher-order terms in the chiral expansion (the latter would
require a complete analysis at order $e^2 p^4$).
For a detailed discussion of the error estimate, we
refer to \textcite{Cirigliano:2008wn}.

The differential decay distribution can be written in the form
\beqa
\frac{d^2 \Gamma}{dy \, d z} &=&
   \frac{G_{\rm F}^2 \, |V_{us}|^2 \, M_K^5 \, C_K^2 }{128 \, \pi^3}
 \, S_{\rm EW}  \,  |f^{K \pi}_{+}(0) |^2
 \nn &&\times\quad
\, \Big[  \bar{\rho}^{(0)} (y,z)  \ + \ \delta \bar{\rho}^{\rm EM} (y,z)
\Big],
\label{differ}
\eeqa
where the Lorentz invariants $y = 2 p_K \cdot p_\ell / M_K^2 =
2 E_{\ell} / M_K$ and $z = 2 p_\pi \cdot p_K / M_K^2 = 2 E_{\pi} / M_K$
are related to the energy of the charged lepton and of the pion,
respectively, measured in the rest frame of the kaon.
Here $\bar{\rho}^{(0)} (y,z)$ represents the Dalitz plot density in absence of
electromagnetic corrections, while $\delta \bar{\rho}^{\rm EM} (y,z)$ accounts for electromagnetic effects.
Illustrative figures displaying the relative size of the electromagnetic
corrections over the Dalitz plot can be found in
\textcite{Cirigliano:2008wn}.
It turns out that the corrections to the Dalitz distributions can
be locally quite large ($\sim 10 \%$) and do not have a definite sign,
implying cancellations in the integrated total electromagnetic
corrections.

\subsubsection{Quark mass ratios and $K_{\ell 3}$ decays}
\label{subsubsec:R}

The isospin-breaking correction to
$K_{\ell 3}$ decays takes the form \cite{Cirigliano:2001mk}
\begin{equation}
\delta_{\rm SU(2)}^{K^\pm \pi^0} = 2 \sqrt{3}\,
\Big( \varepsilon^{(2)} + \varepsilon^{(4)}_{\rm S} +
\varepsilon^{(4)}_{\rm EM} + \ldots \Big) .
\label{deltaSU2}
\end{equation}
It is dominated by the
lowest-order $\pi^0$--$\eta$ mixing angle
$\varepsilon^{(2)}$ defined in Eq.~(\ref{defeps}).
The NLO 
corrections $\varepsilon_{\rm S}^{(4)}$ (order
$p^4$) and $\varepsilon_{\rm EM}^{(4)}$ (order $e^2 p^2$) were computed
in \textcite{Gasser:1984ux} and \textcite{Neufeld:1994eg}, respectively.
The explicit expressions for these quantities can be found in
\textcite{Cirigliano:2001mk}. The dots refer to NNLO
contributions (arising at order $p^6$),
for which the latest results can be found in \textcite{Bijnens:2007xa}.
Working strictly to $\cO(p^4)$ in CHPT and neglecting the tiny contribution
$\varepsilon_{\rm EM}^{(4)}$,  one can relate  $\delta_{\rm SU(2)}^{K^\pm \pi^0}$  to
quark mass ratios by
\begin{equation}
\delta_{\rm SU(2)}^{K^\pm \pi^0}= \frac{3}{2}
\frac{1}{Q^2} \left[ \frac{M_K^2}{M_{\pi}^2} +
\frac{\chi_{p^4}}{2}  \left(1 + \frac{m_s}{\hat{m}} \right) \right] ,
\label{deltaSU2-2}
\end{equation}
where
$
Q^2 =
(m_s^2 - \widehat{m}^2) / (m_d^2 - m_u^2) = R\,
(m_s / \widehat{m} +1) / 2
$,
$M_{K,\pi}$ are the isospin-limit meson masses and
$\chi_{p^4} = 0.219$ is a calculable loop correction. Thus,
$\delta_{\rm SU(2)}^{K^\pm \pi^0}$
is essentially determined by
$Q^2$
(the uncertainty in $m_s/\hat{m}$
affects  $\delta_{\rm SU(2)}^{K^\pm \pi^0}$ negligibly
due to  the smallness of  $\chi_{p^4}$).

The standard strategy up to now has been to use all known information on
light quark masses to predict
$\delta_{\rm SU(2)}^{K^\pm \pi^0}$
using Eq.~(\ref{deltaSU2-2}).
The double ratio $Q^2$
can be expressed in terms of
meson masses and a purely electromagnetic
contribution \cite{Gasser:1984gg},
\begin{equation}
\! Q^2 = \frac{\Delta_{K \pi} M_K^2 \big(1 +
\mathcal{O}(m_q^2)\big)}{M_\pi^2
\big[ \Delta_{K^0 K^+} + \Delta_{\pi^+ \pi^0}
-
( \Delta_{K^0 K^+} + \Delta_{\pi^+ \pi^0})_{\rm EM}
\big]},
\end{equation}
where $\Delta_{PQ} = M_P^2 -M_Q^2$.
The electromagnetic term
$( \Delta_{K^0 K^+} + \Delta_{\pi^+ \pi^0})_{\rm EM}$
vanishes
to lowest order $e^2 p^0$ \cite{Dashen:1969eg}. A calculation at
$\cO(e^2 p^2)$
gives \cite{Urech:1994hd,Neufeld:1994eg}
\begin{eqnarray}
\lefteqn{(\Delta_{K^0 K^+} + \Delta_{\pi^+ \pi^0})_{\rm EM} =
e^2 M_K^2 \Bigg[ 16 Z L_5^r(\mu)
 }&&\nn
&&\mbox{}
+ \frac{4}{3} (K_5 + K_6)^r (\mu) - 8 (K_{10} + K_{11})^r (\mu)
\\
&& \mbox{}
+\frac{1}{4 \pi^2} \bigg(3 \ln \frac{M_K^2}{\mu^2}
- 4 + 2 Z \ln \frac{M_K^2}{\mu^2} \bigg)  \Bigg] + \mathcal{O}(e^2 M_\pi^2)
.
\nonumber
\end{eqnarray}

The current estimates \cite{Ananthanarayan:2004qk}
of the electromagnetic LECs
appearing in this expression imply a
large deviation from Dashen's limit,
$
(\Delta_{K^0 K^+} + \Delta_{\pi^+ \pi^0})_{\rm EM} =
-1.5 \, \Delta_{\pi^+ \pi^0}
$,
which implies $Q = 20.7 \pm 1.2$ \cite{Kastner:2008ch}. Such a small value
of
$Q$ [compared
to $Q = 22.7 \pm 0.8$ given by \textcite{Leutwyler:1996qg}]
was
also supported by other studies [$Q = 22.0 \pm 0.6$ in
\textcite{Bijnens:1996kk} and
$Q \simeq  20 $ in \textcite{Amoros:2001cp}]. It should be noted, however,
that
the
rather large value $Q = 23.2$ was obtained from an analysis of  $\eta \to
3
\pi$ at two loops \cite{Bijnens:2007pr}. On the other hand, the
non-lattice
determinations of the second input parameter $m_s / \widehat{m} \sim 24$
have remained rather stable over the last years. Combining
$Q = 20.7 \pm 1.2$ with $m_s / \widehat{m} = 24.7 \pm 1.1$,
\textcite{Kastner:2008ch}
found $R = 33.5 \pm 4.3$ and finally
$\delta_{\rm SU(2)}^{K^\pm \pi^0} = 0.058 \pm 0.008$.

It is worth stressing that  the present precision of the decay rates and of
the radiative corrections permits also an \lq\lq experimental'' determination of
$\delta_{\rm SU(2)}^{K^\pm \pi^0}$, which can be used as a constraint on the quark mass
ratio $Q$ via the formula (\ref{deltaSU2-2}).
Combining recent $K_{\ell 3}$ data \cite{Antonelli:2009ws} with the expression
\begin{equation}
\delta_{\rm SU(2)}^{K^\pm \pi^0} = \frac{2 \Gamma_{K^+_{\ell 3}}}{\Gamma_{K^0_{\ell 3}}}   \,
\frac{I_{K^0 \ell}}{I_{K^+ \ell}}  \, \left( \frac{M_{K^0}}{M_{K^+}} \right)^5 \,  - \, 1 \, -
\left(
\delta_{\rm EM}^{K^+ \ell}  -  \delta_{\rm EM}^{K^0 \ell}
\right) ,
\end{equation}
one obtains $\delta_{\rm SU(2) \, exp}^{K^\pm \pi^0}  = 0.054 \pm 0.008$,
in perfect agreement with the value obtained from quark mass ratios.

Alternatively, one may use
the $N_f = 2 +1$ lattice average \cite{Colangelo:2010et}
$
m_s /\widehat{m} = 27.4 \pm 0.4
$
being considerably larger than the values obtained with non-lattice
methods.
Combined with $Q= 22.8 \pm 1.2$ from the same data compilation,
Eq.~(\ref{deltaSU2-2}) yields
$\delta_{\rm SU(2)}^{K^\pm \pi^0} = 0.048 \pm 0.006$,
still consistent with the experimentally determined result.

\subsubsection{Form factors and phase space integrals}

Calculation of the phase space integrals  $I_{K \ell}$
requires knowing the momentum dependence of the form factors.
The vector form factor $f_+^{K \pi}(t)$ defined
in Eq.~(\ref{eq:hadronic element}) re\-presents the
p-wave projection of
the crossed-channel
matrix element $ \langle 0 |\bar{s}\gamma^{\mu}u| K \pi \rangle $ whereas
the s-wave projection
is described by the scalar form factor
\begin{equation}
f_0(t)= f_+(t) + \frac{t}{M_K^2-M_\pi^2} f_-(t).
\label{eq:f0def}
\end{equation}
It is convenient to  normalize all  the form factors to $f_+^{K^0 \pi^-}(0)$
(denoted $f_+(0)$ in the following).
In terms of the normalized form factors
$\bar{f}_i(t) \equiv f_i (t)/f_+(0)$, the phase space integrals read
\beqa
I_{K \ell} &=& \frac{2}{3} \, \int_{m^2_\ell}^{t_0}~\frac{dt}{M_K^8}~\bar\lambda^{3/2}~\left(1+\frac{m^2_\ell}{2t}\right)~
\left(1-\frac{m^2_\ell}{2t}\right)^2
\nn &&\times\quad
 \left(\bar f^2_+(t)+\frac{3 m^2_\ell \Delta_{K\pi}^2}{(2t+m^2_\ell)\bar\lambda}\bar f_0^2(t)\right),
\label{eq:IK}
\eeqa
with
$\bar\lambda=(t-(M_K + M_\pi)^2)(t-(M_K - M_\pi)^2)$.

Traditionally, a polynomial
parametrization has been used for the form factors,
\begin{equation}
\bar{f}_{+,0}(t) = 1 + \lambda'_{+,0} \frac{t}{M_{\pi^+}^2} +
\frac{1}{2}\lambda''_{+,0} \left(\frac{t}{M_{\pi^+}^2}\right)^2
+ \ldots,
\label{Taylor}
\end{equation}
where $\lambda'_{+,0}$ and $\lambda''_{+,0}$ are the slope and
curvature, respectively.
Fits to the experimental distributions  of  $K_{\ell 3}$ decays  allow to extract the parameters
$\lambda'_{+}$,  $\lambda''_{+}$, and $\lambda'_{0}$. The resulting uncertainty on the
phase space integrals is at the level of  $0.12 \%$ for $I_{K e}$ and
$0.30 \%$ for $I_{K \mu}$~\cite{Antonelli:2009ws}.
This affects the extraction of $V_{us}$ at the  level of
$0.06 \%$  ($K_{e3}$) and $0.15 \%$  ($K_{\mu 3}$).

Other form factor parametrizations have been proposed, in
which, by using
physical inputs, specific relations between the slope, the curvature
and all the higher-order terms of
the Taylor expansion (\ref{Taylor}) are imposed.
This allows to reduce the correlations between the fitted slope parameters:
only  one parameter is fitted for each form factor.
Explicit examples used to analyze data include
the pole parametrization, dispersive
parametrizations~\cite{Bernard:2006gy,
Bernard:2009zm,
Abouzaid:2009ry},
and the so-called $z$-parametrization~\cite{Hill:2006bq}.

\subsubsection{The $K_{\ell 3}$ scalar form factor}

SM predictions for the slope parameter $\lambda_0^\prime$ of the scalar
form factor of $K_{\ell 3}$ decays were obtained by using different
approaches. In the isospin limit, the combination of a two-loop result in
chiral perturbation theory \cite{Bijnens:2003uy} with an updated estimate
of the relevant $p^6$ low-energy couplings based on
\textcite{Cirigliano:2005xn} and \textcite{Cirigliano:2003yq} gave the
result
$
\lambda_0^\prime = (13.9^{+1.3}_{-0.4} \pm  0.4) \times 10^{-3}
$
\cite{Kastner:2008ch}.

Dispersive methods were employed by several
authors.\footnote{See for instance \textcite{Jamin:2001zq},
\textcite{Jamin:2004re},
\textcite{Jamin:2006tj}, \textcite{Bernard:2006gy},
\textcite{Bernard:2007tk},
\textcite{Bernard:2009zm},
\textcite{Bernard:2011ae},
\textcite{Abbas:2010ns}.}
Typical numbers for the resulting scalar slope parameter are:
$
\lambda_0^\prime = (14.7 \pm 0.4)\times 10^{-3},
$
\cite{Jamin:2006tj}, and
$
\lambda_0^\prime = 13.71 \times 10^{-3}
$
\cite{Bernard:2011ae}.

The low-energy theorem of Callan and Treiman
\cite{Callan:1966hu,Dashen:1969bh} predicts the size of the scalar
$K_{\ell 3}$ form factor at the (unphysical) momentum transfer $t =
\Delta_{K \pi}$,
\begin{equation}
f_0(\Delta_{K \pi}) = F_K / F_\pi + \Delta_{\rm CT},
\end{equation}
with a correction term of $\cO(m_u, m_d, e^2)$. In the isospin limit ($m_u
=
m_d, \, e=0$), and at first non-leading order, the tiny value $\Delta_{\rm
CT} =
-3.5 \times 10^{-3}$ was computed by \textcite{Gasser:1984ux}.
A discussion of higher-order effects on this quantity can be found in
\textcite{Bijnens:2003uy} and \textcite{Kastner:2008ch}. Note that the
constraint at the Callan-Treiman point plays an essential role in the
dispersive analysis of the scalar form factor
\cite{Jamin:2004re,Jamin:2006tj,Bernard:2006gy,Bernard:2007tk}.

The effect of isospin violation and electromagnetic corrections has also
been considered
\cite{Kastner:2008ch}. This introduces an additional uncertainty for the
values of the slope parameters of at most $\pm 10^{-3}$, mainly due to not
yet fully determined low-energy couplings. Combining the loop results
given by \textcite{Bijnens:2007xa} with an estimate of the relevant
low-energy couplings, the difference of the slope
parameters of the neutral and charged kaons
was found \cite{Kastner:2008ch}
to be confined to the rather small
range $
0 \stackrel{<}{_\sim}
\lambda_0^\prime (K^0_{\ell 3}) - \lambda_0^\prime (K^\pm_{\ell 3})
 \stackrel{<}{_\sim}
10^{-3}
$.

The present experimental situation is displayed in Table \ref{tab:lambda0}.
Note that the numbers\footnote{The ISTRA$+$ result has been rescaled by
$M_{\pi^+}^2/ M_{\pi^0}^2$.} shown here are those where the quadratic
parametrization (\ref{Taylor}) has been used for the simultaneous
determination of the vector form factor. The KTeV data were also reanalyzed
by \textcite{Abouzaid:2009ry} using the dispersive parametrization. The results
$\lambda_0^\prime = (13.22 \pm 1.39) \times 10^{-3}$ ($K_{L \mu 3}$ only) and
$\lambda_0^\prime = (12.95 \pm 1.17) \times 10^{-3}$
($K_{L\mu 3} + K_{L e 3}$)
are rather close to the corresponding numbers given in Table \ref{tab:lambda0}.
The analogous procedure for the KLOE data \cite{Ambrosino:2007yza} gave
$\lambda_0^\prime =(14.0 \pm 2.1) \times 10^{-3}$
($K_{L\mu 3} + K_{L e 3}$).

The experimental results for the scalar slope parameter found by ISTRA$+$,
KTeV and KLOE are in agreement with the predictions of the SM. On the
other hand, the value found by NA48 can hardly be reconciled with the
theoretical
numbers. Furthermore, an isospin violation of the size as it would be
suggested by the simultaneous validity of the ISTRA$+$ and NA48 results
cannot be explained within the SM \cite{Kastner:2008ch}. A clarification
of the origin of these
puzzling results in $K_{\mu 3}$ decays would be
highly welcome \cite{Leutwyler:2009jg}.

\renewcommand{\arraystretch}{1.2}
\begin{table}[floatfix]
\centering

\caption{\label{tab:lambda0}
Experimental results for the slope parameter of the scalar $K_{\ell 3}$ form factor,
in units of $10^{-3}$.
\\  }

\begin{tabular}{|c|c|c|}
\hline
& & \\[-.4cm]
Experiment  &  $\lambda_0^{\prime}$ &  Ref. \\[4pt]
\hline
& & \\[-.3cm]
ISTRA$+$ ($K^-_{\mu 3})$ & $17.1 \pm 2.2$ &
\textcite{Yushchenko:2003xz}   \\[.1cm]
KTeV ($K_{L \mu 3}$) & $12.8 \pm 1.8$ &
\textcite{Alexopoulos:2004sy}  \\[.1cm]
KTeV ($K_{L \mu 3} + K_{L e 3})$ & $13.7 \pm 1.3$ &
\textcite{Alexopoulos:2004sy}  \\[.1cm]
NA48 ($K_{L \mu 3}$) & $9.5 \pm 1.4$ &
\textcite{Lai:2007dx} \\[.1cm]
KLOE ($K_{L \mu 3}$) & $9.1 \pm 6.5$ &
\textcite{Ambrosino:2007yza} \\[.1cm]
KLOE ($K_{L \mu 3} + K_{L e 3})$ & $15.4 \pm 2.2$ &
\textcite{Ambrosino:2007yza} \\[.1cm]
\hline
\end{tabular}

\end{table}

\subsubsection{$\rm SU(3)$ breaking effects in $f_{+}^{K^0 \pi^-}(0)$}

The value of the $K^0_{\ell 3}$ form factor at zero momentum transfer,
$f_{+}^{K^0 \pi^-}(0) \equiv f_{+}(0)$, is
the missing  theoretical ingredient  for the extraction of $V_{us}$.
Within CHPT we can  break up the form factor according to its expansion in quark masses:
\begin{equation}
f_{+}(0) = 1 + f_{p^4} + f_{p^6} + \dots \ .
\end{equation}
Deviations from unity (the octet symmetry limit) are of second order
in SU(3) breaking~\cite{Behrends:1960nf,Ademollo:1964sr}.  The first
correction arises at $\mathcal{O}(p^4)$ in CHPT: a finite one-loop
contribution~\cite{Gasser:1984ux,Leutwyler:1984je} determines
$f_{p^4}= - 0.0227$ in terms of $F_\pi$, $M_K$ and $M_\pi$, with
essentially no uncertainty.
The $p^6$ term was first estimated
by \textcite{Leutwyler:1984je} in the quark model framework,
leading to
\begin{equation}
f_{+}(0)_{\rm LR} = 0.961 \pm 0.008.
\label{eq:LRoss}
\end{equation}

Within CHPT,  $f_{p^6}$ receives contributions from
two-loop diagrams, one-loop diagrams with insertion of one
vertex from the $p^4$ effective Lagrangian, and
tree-level diagrams with two insertions from
${\cal L}_{p^4}$ or one from ${\cal L}_{p^6}$
\cite{Post:2001si,Bijnens:2003uy}:
\begin{equation}
f_{p^6} = f_{p^6}^{\rm 2-loop} (\mu) +
f_{p^6}^{L_i \times {\rm loop}} (\mu)
+  f_{p^6}^{\rm tree} (\mu)    \ .
\end{equation}
Individual components depend on the chiral renormalization scale
$\mu$, their sum being scale independent.
Using  $\mu=M_\rho=0.77$ GeV, one has~\cite{Bijnens:2003uy}
$f_{p^6}^{\rm 2-loop} (M_\rho) +
f_{p^6}^{L_i \times {\rm loop}} (M_\rho) = + 0.0093 \pm 0.0005$.
The $p^6$ constants
appearing in  $f_{p^6}^{\rm tree}$
could be determined
phenomenologically~\cite{Bijnens:2003uy},
provided the experimental errors  on the slope and curvature of the scalar form factor
reach the level
$\Delta \lambda_0^\prime \sim 0.001$
and
$\Delta \lambda_0'' \sim 0.0001$,  which is unfortunately not achievable
in the near future.
Therefore,  further  theoretical input on $f_+(0)$ is needed.

In \textcite{Cirigliano:2005xn} a (truncated) large-$N_C$ estimate of
$f_{p^6}^{\rm tree}$ was performed.  It was based on matching a
meromorphic approximation to the $\langle S P P \rangle$ Green
function (with poles corresponding to the lowest-lying scalar and
pseudoscalar resonances) onto QCD by imposing the correct
large-momentum fall-off, both off-shell and on one- and two-pion mass
shells.  The uncertainty was estimated by
varying the matching scale in the range $\mu \in [M_\eta, 1 {\rm
    GeV}]$, leading to
\begin{equation}\label{eq:f+_chpt}
f_{+}(0)_{\rm  CHPT +1/N_C} = 0.984 \pm 0.012.
\end{equation}

Finally,   starting with the work of \textcite{Becirevic:2004ya}
it has been realized that lattice QCD
is a powerful tool to estimate $f_{+}(0)$ at a level of
accuracy interesting for phenomenological purposes.
Unquenched results are now available with both $N_f = 2+1$  and $N_f=2$
\cite{Dawson:2006qc,
Boyle:2007qe,
Lubicz:2009ht,
Boyle:2010bh,
Brommel:2007wn,
Tsutsui:2005cj}.
The  lattice results agree quite well with the Leutwyler-Roos estimate,
while the analytic approaches tend to be higher as a consequence of
including the large and positive ($\sim 0.01$) two-loop effects
\cite{Bijnens:2003uy}.
In coming years lattice calculations will be performed
closer and closer to the physical light quark masses
and improved chiral extrapolations can be expected. For our
phenomenological extraction of  $V_{us}$
we take as reference value the most recent lattice result with
 $N_f =2+1$ dynamical flavors~\cite{Boyle:2010bh},  which is
also the most precise result for $f_+(0)$,
\begin{equation}
f_{+}(0)_{\rm  RBC/UKQCD} = 0.9599(34)(^{+31}_{-47}) (14),
\end{equation}
where the first error is statistical,  the second is due to the uncertainties in the chiral
extrapolation and the third is an estimate of  discretization effects.
Currently the dominant systematic uncertainty arises from the extrapolation
of lattice results, obtained with unphysical quark masses,  to physical light quark masses
\cite{Bernard:2009ds}.
More progress on $f_+(0)$ is expected soon from other lattice collaborations.

\begin{flushleft}
\begin{figure}[floatfix]
\leavevmode
\includegraphics[height=6.5cm]{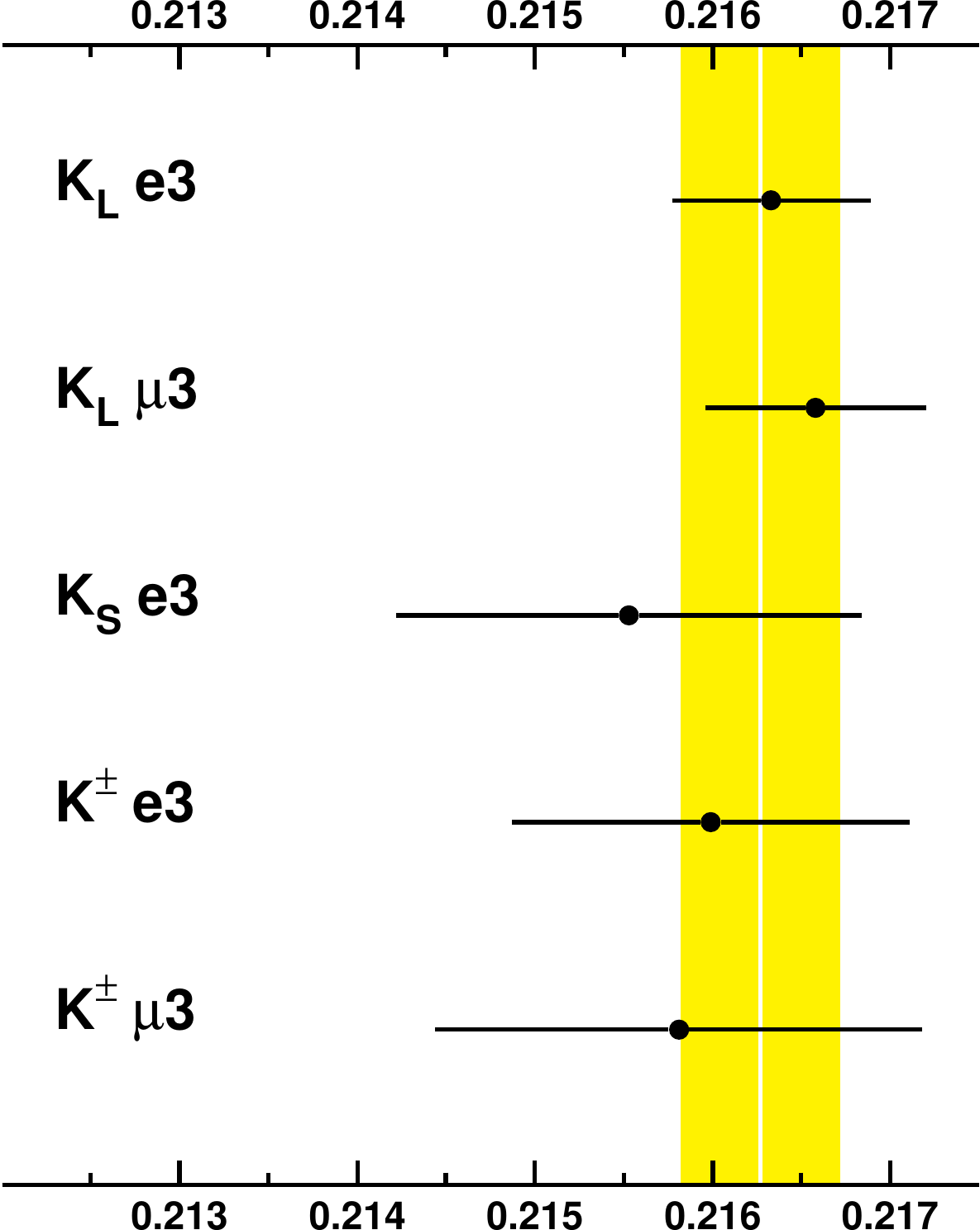}
\caption{\label{fig:f0Vus}
Compilation of values for $|V_{us}| f_+(0)$ extracted from
all $K_{\ell 3}$ channels. The vertical band denotes the average. From \textcite{Antonelli:2010yf}.}
\end{figure}
\end{flushleft}

\subsubsection{Determination of $V_{us}$ and  CKM unitarity tests}

\begin{figure}[floatfix]
\leavevmode
\hskip -.3cm
\includegraphics[width=8.75cm,clip]{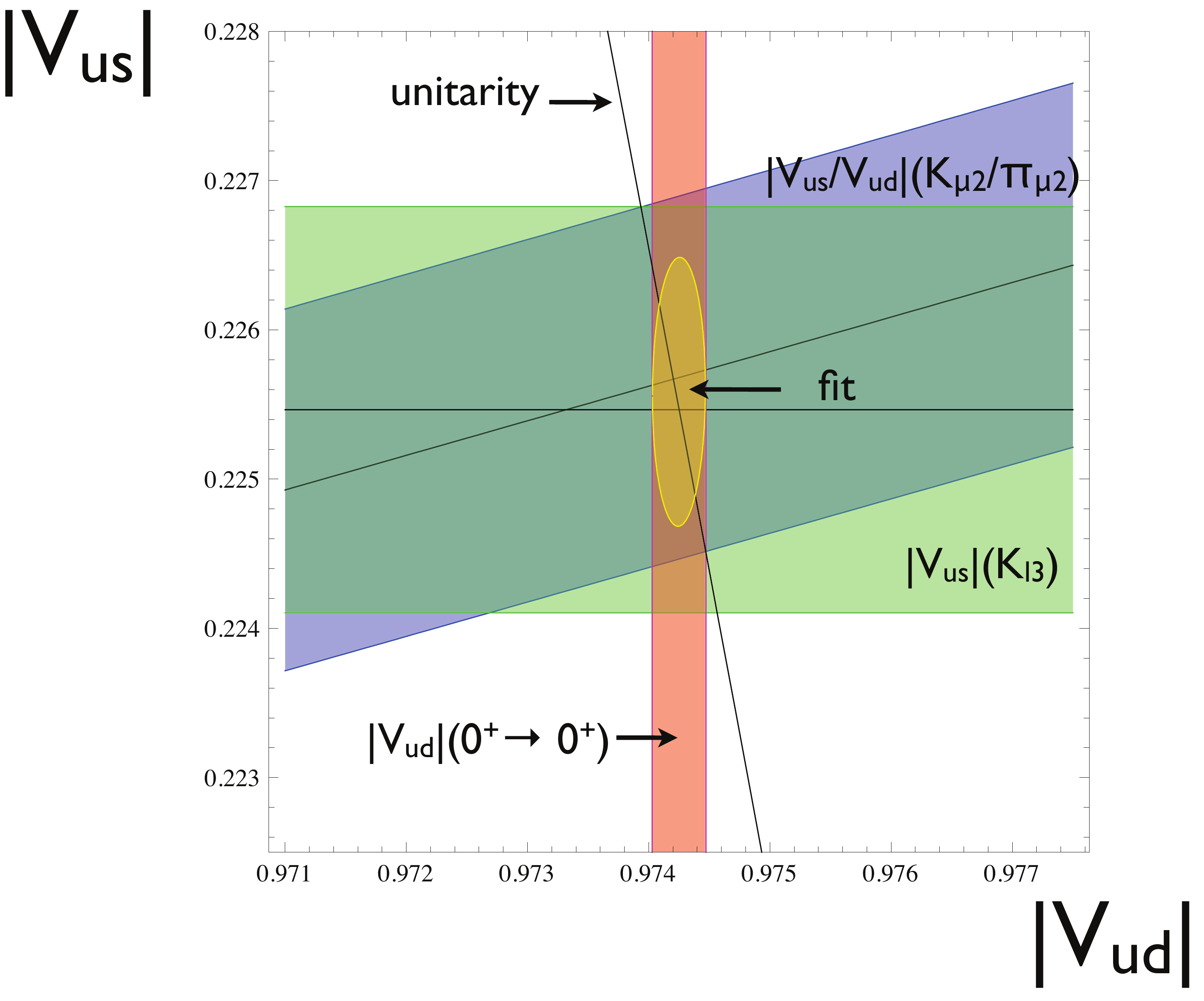}
\caption{\label{fig:unitarity}  Graphical representation of the current
status
of $|V_{ud}|$, $|V_{us}|$ and the corresponding CKM unitarity test.
The horizontal band represents the constraint from $K_{\ell 3}$ decays,
the thin vertical band the constraint from $0^+ \to 0^+$ nuclear decays,
the oblique band the constraint from $K_{\mu 2}/\pi_{\mu 2}$,
and the ellipse is the $1 \sigma$  fit region. From \textcite{Cirigliano:2011tm}.
}
\end{figure}

The combination $|V_{us}| f_+(0)$ can be extracted from both charged and
neutral $K$ decays and its value is dominated by $K^0$ modes (see Fig.~\ref{fig:f0Vus}).
Using  the experimental averages  from
\textcite{Antonelli:2010yf}, one obtains
\begin{equation}
| V_{us}|  f_{+} (0) = 0.2163 \pm 0.0005.
\end{equation}
Taking as a reference value  for $f_+(0)$
the  lattice result of  \textcite{Boyle:2010bh},
$|V_{us}|$ is given by
\begin{equation}
|V_{us}| (K_{\ell 3}) = 0.2255 \pm 0.0005_{\mathrm{exp}}\pm 0.0012_{\mathrm{th}},
\label{eq:VusKl3}
\end{equation}
where we have explicitly displayed the current experi\-mental and theoretical
uncertainties [dominated by $f_+(0)$].
A smaller value would be obtained with the analytical $f_+(0)$ value in Eq.~(\ref{eq:f+_chpt}).
As discussed by \textcite{Antonelli:2010yf},
one can perform a fit to $|V_{ud}|$ and $|V_{us}|$
using as input  the values of
$|V_{us}|$  from  $K_{\ell 3}$ decays [Eq.~(\ref{eq:VusKl3})],
$|V_{us}/V_{ud}|$  from $K_{\ell 2} / \pi_{\ell 2}$
[Eq.~(\ref{eq:VusKl2})],
and $|V_{ud}| =   0.97425 \pm 0.00022$  from
superallowed nuclear $\beta$  transitions \cite{Hardy:2008gy}.
The outcome is \cite{Cirigliano:2011tm}
\begin{eqnarray}
|V_{ud} | &=& 0.97425 \pm 0.00022, \nn
|V_{us}| &=&  0.2256 \pm 0.0009,
\label{eq:Vus}
\end{eqnarray}
with $\chi^2/{\rm ndf} = 0.012$ and negligible correlations between
$|V_{ud}|$ and $|V_{us}|$. Fig.~\ref{fig:unitarity} provides a graphical
representation of the various constraints in the $|V_{us}|$ -- $|V_{ud}|$ plane
and the $1 \sigma$ fit region.
These values [together with the negligible contribution from
$|V_{ub}| = 0.00393(36)$ \cite{Antonelli:2009ws}]
can be used to perform a very stringent test of CKM unitarity
or, equivalently, of
the universality of quark and lepton weak charged-current couplings.
For the  first-row unitarity sum we find
\beq
\Delta_{\rm CKM} = |V_{ud}|^2  +  |V_{us}|^2  +  |V_{ub}|^2  - 1  =
0.0001(6).
\eeq
This constraint allows one to set bounds on the effective scale
of operators that parametrize new-physics
contributions to $\Delta_{\rm CKM}$  \cite{Cirigliano:2009wk}.
The effective scale is constrained to be $\Lambda >  11$ TeV ($90 \%$ C.L.),
which puts this low-energy constraint at the
same level as the bounds from $Z$-pole measurements.

\subsubsection{T violation in $K_{\mu 3}$ decays}
\label{sect:Tviol}

The transverse muon polarization in $K_{\mu 3}$ decays
\begin{equation}
P_T = \frac{\vec{s}_\mu \cdot (\vec{p}_\pi \times \vec{p}_\mu)}{
|\vec{p}_\pi \times \vec{p}_\mu|}
\end{equation}
violates T in the absence of final-state interactions (FSI)
\cite{D'Ambrosio:1996nm}. In the case of $K_L \to \pi^- \mu^+ \nu$, with
two charged particles in the final state, the electromagnetic interaction
generates $\langle P_T \rangle_{\rm FSI} \sim 10^{-3}$ \cite{Okun:1967ww}.
In $K^\pm_{\mu 3}$ decays, this effect does not exceed $10^{-5}$
\cite{Zhitnitsky:1980he,Efrosinin:2000yv} and T-violating effects
could be important.

The SM CP-violating contribution to $P_T$ is
very small $\sim 10^{-7}$
\cite{Cheng:1983mh,Bigi:2000yz}. Therefore, the measurement of the transverse
polarization of muons in $K^\pm_{\mu 3}$ is regarded as a sensitive probe
for physics beyond the SM \cite{Paton:2006xxx,Kohl:2010zz}.

The present experimental value \cite{Abe:2006de}
\begin{equation}
P_T = -0.0017 \pm 0.0023_{\rm stat} \pm 0.0011_{\rm syst}
\end{equation}
is consistent with no T violation and corresponds to the limit
$|P_T| < 0.0050$  $(90 \% {\rm C.L.})$. The sensitivity of TREK
\cite{Paton:2006xxx,Kohl:2010zz} will be able to improve this current
upper limit by at least a factor 20.

\subsection{$K_{\ell 3 \gamma}$}
\label{sect:kl3g}

\begin{flushleft}
\begin{figure}[floatfix]
\leavevmode
\includegraphics[width=7cm]{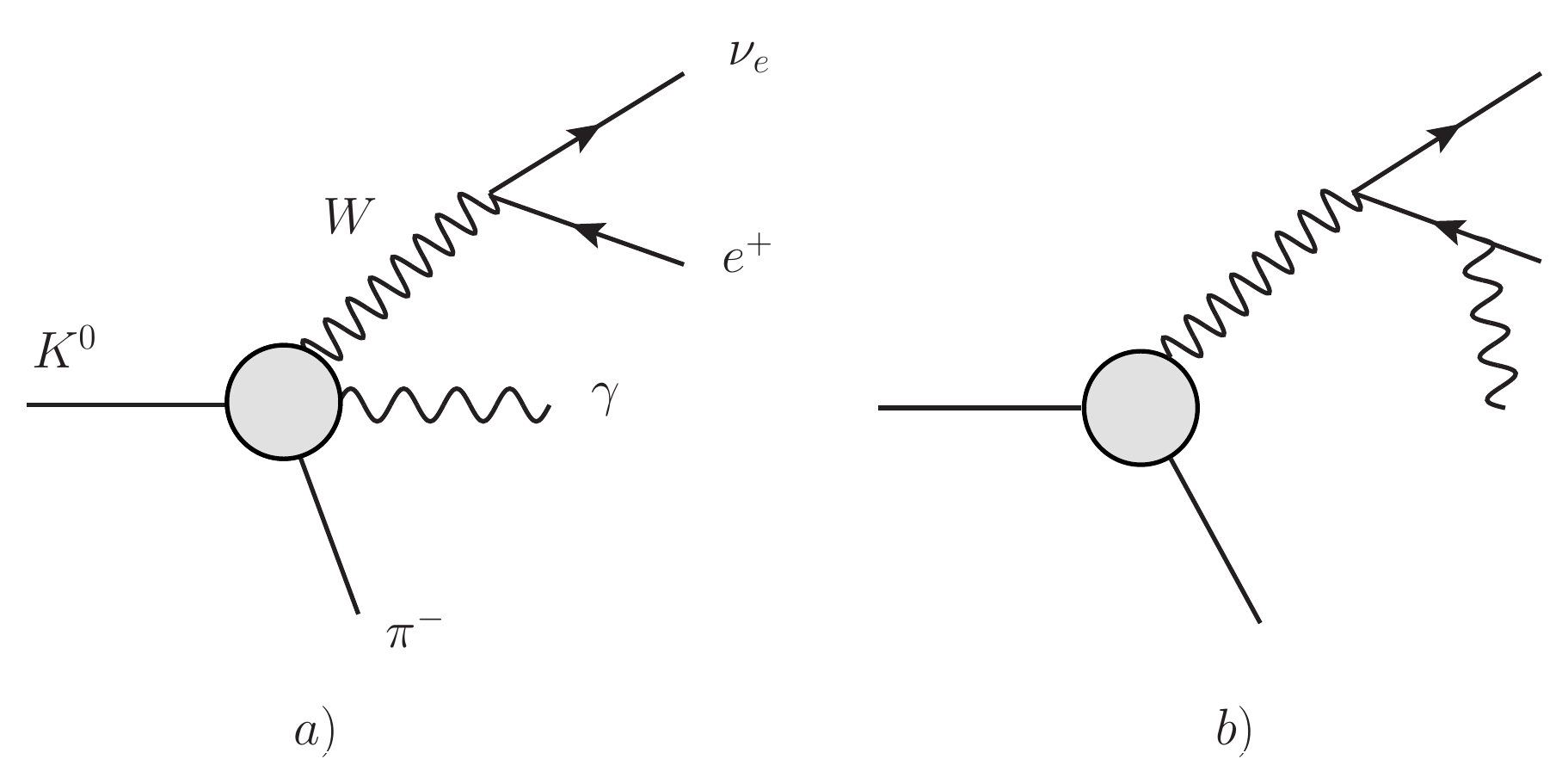}
\caption{\label{fig:kl3g}  Diagrams describing the
$K^0_{\ell 3 \gamma}$ amplitude.}
\end{figure}
\end{flushleft}

The  radiative $K_{\ell 3 \gamma}$  decays ($\ell = e,\mu$)
\begin{eqnarray}
K^+ (p) & \to & \pi^0 (p') \ell^+ (p_\ell) \nu_\ell (p_\nu) \gamma (q), \nn
K^0 (p) & \to & \pi^- (p') \ell^+ (p_\ell) \nu_\ell (p_\nu) \gamma (q)
\end{eqnarray}
allow us to perform quantitative tests of CHPT,
thanks to theoretical  developments over the past couple of decades as well as
recent and ongoing high-statistics experimental studies.
The decay amplitude can be written as
[we focus for definiteness on
$K^0_{e 3 \gamma}$, the generalization is straightforward and can be
found in \textcite{Bijnens:1992en}]:
\beqa\label{eq:Tkl3g}
\lefteqn{
 T (K^0_{e 3 \gamma})
 =  \frac{G_F}{\sqrt{2}} e V_{us}^*   \epsilon_\mu (q)^* \,}\quad &&
\\
& \times & \Big[  \left( V_{\mu \nu} - A_{\mu \nu} \right) \bar{u} (p_\nu)
\gamma^\nu ( 1- \gamma_5) v(p_e)
\nn
&+ &\frac{F_\nu}{2 p_e \cdot q}
\bar{u} (p_\nu)  \gamma^\nu(1- \gamma_5)
\left( m_e - \slashed{p}_e - \slashed{q} \right) \gamma_\mu
v(p_e)  \Big] ,
\nonumber
\eeqa
where the first and second terms correspond to diagrams a) and b), respectively, in  Fig.~\ref{fig:kl3g}.
The hadronic matrix elements are defined by ($J=V,A$)
\begin{eqnarray}
 J_{\mu \nu}  &=& i \int \! dx \, e^{iq x} \langle \pi^-(p') | T(
V_\mu^{\rm em} (x) J_\nu^{\rm w} (0)) | K^0(p) \rangle ,
\nn
 F_\mu  &=&   \langle \pi^-(p') | V_\mu^{\rm w} (x)  | K^0(p) \rangle,
\end{eqnarray}
with the weak and electromagnetic currents
defined in Eq.~(\ref{eq:wemcurrents}).
The Ward identities
$q^\mu \, V_{\mu \nu} = F_\nu$ and $q^\mu \, A_{\mu \nu} = 0$
guarantee the gauge invariance of the total amplitude.

The total amplitude can be decomposed into an
``inner-bremsstrahlung" (IB)
and a ``structure-dependent" (SD) part,  both gauge invariant.
IB captures the infrared singularities according to the
\textcite{Low:1958sn} theorem
and the SD
part contains terms of $\mathcal{O}(q)$ and higher.
In \textcite{Gasser:2004ds} this decomposition was  performed
in such a way as to guarantee that  the SD
amplitude is regular in the
Mandelstam plane, except for the branch points required  by unitarity.
In this treatment the tensor $V_{\mu \nu}$
has an IB component calculable in terms of $K_{\ell 3}$ form factors $f_{\pm}$
and a purely SD component, while the tensor
$A_{\mu \nu}$ is purely SD.
The SD amplitudes can be parametrized in terms of eight
structure
functions  $V_{i}, A_{i}$ ($i=1,\dots,4$).
Defining $W = p - p' - q$, one has \cite{Gasser:2004ds}:
\begin{eqnarray}
A_{\mu \nu}^{\mathrm{SD}} &=& i \epsilon_{\mu \nu \rho \sigma}  \left(
A_1 \, p'^{\rho} q^\sigma
+ A_2 \, q^\rho W^\sigma \right)
\nonumber \\
&+&
 i \epsilon_{\mu \lambda  \rho \sigma}   \, p'^\lambda q^\rho W^\sigma
 \left( A_3  \, W_\nu + A_4 \, p'_\nu
 \right),
\end{eqnarray}
\begin{eqnarray}
V_{\mu \nu}^{\mathrm{SD}} &=&
V_1 \left(p'_\mu  q_\nu - g_{\mu \nu} p' \cdot q\right)  +
V_2 \left(W_\mu  q_\nu - g_{\mu \nu} W \cdot q\right)
\nonumber \\
&+&
V_3 \left(q \cdot W p'_\mu W_\nu  -  p' \cdot q W_\mu W_\nu \right)
\nonumber \\
&+& V_4 \left(q \cdot W p'_\mu p'_\nu  -  p' \cdot q W_\mu p'_\nu \right).
\end{eqnarray}

Early theoretical calculations of $K_{\ell 3 \gamma}$
were based on
Low's theorem and  current algebra \cite{Fearing:1970zz}.
Modern calculations \cite{Holstein:1990gu,Bijnens:1992en,Gasser:2004ds,Kubis:2006nh}
have been performed  within  CHPT,
which provides a natural framework to systematically expand the hadronic amplitudes
$F_\mu$, $A_{\mu \nu}$, and $V_{\mu \nu}$.
The chiral expansion for  $V_{\mu \nu}$ contains both IB and SD terms
and starts at $\mathcal{O}(p^2)$.
$A_{\mu \nu}$ starts at $\mathcal{O}(p^4)$, with the leading contribution
generated by the WZW functional
\cite{Wess:1971yu, Witten:1983tw} accounting for the chiral anomaly.

The first complete  analysis to $\cO(p^4)$ was performed by
\textcite{Bijnens:1992en} who calculated the  branching ratios
for all $K_{\ell 3 \gamma}$ modes for given cuts
in the photon energy and in the photon-electron opening angle in
  the kaon rest frame:
$E_\gamma^{*} > E_\gamma^{\rm cut}$,
$\theta_{e \gamma}^{*} > \theta_{e \gamma}^{\rm cut}$.
Recently, the CHPT analysis was revisited and extended to
$\mathcal{O}(p^6)$
for $K^0_{e 3 \gamma}$ \cite{Gasser:2004ds} and $K^\pm_{e 3 \gamma}$ decays
\cite{Kubis:2006nh},
which represent the theoretical  state of the art.
To $\cO(p^4)$ the axial form factors $A_i$ are constant,  while
the vector form factors $V_i$ receive contributions from both
LECs ($L_{9,10}^r$) and loops.
Since all cuts in loop functions lie far outside the physical region,
the $V_i$ are  constant  to good accuracy.
In order to gain control on the size of higher-order corrections,
\textcite{Gasser:2004ds} and \textcite{Kubis:2006nh} performed
a complete $p^6$
analysis of the axial terms $A_i$ (one loop) and determined
the $\mathcal{O}(p^6)$  $L_i \times L_j$ contributions to the vector terms $V_i$.
Despite the appearance of cuts in the physical region, to $\mathcal{O}(p^6)$ the real parts of
$V_i$ and $A_i$ are well approximated by smooth functions,
with dominant uncertainties coming from the $\mathcal{O}(p^6)$ LECs.

The updated theoretical analysis \cite{Gasser:2004ds,Kubis:2006nh}
leads to very stable predictions for the relative branching ratios
defined by
\begin{equation}
R_{K_{e3 \gamma}} (E_\gamma^{\rm cut}, \, \theta_{e \gamma}^{\rm cut}) \equiv
\frac{\Gamma_{K_{e 3 \gamma}}
(  E_\gamma^* > E_\gamma^{\rm cut}, \, \theta_{e \gamma}^* >  \theta_{e \gamma}^{\rm cut})}
{\Gamma_{K_{e3}}}.
\end{equation}
It turns out that this ratio is very insensitive to the details of the  non-radiative $K_{\ell 3}$ form factor $f_+ (t)$.
Moreover,  the SD terms contribute  only a  $\sim 1 \%$ correction to the IB result for $R$.
The final predictions are (for representative cuts\footnote{For more choices
see \textcite{Gasser:2004ds} and \textcite{Kubis:2006nh}.} for which experimental
data are available):
\begin{eqnarray}
& & \! \! \! R_{K^L_{e 3 \gamma}} \!
(E_\gamma^{\rm cut} \! = 30 \, {\rm MeV},
\, \theta_{e \gamma}^{\rm cut} \! = 20^\circ)
 = 0.0096(1),
\label{eq:RkLe3gth} \\
& & \! \! \! R_{K^\pm_{e 3 \gamma}} \! (E_\gamma^{\rm cut} \! = 10 \, {\rm
MeV}, \,   26^\circ \! <  \theta_{e \gamma}^* \! < 53^\circ)
= 0.00559(6). \nonumber \\
\label{eq:Rkpe3gth}
\end{eqnarray}

In order to measure the SD
terms one needs to resort to differential decay distributions.
A natural first observable to consider is the photon spectrum
$d\Gamma/ d E_\gamma^*$.
It was found
\cite{Gasser:2004ds,Kubis:2006nh} that in
both charged and neutral modes
the SD terms correct the IB spectrum by one single
function $f(E_\gamma^*)$:
\begin{eqnarray}
\frac{d \Gamma}{d E_\gamma^*}  &  \simeq & \frac{d \Gamma_{\rm IB}}{d E_\gamma^*}   +
\langle X \rangle \, f(E_\gamma^*),
\end{eqnarray}
where $\langle X \rangle$ is a mode-dependent linear combination
of phase space averages  of the SD
terms $\langle V_i \rangle$ and $\langle A_i \rangle$.
For the $K^0_{e 3 \gamma}$ mode,
$\langle X \rangle = -1.2 \pm 0.4$  is dominated by $\langle V_1 \rangle$.
In contrast,
in the $K^\pm_{e 3 \gamma}$ mode
$\langle X \rangle = -2.2 \pm 0.7$  receives contributions of similar size
from $\langle V_1 \rangle$ and   $\langle A_1 \rangle$, thus making
possible a detection of
the effect of the chiral anomaly.
More complicated angular distributions might be used to disentangle the
dominant SD terms  \cite{Gasser:2004ds,Kubis:2006nh}.

In parallel to new theoretical developments,
there has been considerable experimental progress in
these modes in the past decade. While we refer to
\textcite{Nakamura:2010zzb} for a complete experimental summary
we focus here on the
$K_{e 3 \gamma}$ decays.
For the neutral mode, the world average
is dominated by
four recent measurements
\cite{Ambrosino:2007jc,Alexopoulos:2004up, Lai:2004fq, AlaviHarati:2001wda}
that are not fully consistent (scale factor $S=1.9$):
\begin{eqnarray}
R_{K^L_{e 3 \gamma}}(E_\gamma^{\rm cut} \! =  30 \, {\rm MeV}, \,
\theta_{e \gamma}^{\rm cut} \!= 20^\circ)
&=& 0.935(15)\times 10^{-2}.\nn
&&
\label{eq:RkLe3gexp}
\end{eqnarray}
This PDG  average \cite{Nakamura:2010zzb} is about one sigma below the
theoretical prediction
given in Eq.~(\ref{eq:RkLe3gth}).
The most recent individual measurement from  KLOE \cite{Ambrosino:2007jc},
\begin{equation}
R_{K_{e3 \gamma}^L} = (0.924 \pm 0.023_{\rm stat} \pm
0.016_{\rm syst})\times 10^{-2},
\end{equation}
does not have sufficient precision
to resolve the tension between the NA48  result
$R = (0.964  \pm 0.013)\times 10^{-2}$
\cite{Lai:2004fq}
and the KTeV result
$R = (0.916  \pm 0.017)\times 10^{-2}$
\cite{Alexopoulos:2004up}.
By measuring the photon spectrum, the KLOE collaboration
\cite{Ambrosino:2007jc}
performed a fit to the SD contribution $\langle X \rangle$, finding
\begin{equation}
\langle X \rangle  =  -2.3 \pm 1.3_{\rm stat} \pm 1.4_{\rm syst},
\end{equation}
in agreement (within the large
uncertainty) with the CHPT prediction $\langle X \rangle = -1.2 \pm 0.4$.

For the charged mode, the PDG average
\cite{Nakamura:2010zzb}
is dominated by three measurements
\cite{Barmin:1991zn,Bolotov:1986zz,Akimenko:2007zz},
again  not fully consistent (scale factor $S=1.3$):
\begin{eqnarray}
R_{K^\pm_{e 3 \gamma}} \!
(E_\gamma^{\rm cut} \!  = 10 \, {\rm MeV}, \, 26^\circ \! <  \theta_{e
\gamma}^* \! < 53^\circ)
= 0.00505(32). &&\nn
\label{eq:Rkpe3gexp}
\end{eqnarray}
This value is about one sigma below the theoretical prediction
given in Eq.~(\ref{eq:Rkpe3gth}).

Despite the tremendous progress in $K_{\ell 3 \gamma}$ decays,
from the comparison of theory and experiment we conclude that
more accurate experimental data are desirable
in order to perform definite tests of the CHPT predictions
and to detect SD contributions.

\subsection{$K_{\ell 4}$}

$K_{\ell 4}$ is the shorthand notation for the decays
\begin{eqnarray}
K^+(p) &\to& \pi^+(p_1) \pi^-(p_2) \ell^+(p_\ell) \nu_\ell (p_\nu),
\label{Kpluscharged}
\\
K^+(p) &\to& \pi^0(p_1) \pi^0(p_2) \ell^+(p_\ell) \nu_\ell (p_\nu),
\label{Kplus0}
 \\
K^0(p) &\to& \pi^0(p_1) \pi^-(p_2) \ell^+(p_\ell) \nu_\ell (p_\nu)
\label{Kneutral}
\end{eqnarray}
and their charge-conjugate modes. In the isospin limit ($m_u = m_d$,
$e=0$), the amplitude for (\ref{Kpluscharged}) is given by
\begin{equation}
T = \frac{G_F}{\sqrt{2}} V_{us}^*\; \bar{u} (p_\nu) \gamma^\mu (1-\gamma_5)
v(p_\ell)\; (V_\mu - A_\mu),
\end{equation}
where the last factor contains the hadronic matrix elements of the
strangeness-changing vector and axial-vector currents,
\begin{equation}
V_\mu - A_\mu = \langle\pi^+(p_1) \pi^-(p_2) \, | 
\bar{s} \gamma_\mu u - \bar{s} \gamma_\mu \gamma_5 u | K^+(p) \rangle,
\qquad
\end{equation}
with the form factor decompositions
\begin{eqnarray}
V_\mu &\!\! =&\!\! - \frac{H}{M_K^3}\, \varepsilon_{\mu \nu \rho \sigma}
(p_\ell + p_\nu)^\nu (p_1+p_2)^\rho (p_1-p_2)^\sigma , \\
A_\mu &\!\! =&\!\! - \frac{i}{M_K} \big[ F (p_1+p_2)_\mu +G (p_1-p_2)_\mu +
R (p_\ell +p_\nu)_\mu \big]. \nonumber
\end{eqnarray}
The matrix elements for the
channels (\ref{Kplus0}) and (\ref{Kneutral})
can be obtained by isospin symmetry. The
form factors $F$, $G$, $R$ and $H$ depend on the variables
$s=(p_1+p_2)^2$, $t =(p_1 - p)^2$ and $u=(p_2-p)^2$.
Quite often,
\begin{equation}
s=(p_1+p_2)^2, \quad s_\ell =(p_\ell + p_\nu)^2, \quad
\cos \theta_\pi
\end{equation}
are used instead, where $\theta_\pi$ is the angle of the $\pi^+$ in the
center-of-mass system of the two charged pions relative to the dipion line
of flight in the rest system of the kaon
\cite{Cabibbo:1965zz,Cabibbo:1968zz}.

The chiral expansion of the $K_{\ell 4}$ form factors was studied in
CHPT at the one-loop level \cite{Bijnens:1989mr,Riggenbach:1990zp} and
beyond \cite{Bijnens:1994ie,Amoros:1999qq,Amoros:2000mc}. Comparison
with experimental data yields information on several LECs.

The analysis of $K^\pm  \to \pi^+ \pi^- e^\pm \nu$ is a very efficient
and clean approach to study pion-pion scattering at low energies. In the
limit of isospin symmetry, one identifies the $\pi \pi$ phase shifts in
the matrix element in a standard manner by performing a partial-wave
expansion using unitarity and analyticity. Details of this procedure can
be found in \textcite{Pais:1968zz} and
\textcite{Berends:1967xxx,Berends:1968zz}.
At the end, it boils down to the following parametrization of the form
factors,
\begin{eqnarray}
F &=& F_s e^{i \delta_s} +  F_p e^{i \delta_p} \cos \theta_\pi + \ldots,
\nonumber \\
G &=& G_p e^{i \delta_p} + \ldots,
\nonumber \\
H &=& H_p e^{i \delta_p} + \ldots,
\end{eqnarray}
where the dots refer to neglected d-wave contributions.
Note that the third axial form factor $R$ gets multiplied by a factor
$m_e^2/s_e$
and cannot be measured in $K_{e 4}$ decays.
One is therefore left with one phase difference $\delta = \delta_s -
\delta_p$ and the four real form factors $F_s, \, F_p, \, G_p, \, H_p$.

Under the assumption of isospin symmetry, the form factors can be expanded
in a series of the dimensionless invariants
$q^2 =s / 4 M_\pi^2 -1$ and $s_e/4 M_\pi^2$
\cite{Amoros:1999mg}. Within the currently
available statistics \cite{Batley:2010zza}, a constant term, two slope
parameters and one curvature parameter are sufficient to describe the
variation of the form factor $F_s$,
\begin{equation}
F_s = f_s  + f_s^\prime q^2 +f_s^{\prime \prime} q^4 + f_e^\prime \,
\frac{s_e}{4 M_\pi^2},
\end{equation}
while two terms are needed to describe 
$G_p$,
\begin{equation}
G_p = g_p + g_p^\prime q^2.
\end{equation}
$F_p$ and $H_p$ can be described by two constants. Based on an
analysis of 1$.13 \times 10^{6}$ decays, the NA48/2 collaboration reports the
following results for the form factor measurements \cite{Batley:2010zza}:
\begin{eqnarray}
f_s^\prime / f_s &=& 0.152 \pm 0.007 \pm 0.005, \nonumber \\
f_s^{\prime \prime} / f_s &=& -0.073 \pm 0.007 \pm 0.006, \nonumber \\
f_e^\prime / f_s &=& 0.068 \pm 0.006 \pm 0.007, \nonumber \\
f_p / f_s &=& -0.048 \pm 0.003 \pm 0.004, \nonumber \\
g_p / f_s &=& 0.868 \pm 0.010 \pm 0.010, \nonumber \\
g_p^\prime / f_s &=& 0.089 \pm 0.017 \pm 0.013, \nonumber \\
h_p / f_s &=& -0.398 \pm 0.015 \pm 0.008,
\end{eqnarray}
with statistical (first) and systematic (second) errors.

A preliminary branching ratio based on the NA48/2 data has recently
been presented \cite{BlochDevaux:2011xx},
\begin{equation}
{\rm BR}(K^\pm \to \pi^+ \pi^- e^\pm \nu) = (4.279 \pm 0.035)
\times 10^{-5},
\end{equation}
which improves by a factor of $3$ the PDG average
$(4.09 \pm 0.10) \times {10}^{-5}$
\cite{Nakamura:2010zzb}.
The final analysis will allow to give precise absolute values for all
decay form factors \cite{BlochDevaux:2011xx}.

The extraction of the $\pi \pi$ scattering lengths from the phase shift
measurements of $\delta = \delta_s - \delta_p$ requires additional
theoretical ingredients. Forty years ago, \textcite{Roy:1971tc}
established
an integral equation based on analy\-ticity, unitarity and crossing that
allows to predict the $\pi \pi$ phase values close to threshold by
using experimental data above the matching point ($\sqrt{s} = 0.8 \,
\rm
GeV$) and two subtraction constants $a_0$ and $a_2$, the
$\mathrm I=0,2$
s-wave scattering lengths. On the other hand, using measurements and the
Roy equations, one can determine the corresponding values of the
scattering lengths. Numerical solutions of the Roy equations obtained
by two groups \cite{Ananthanarayan:2000ht,DescotesGenon:2001tn} were
employed for this purpose \cite{Batley:2010zza}.

Isospin-breaking turns out to be quite substantial in $K_{\ell 4}$ decays
\cite{Cuplov:2003bj,Nehme:2003bz,Nehme:2004ui,Nehme:2004xy}.
Triggered by the precise results of NA48/2
\cite{Batley:2007zz}, a new theoretical procedure for the treatment of
isospin-breaking on $K_{e4}$ phase measurements
was suggested by \textcite{Colangelo:2008sm}.
The measured
phase of the $\mathrm I=0$ s-wave is no longer $\delta_0^0$ but
\begin{eqnarray}
\psi_0 &\!=&\! \frac{1}{32 \pi F^2} \Big[(4 \Delta_{\pi^+\pi^0} + s)\,
\sigma_\pm +(s-M_{\pi^0}^2) \big( 1 +\frac{3}{2R} \big) \sigma_0 \Big]
\nonumber \\
 & &\! + \; \mathcal{O}(p^4).
\end{eqnarray}
In this expression, $F$ is the pion decay constant in the limit of
chiral SU(2),
$\Delta_{\pi^+\pi^0} = M_{\pi^\pm}^2 - M_{\pi^0}^2$, $R$ is the quark-mass ratio
defined in Eq.~(\ref{defeps}) and
\beq
\sigma_x = \sqrt{1 - 4 M_{\pi^x}^2 /s} \qquad\quad (x=\pm, 0).
\eeq

\begin{flushleft}
\begin{figure}[floatfix]
\leavevmode
\includegraphics[width=7.5cm]{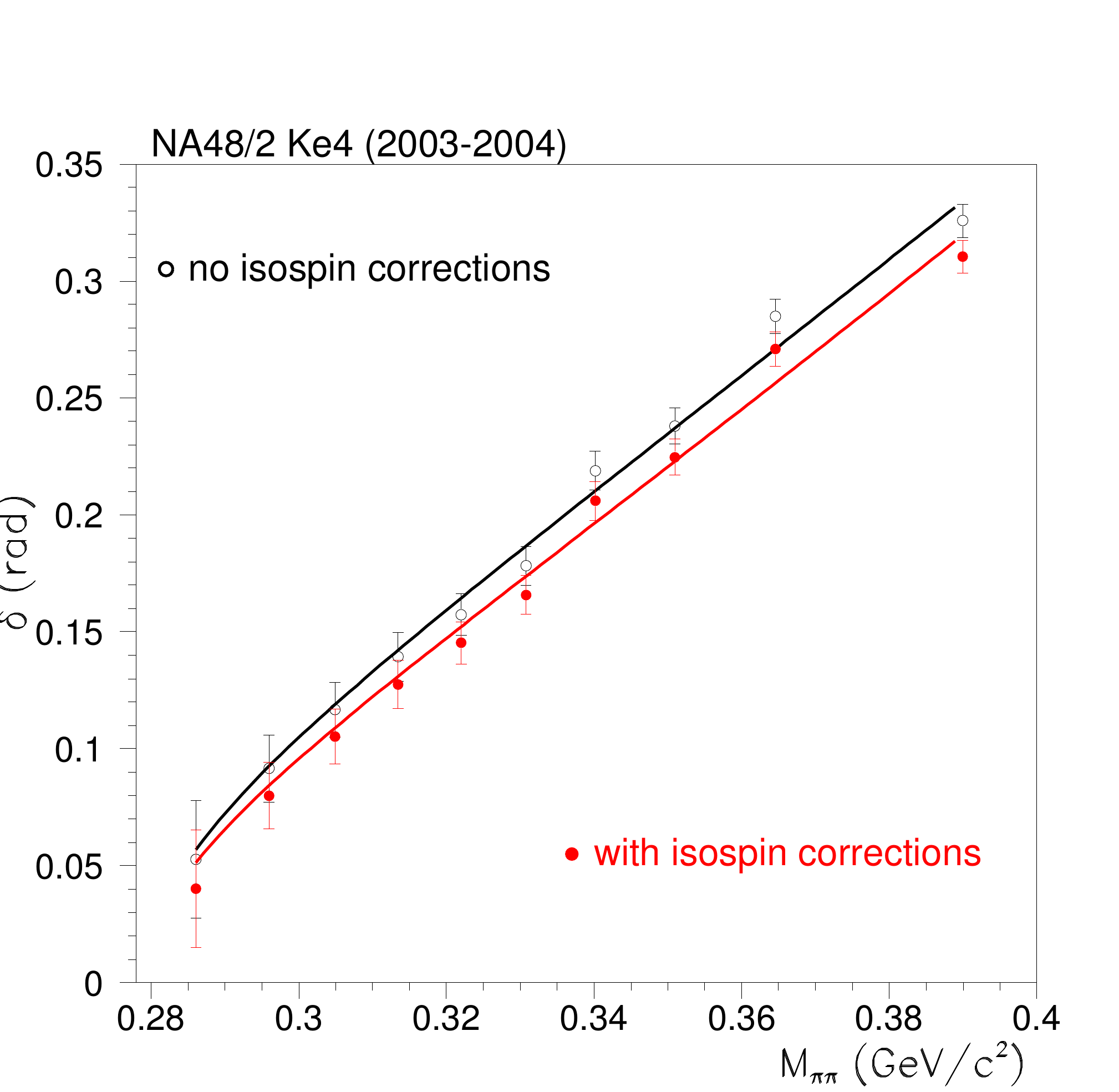}  
\caption{\label{fig:delta}
Measurements of the phase shift $\delta$ without (open
circles) and with (full circles) isospin mass effect
  corrections from NA48/2 $K_{e 4}$ data. From
\textcite{Batley:2010zza}, Copyright CERN for the benefit of the NA48/2
Collaboration 2010.}
\end{figure}
\end{flushleft}

\begin{flushleft}
\begin{figure}[floatfix]
\leavevmode
\includegraphics[width=7.5cm]{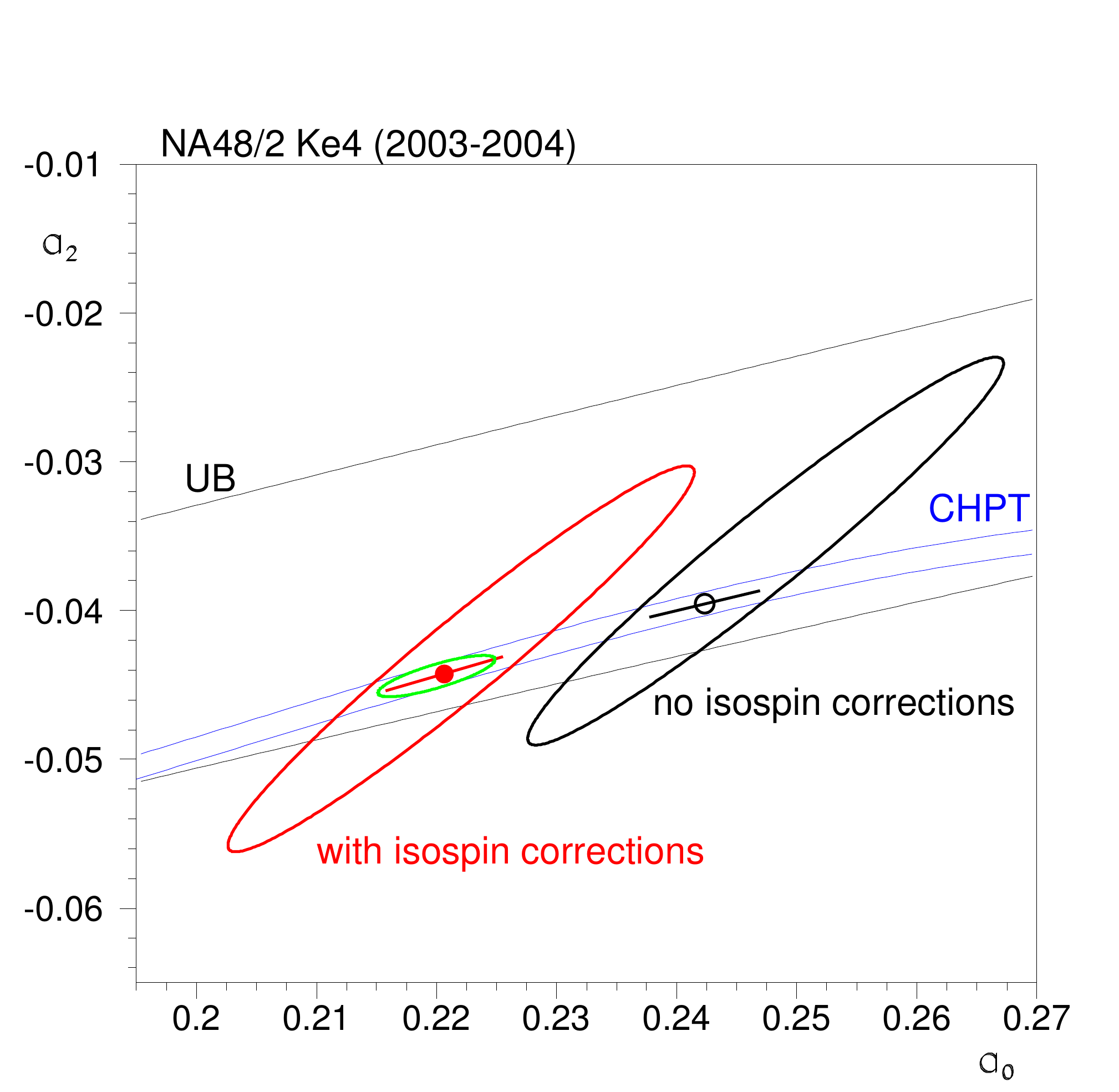} 
\caption{\label{fig:a0a2fit}
Fits of the NA48/2 $K_{e4}$ data in the ($a_0, a_2$) plane without and
with isospin mass effects.
The wide band
(UB) refers to the \lq\lq universal band\rq\rq {} [see
\textcite{Batley:2010zza} and \textcite{DescotesGenon:2001tn} for
details].
The large ellipses are $\rm 68\% \, C.L.$ contours of
the two-parameter fit leading to Eq.~(\ref{twoparfit}) and the circles
the one-parameter fit of
Eq.~(\ref{oneparfit}), imposing the
CHPT constraint of Eq.~(\ref{constraint}). The small ellipse
corresponds
to Eq. (\ref{eq:a0a2chpt}).
 Adapted from \textcite{Batley:2010zza}.}
\end{figure}
\end{flushleft}

Although the difference between the mass-symmetric angle ($\Delta_{\pi^+\pi^0} =
0, \, 1/R=0, \, \sigma_\pm = \sigma_0$) and $\psi_0$ is rather small (10
to 15
mrad) over the whole range accessible in $K_{e 4}$ decays, there are
non-negligible effects on the extraction of the scattering lengths
\cite{Batley:2010zza},
as can be seen in Figs. \ref{fig:delta}
and \ref{fig:a0a2fit}. Note that these effects had not been taken into
account in the analyses of the older Geneva-Saclay \cite{Rosselet:1976pu}
and BNL E865 \cite{Pislak:2001bf,Pislak:2003sv} experiments.


A two-parameter fit of the NA48/2 data leads to the result
\cite{Batley:2010zza}
\begin{eqnarray} \label{twoparfit}
a_0 &=& 0.2220 \pm 0.0128  \pm 0.0050 \pm 0.0037,
\nn
a_2 &=& -0.0432 \pm 0.0086 \pm 0.0034 \pm 0.0028,
\end{eqnarray}
where the errors refer to statistics, systematics and theory.
Alternatively, using the additional theoretical constraint
\cite{Colangelo:2000jc,Colangelo:2001sp,Colangelo:2001df}
\begin{eqnarray}
a_2 &=& -0.0444+0.236(a_0-0.220)-0.61(a_0-0.220)^2
\nonumber \\
 & & -9.9(a_0-0.220)^3 \label{constraint}
\end{eqnarray}
(with a width of $\pm 0.0008$), a one-parameter fit was performed
giving \cite{Batley:2010zza}
\begin{equation} \label{oneparfit}
a_0 = 0.2206 \pm 0.0049 \pm 0.0018 \pm 0.0064,
\end{equation}
corresponding to the value $a_2 = -0.0442$ from
Eq.~(\ref{constraint}).
These results should be compared with the most precise
prediction to NNLO in CHPT
\cite{Colangelo:2001sp,Colangelo:2001df}:
\begin{equation}
\label{eq:a0a2chpt}
a_0=0.220 \pm 0.005, \qquad a_2 =-0.0444\pm0.0010.\quad
\end{equation}

A measurement of $K^+ \to \pi^0 \pi^0 e^+ \nu$ by the
KEK-E740 Collaboration with a data sample of $216$ events was published
in \textcite{Shimizu:2004it}. However, due to large systematic errors,
this result was not included in the PDG fit
${\rm BR} (K^+ \to \pi^0 \pi^0 e^+ \nu) = (2.2 \pm 0.4) \times 10^{-5}$
\cite{Nakamura:2010zzb}, which
uses data from an old low-statistics experiment \cite{Barmin:1997nx}.
Based on the analysis of $4.4 \times 10^5$ $K_{e4}^{00}$ events,
the NA48/2 Collaboration has recently reported the preliminary
branching ratio
\begin{equation}
{\rm BR}(K^\pm \to \pi^0 \pi^0 e^\pm \nu) = (2.595 \pm 0.042) \times
10^{-5},
\end{equation}
which corresponds to a factor $10$ improvement compared to the PDG value.
They could also show the consistency of the $F_s$ form factors in the
$K_{e4}^{00}$ and $K_{e4}^{+-}$ modes.

Experimental results for
$K_L \to \pi^\pm \pi^0 e^\mp \nu_e$
are also available
\cite{Batley:2004cp,Carroll:1980kw,Makoff:1993xb}.
Although the present statistics [5464 events
in the experiment of \textcite{Batley:2004cp}] is not
comparable with the charged mode, the branching ratio and the form factors
have been determined and a value for the chiral LEC $L_3$ was obtained:
\begin{equation}
L_3 = (-4.1 \pm 0.2) \times 10^{-3}.
\end{equation}
The contribution of the decay chain $K^0 \to (K^+ e^-
  \overline{\nu_e}) \to \pi^+ \pi^0 e^- \overline{\nu_e}$ is
  suppressed by at least four orders of magnitude in the branching
  ratio \cite{Chizhov:2006ze}.

\subsection{$K_{e5}$}
Except for the charge-conjugate modes, there are four kinematically
allowed $K_{e5}$ decays:
\begin{eqnarray}
K^+ \to \pi^+ \pi^- \pi^0 e^+ \nu_e, & \! \! \quad K^+ \to \pi^0 \pi^0
\pi^0
e^+ \nu_e, \nn
K^0 \to \pi^0 \pi^0 \pi^- e^+ \nu_e, & \! \! \quad K^0 \to \pi^+ \pi^-
\pi^-
e^+ \nu_e, \quad
\end{eqnarray}
related by the isospin relation \cite{Blaser:1994fb}
\begin{eqnarray}
2\,\Gamma(K^+ \to \pi^0 \pi^0 \pi^0 e^+ \nu_e) &=& \\[.2cm]
\Gamma(K^0_L \to \pi^\pm \pi^\mp \pi^\pm e^\mp \nu_e) &-& \Gamma(K^0_L \to \pi^0
\pi^0 \pi^\pm e^\mp \nu_e). \no
\end{eqnarray}
At lowest order in CHPT, only the vector current contributes.
Because of the small phase space, the branching
ratios are suppressed: $\mathcal{O}(10^{-12})$ for the $K^+$
decays and $\mathcal{O}(10^{-11})$ for the $K^0_L$ decays at leading order
\cite{Blaser:1994fb}. Experimentally, there is only an upper bound
\cite{Nakamura:2010zzb}
\begin{equation}
{\mathrm {BR}}(K^+ \to \pi^0 \pi^0 \pi^0 e^+ \nu_e) < 3.5 \times 10^{-6}.
\end{equation}

\section{Nonleptonic decays}
\label{sec:nonlep}
In this section, we first review the status of $K \to 2 \pi$ decays.
NLO calculations including strong isospin breaking and
electromagnetic corrections allow for the extraction of the
lowest-order couplings $G_8, G_{27}$ in the nonleptonic weak
Lagrangian (\ref{eq:Lweak}) from the decay rates. A careful
treatment of isospin violation is essential for a reliable
determination of the s-wave $\pi\pi$ phase shift difference
$\delta_0(M_K)- \delta_2(M_K)$. Despite a lot of efforts, the
accuracy of theoretical predictions of the ratio $\epsilon'/\epsilon$
still does not match the experimental precision. A NLO calculation of
$K \to 3 \pi$ decays significantly improves the compatibility between
theory and experiment even though some assumptions must be made about
NLO LECs. Both isospin-violating and CP-violating effects
seem to be too small to be detected experimentally at this
time. Somewhat unexpected, investigation of a cusp effect in the
$M_{\pi^0\pi^0}$ distribution near threshold in $K \to 3 \pi$ decays
has led to very precise values for the s-wave $\pi\pi$ scattering
lengths.

\subsection{$K \to \pi\pi$}
\label{subsec:2pi}
The amplitudes for the two-pion decay modes can be parametrized in the
form
\begin{eqnarray}  A(K^0 \to \pi^+ \pi^-) &=&  A_{+-}= A_{1/2} +
{1 \over \sqrt{2}} \left(A_{3/2} + A_{5/2} \right) \nn
 &=& A_{0}  e^{i \chi_0}  + { 1 \over \sqrt{2}}    A_{2}  e^{i\chi_2 },
\nn
A(K^0 \to \pi^0 \pi^0) &=& A_{00}=A_{1/2} - \sqrt{2} \left(A_{3/2} +
A_{5/2}  \right) \nn
 &=& A_{0}  e^{i \chi_0}  - \sqrt{2}    A_{2}  e^{i\chi_2 },\nn
A(K^+ \to \pi^+ \pi^0) &=& A_{+0}={3 \over 2}  \left(A_{3/2} -
{2 \over 3} A_{5/2} \right) \nn
 &=& {3 \over 2}
A_{2}^{+}   e^{i\chi_2^{+}}.
\label{eq:2pipar}
\end{eqnarray}
This parametrization holds for the infrared-finite amplitudes where the
Coulomb and infrared parts are removed from $A_{+-},A_{+0}$
\cite{Cirigliano:2003gt}.
The amplitudes $A_{\Delta I}$ ($\Delta I=1/2,3/2,5/2$) are generated
by the $\Delta I$
component of  the electroweak effective Hamiltonian in the limit of
isospin conservation. In the SM, the $\Delta I = 5/2$
piece is absent in the isospin limit. More precisely, $A_{5/2}=0$ in
the absence of electromagnetic interactions and therefore $A_2 = A_2^+$.
In the limit of CP conservation, the amplitudes $A_{0}, A_{2}$ and
$A_{2}^+$ are real and positive by definition. In the isospin limit,
the phases $\chi_I$ ($I=0,2$) can then be identified with the
s-wave $\pi \pi$ scattering phase shifts $\delta_I (\sqrt{s} = M_K)$.

In the isospin limit (phase space is calculated with physical meson
masses), the amplitudes $A_0, A_2$ and the phase
difference $\chi_0 - \chi_2$ can be obtained directly from
the three $ K \rightarrow \pi \pi $ branching ratios [we use the recent
compilation of \textcite{Antonelli:2010yf}]:
\begin{eqnarray}
A_0 &=& (2.704 \pm 0.001) \times 10^{-7} \mbox{ GeV}, \nn
A_2 &=& (1.210 \pm 0.002) \times 10^{-8} \mbox{ GeV}, \nn
\chi_0 - \chi_2 &=& (47.5 \pm 0.9)^{\circ}.
\label{eq:isoamps}
\end{eqnarray}

In the presence of isospin violation and to lowest order in the chiral
expansion, the amplitudes $A_{\Delta I}$ are given by (omitting
isospin violation in the 27-plet amplitudes)
\begin{eqnarray}
A_{1/2} & = &
{\sqrt{2} \over 9} G_{27} F_0 \left( M_{K^0}^2 - M_{\pi^0}^2 \right)  \nn
&&{} +
\sqrt{2} G_8 F_0 \bigg[  \left( M_{K^0}^2 - M_{\pi^0}^2 \right)
 \left(1 - {2 \over 3 \sqrt{3}} \varepsilon^{(2)} \right)  \nn
&&{} \qquad \qquad  - {2 \over 3} e^2 F_0^2 \left( g_{\rm ewk} + 2 Z
\right)  \bigg], \nn
A_{3/2} & = &
{10 \over 9}  G_{27} F_0 \left( M_{K^0}^2 - M_{\pi^0}^2 \right) \nn
&&{} + G_8 F_0 \bigg[  \left( M_{K^0}^2 - M_{\pi^0}^2 \right) {4 \over 3
\sqrt{3}} \varepsilon^{(2)}  \nn
&&{} \qquad \quad \; \; -  {2 \over 3} e^2 F_0^2 \left( g_{\rm ewk} + 2 Z
\right)
\bigg], \nn
A_{5/2} & = & 0. \label{eq:LOamp}
\end{eqnarray}
The various couplings are defined in
Eqs.~(\ref{eq:Lweak}) and (\ref{eq:Lelweak}). Following
\textcite{Cirigliano:2003gt},
we have expressed all amplitudes in terms of
the neutral pion and kaon masses.
The parameter $F_0$ can be identified with the pion
decay constant $F_\pi$ at this order.  The effect
of strong isospin breaking
is entirely due to the lowest-order $\pi^0$--$\eta$ mixing angle
$\varepsilon^{(2)}$ given in Eq.~(\ref{defeps}).
Electromagnetic interactions contribute through mass splitting (terms
proportional to $Z$) and insertions of $g_{\rm ewk}$.
As a consequence of CPS symmetry \cite{Bernard:1985wf},
electromagnetic corrections to the octet weak Hamiltonian do not
generate a $\Delta I= 5/2$ amplitude at lowest chiral order.

The NLO amplitudes in the isospin limit for both $K \to \pi\pi$ and $K
\to 3\pi$ were first calculated by \textcite{Kambor:1991ah}. Their
main conclusion was that the octet amplitude $A_0$ is strongly
enhanced by final state interactions at NLO whereas the 27-plet
amplitude $A_2$ is only mildly reduced. Since the generically small
isospin-violating effects are enhanced in subdominant amplitudes with
$\Delta I > 1/2$ because of the $\Delta I=1/2$ rule, a systematic
treatment of isospin violation is called for.

The first complete analysis of isospin breaking in $K \to \pi\pi$
amplitudes was carried out by \textcite{Cirigliano:2003gt}.
References to earlier work can be found there. The analysis was
repeated and extended to $K \to 3\pi$ amplitudes by
\textcite{Bijnens:2004ai}. In the following, we update the analysis of
\textcite{Cirigliano:2003gt} with new experimental input
\cite{Antonelli:2010yf} and with new information on the LECs involved,
both in the strong \cite{Cirigliano:2006hb} and in the electromagnetic
sector \cite{Ananthanarayan:2004qk}.

As usual at NLO in the chiral expansion, both one-loop diagrams with
LO couplings and tree diagrams with a single insertion of NLO
couplings must be taken into account. Including the leading
isospin-breaking corrections (proportional to $G_8$), the amplitudes
$A_{\Delta I}$ have the following form  \cite{Cirigliano:2003gt}:
\begin{eqnarray}
A_{\Delta I} &=& G_{27}  F_\pi ( M_{K^0}^2 - M_{\pi^0}^2 ) A_{\Delta I}^{(27)}
   \nn
&&{} +
G_{8} F_\pi \Bigg\{ \left( M_{K^0}^2 - M_{\pi^0}^2 \right)
\bigg[ A_{\Delta I}^{(8)} +  \varepsilon^{(2)}
A_{\Delta I}^{(\varepsilon)} \bigg]
\nn
&&{}  -   e^2  F_\pi^2 \bigg[
A_{\Delta I}^{(\gamma)} + Z  A_{\Delta I}^{(Z)} +
 g_{\rm ewk}  A_{\Delta I}^{(g)} \bigg]  \Bigg\}.
 \label{eq:structure1}
\end{eqnarray}
The meaning of the amplitudes $A_{\Delta I}^{(X)}$ can be inferred
from the superscript $X$.  $A_{\Delta I}^{(8)}, A_{\Delta I}^{(27)}$
represent the octet and 27-plet amplitudes in the isospin limit.
$A_{\Delta I}^{(\varepsilon)}$ represents the effect of strong
isospin breaking, while the electromagnetic contribution is split
into a part induced by photon loops $A_{\Delta I}^{(\gamma)}$ and the
parts induced by insertions of $Z$ and $g_{\rm ewk}$ vertices
($A_{\Delta I}^{(Z)}$ and $A_{\Delta I}^{(g)}$, respectively). The
photon loops are actually infrared divergent:
$A_{\Delta I}^{(\gamma)}$ are the infrared-finite
(structure-dependent) terms whereas the divergent ``infrared components" must be
treated in combination with real photon emission. Details on this
decomposition can be found in \textcite{Cirigliano:2003gt}.

Confronting the amplitudes in Eq.~(\ref{eq:structure1}) with
experimental rates \cite{Antonelli:2010yf}, one can extract the
LO couplings $G_8, G_{27}$ and the phase difference $\chi_0 -
\chi_2$. Using instead the dimensionless couplings $g_8, g_{27}$
defined in Eq.~(\ref{eq:defgn}),
an update of the analysis of \textcite{Cirigliano:2003gt} produces
the results in Table \ref{tab:fitcouplings}.
\renewcommand{\arraystretch}{1.2}
\begin{center}
\begin{table*}[floatfix]

\caption{\label{tab:fitcouplings} Weak couplings $g_8, g_{27}$ and phase
  difference $\chi_0 - \chi_2$ at LO and NLO, with (IV) and without
  (IC) isospin violation. The theoretical uncertainties are twice the
  ones assigned by \textcite{Cirigliano:2003gt}.
\\}

{\begin{tabular}{|c|c|c|c|c|}
\hline
& & & & \\[-.4cm]
\hspace*{.1cm} chiral order \hspace*{.1cm} & \hspace*{.1cm} isospin
\hspace*{.1cm} & \hspace*{.5cm} $g_8$ \hspace*{.5cm} &
\hspace*{.5cm} $g_{27}$ \hspace*{.5cm} & \hspace*{.5cm} $\chi_0 -
\chi_2$ \hspace*{.5cm}   \\[4pt]
\hline
& & & & \\[-.3cm]
LO & IC & $4.96$ & $0.285$ & $47.5^\circ$ \\[.1cm]
LO & IV & $4.99$ & $0.253$ & $47.8^\circ$ \\[.1cm]
NLO & IC & \hspace*{.1cm} $3.62 \pm 0.002_{\rm exp} \pm  0.28_{\rm th}$
 \hspace*{.1cm} & \hspace*{.1cm} $0.286 \pm 0.0006_{\rm
  exp} \pm  0.028_{\rm th}$ \hspace*{.1cm}  &
 \hspace*{.1cm} $(47.5 \pm 0.9_{\rm exp})^ \circ$
\hspace*{.1cm}\\[.1cm]
NLO & IV & \hspace*{.1cm} $3.61 \pm 0.002_{\rm exp} \pm  0.28_{\rm th}$
 \hspace*{.1cm}
& \hspace*{.1cm} $0.297 \pm 0.0006_{\rm exp} \pm  0.028_{\rm th}$
\hspace*{.1cm}  & \hspace*{.1cm} $(51.3 \pm 0.8_{\rm exp})^ \circ$
\hspace*{.1cm}  \\[5pt]
\hline
\end{tabular}}

\end{table*}
\end{center}
\noindent
Comments.
\begin{itemize}
\item[i.] The reduction of the octet coupling $g_8$ at $\mathcal{O}(G_F p^4)$
  corresponds to the
  enhancement of the $\Delta I=1/2$ amplitude by roughly 30 $\%$ at
  NLO \cite{Kambor:1991ah}. The (rather generous) theory errors of
  $g_8$, $g_{27}$ at NLO account for the uncertainties of the various
  LECs involved. They should not be interpreted to include also
  effects of NNLO or higher. In fact, a calculation of the
  leading (double) chiral logarithms at NNLO \cite{Buchler:2005xn}
  confirmed that $\pi\pi$ rescattering generates an
  additional enhancement of $A_0$ at higher orders
  \cite{Pallante:1999qf,Pallante:2000hk}.
  However, the values of $g_8$, $g_{27}$ displayed in Table
  \ref{tab:fitcouplings} are the appropriate values for nonleptonic
  weak amplitudes at LO or NLO, respectively.
\item[ii.] At NLO, both $g_8$ and $g_{27}$ receive small
shifts from isospin-violating corrections. While this could be expected
for $g_8$, it results from a cancellation of different effects in the
case of $g_{27}$. On the other hand, as indicated in Table
\ref{tab:fitcouplings}, the inclusion of isospin breaking reduces
$g_{27}$ by roughly $10 \%$ at LO.
\item[iii.] The phase difference $\chi_0 - \chi_2$ is
  taken as a fit parameter because the phases cannot be reliably
  calculated to NLO in CHPT. At NLO, the fitted phase difference is
  rather sensitive to isospin violation. The more interesting
  difference of  s-wave $\pi \pi$ scattering phase shifts in the
  isospin limit $\delta_0 - \delta_2$ at  $\sqrt{s} = M_K$ will be
  discussed below.
\item[iv.] At NLO, electromagnetic corrections
  induce a $\Delta I=5/2$ amplitude:
\begin{eqnarray}
f_{5/2} &\equiv& {\rm Re}A_2/{\rm Re}A_2^+ - 1 \nn
&=& \big( 8.44 \pm 0.02_{\rm exp} \pm 2.5_{\rm th} \big)
\times 10^{-2}.
\label{eq:f52}
\end{eqnarray}
\end{itemize}

\subsubsection{$\pi\pi$ phase shifts from $K \to \pi\pi$ decays}
\label{subsec:pipifromK}
In the isospin limit, the phase difference $\chi_0 - \chi_2$
accessible in $K \to \pi\pi$ decays equals the s-wave $\pi\pi$
phase shift difference $\delta_0(M_K) -\delta_2(M_K)$ (Watson's theorem).
It has been a long-standing problem to reconcile the phase shift
difference extracted from $K \to \pi\pi$ decays with other
determinations of pion-pion phase shifts. This problem became
especially acute after the precise determination of $\pi\pi$ phase
shifts from combining dispersion theory with CHPT
\cite{Colangelo:2001df}. We review here the
present status of this problem \cite{Cirigliano:2009rr}.

\begin{itemize}
\item The experimental situation has substantially improved in recent
 years for both the $K^+$ and $K_S$ lifetimes and for the branching ratios
 of $K \to \pi\pi$ decays \cite{Antonelli:2010yf}. Compared to the
 analysis of \textcite{Cirigliano:2003gt}, the new experimental
 information reduces the phase shift difference by more than three
 degrees (with higher statistical significance), bringing it
 closer to the dispersion theoretical value \cite{Cirigliano:2008kn}.
\item In the original analysis of \textcite{Cirigliano:2003gt} the
 differences $\gamma_I  = \chi_{I} - \delta_{I}(M_K)$ were
 calculated in CHPT although the $\chi_I$ and $\delta_I$ separately
 come out much too small at NLO in the chiral expansion. The
 analysis can be improved by  relying to a lesser  extent on the NLO
 calculation of $K \to \pi\pi$ amplitudes altogether. Using less
 information potentially increases the uncertainty but this is
 compensated by a less biased comparison with the data. The resulting
 estimate of isospin violation is more robust
 and it leads to a further decrease of the
 phase shift difference by nearly two degrees.
\end{itemize}

The main idea of the alternative procedure of
\textcite{Cirigliano:2009rr} is to use only the isospin-violating
parts of the NLO amplitudes as theory input and to determine
$\delta_0(M_K) - \delta_2(M_K)$ directly from the data. In
contrast to the chiral corrections for the full amplitudes, the
isospin-violating (IV) corrections are much smaller and therefore more
suitable for a perturbative estimate.

The amplitudes are now parametrized as
\begin{eqnarray}
A_{+-} &=&
\overline{A}_{0} \, e^{i \delta_0(M_K)}  + { 1 \over \sqrt{2}} \,
\overline{A}_{2}\,  e^{i\delta_2(M_K) } + \Delta A_{+-}^{\rm IV},
\nn
A_{00} &=&
\overline{A}_{0} \, e^{i \delta_0(M_K)}  - \sqrt{2} \,   \overline{A}_{2}\,
e^{i\delta_2(M_K) } + \Delta A_{00}^{\rm IV}, \nn
A_{+0} &=& {3 \over 2} \,
\overline{A}_{2} \,  e^{i\delta_2(M_K)} + \Delta A_{+0}^{\rm IV}.
 \label{eq:amps}
\end{eqnarray}
All isospin violation is contained in
$\Delta A_{+-}^{\rm IV}, \,\Delta A_{00}^{\rm IV}, \,\Delta A_{+0}^{\rm IV}$ that
can be extracted from the NLO amplitudes of
\textcite{Cirigliano:2003gt}. Since isospin violation was neglected in
the 27-plet amplitudes in view of the $\Delta I=1/2$ rule,
$\Delta A_{n}^{\rm IV}$ ($n=+-,00,+0$) scale linearly with
the lowest-order octet coupling $g_8$.

The moduli of the amplitudes in the isospin limit
are denoted as $\overline{A}_{0}, \,\overline{A}_{2}$.
These amplitudes together with the phase shift difference $\delta_0(M_K) -
\delta_2(M_K)$ are then determined directly from the rates. For this
purpose, the moduli are written as
\begin{eqnarray}
|A_{+-}| &=&
\left| \overline{A}_{0}  + { 1 \over \sqrt{2}} \,
\overline{A}_{2}\,  e^{i(\delta_2(M_K)- \delta_0(M_K))}\right. \nn
&& ~+ \left.\Delta A_{+-}^{\rm IV}\,  e^{- i \delta_0(M_K)}\right|,
\nn
|A_{00}| &=&
\left| \overline{A}_{0}   - \sqrt{2} \, \overline{A}_{2}\,
e^{i(\delta_2(M_K)- \delta_0(M_K))}\right.  \nn
&& ~+ \left.\Delta A_{00}^{\rm IV} \,  e^{- i
 \delta_0(M_K)}\right|,
\label{eq:moduli}
\\
|A_{+0}| &=& \left| {3 \over 2} \,
\overline{A}_{2}\, e^{i(\delta_2(M_K)- \delta_0(M_K))}  + \Delta
A_{+0}^{\rm IV} \, e^{- i \delta_0(M_K)}\right|.
\no
\end{eqnarray}
In order to determine $\overline{A}_{0}, \overline{A}_{2}$ and
$\delta_0(M_K)- \delta_2(M_K)$ from the three rates, also
the $I=0$ phase $\delta_0(M_K)$ is needed as input
\cite{Colangelo:2001df}:
\begin{equation}
\delta_0(M_K) = (39.2 \pm 1.5)^\circ.
\label{eq:d0}
\end{equation}
From the structure of the moduli in Eq.~(\ref{eq:moduli}) it is obvious
that the precise value of $\delta_0(M_K)$ has little
impact on the phase shift difference.

With the (updated) values of $\Delta A_{n}^{\rm IV}$
\cite{Cirigliano:2003gt} and with $g_8=3.6$ (see Table
\ref{tab:fitcouplings}), the
experimental rates \cite{Antonelli:2010yf} give rise to
\begin{eqnarray}
\overline{A}_{0} = (2.7030\pm 0.0008) \times 10^{-7} ~~{\rm GeV}, \nn
\overline{A}_{2} = (0.1249\pm 0.0003) \times 10^{-7} ~~{\rm GeV}, \nn
\delta_0(M_K)- \delta_2(M_K) = (52.54 \pm 0.83)^\circ,
\label{eq:fit}
\end{eqnarray}
\noindent
where the errors are purely experimental.
The octet enhancement in $K \to \pi\pi$ decays is characterized by the
amplitude ratio
\begin{equation}
\displaystyle\frac{\overline{A}_{0}}{\overline{A}_{2}} = 21.63\pm 0.04,
\label{eq:ratio}
\end{equation}
again with experimental error only.

There are various sources of theoretical uncertainties associated with
the phase shift difference. Whereas the error of $\delta_0$ in
Eq.~(\ref{eq:d0}) is completely negligible, a generous error of 20\% for
the overall scale $g_8$ of the isospin-violating amplitudes
gives rise to an uncertainty $\pm
1.1^\circ$ for the phase difference. The major part of the error is
due to unknown effects of $\mathcal{O}(e^2 p^4)$ that were estimated by
\textcite{Cirigliano:2009rr} in two different ways. Altogether, the
final value for the phase shift difference in the isospin limit from
$K \to \pi\pi$ decays is
\begin{eqnarray}
&& \hspace*{-2.cm} \left[ \delta_0(M_K)- \delta_2(M_K)
    \right]_{K \to \pi\pi}\nn
&=&   (52.5 \pm  0.8_{\mathrm{exp}} \pm 2.8_{\mathrm{th}})^\circ.
\label{eq:result}
\end{eqnarray}
Before comparing this result with other
determinations, one should recall that
the isospin limit is defined in terms of the neutral meson
masses. However, a LO estimate \cite{Cirigliano:2009rr} suggests that
expanding instead around the charged pion mass modifies the phase shift
difference only by much less than a degree.

The most recent determinations of the phase shift difference from
$\pi\pi$ scattering data are
\begin{eqnarray}
&& \hspace*{-1.2cm} \left[ \delta_0(M_K) -
 \delta_2(M_K)\right]_{\pi\pi}\nn
 &=&\left\{
\begin{array}{cc}
 (47.7 \pm 1.5)^\circ  &  \mbox{\protect\cite{Colangelo:2001df}},
 \\[7pt]
 (50.9 \pm 1.2)^\circ  &  \mbox{\protect\cite{Kaminski:2006qe}},
 \\[7pt]
 (47.7 \pm 0.4)^\circ  &  \mbox{\protect\cite{Batley:2010zza}},
\\[7pt]
(47.3 \pm 0.9)^\circ  &  \mbox{\protect\cite{Garcia:2011cn}}.
 \end{array}\right.\quad
\label{eq:d02_pipi}
\end{eqnarray}
These values would agree perfectly well with the phase difference
$\chi_0 - \chi_2$ in Eq.~(\ref{eq:isoamps}), obtained in the isospin
limit. However, due to the large ratio $A_0/A_2$ in $K\to 2 \pi$
decays, isospin-breaking corrections to the dominant $\Delta I=1/2$
amplitude generate sizable contributions to $A_2$, modifying also the
amplitude phases. The
updated determination of the phase shift difference from $K\to 2 \pi$
decays in Eq.~(\ref{eq:result}) \cite{Cirigliano:2009rr} turns out to be
in reasonable agreement with the $\pi\pi$ results in
Eq.~(\ref{eq:d02_pipi}), although with a larger uncertainty.

\subsubsection{$\epsp/\eps$}
\label{subsec:epsprime}

The CP-violating ratio  $\epsilon'/\epsilon$  constitutes
a fundamental test for our understanding of flavor-changing
phenomena.
$\epsilon$ and $\epsilon'$ parametrize
different sources of CP violation in $K_L\to\pi\pi$:
\beqa\label{eq:epsilon_def}
\eta_{_{+-}} &\equiv & {A(K_L\to\pi^+\pi^-)\over A(K_S\to\pi^+\pi^-)}
\, = \, \epsilon + \epsilon'\, ,
\nn
\eta_{_{00}} &\equiv & {A(K_L\to\pi^0\pi^0)\over A(K_S\to\pi^0\pi^0)}
\, = \, \epsilon -2 \,\epsilon'\, .
\eeqa
The dominant effect from CP violation in $K^0$-$\overline{K^0}$ mixing
is contained in $\epsilon$, while $\epsilon'$ accounts for direct
CP violation in the decay amplitudes.
The present experimental world average
\cite{Abouzaid:2010ny,Batley:2002gn,Lai:2001ki,Fanti:1999nm,
AlaviHarati:2002ye,AlaviHarati:1999xp,
Gibbons:1993zq,Barr:1993rx,Burkhardt:1988yh},
\beq\label{eq:exp}
{\rm Re} \left(\epsilon'/\epsilon\right) =
\frac{1}{3} \left( 1
  -\left|\frac{\eta_{_{00}}}{\eta_{_{+-}}}\right|\right) =\,
(16.8 \pm 1.4) \times
 10^{-4} \, ,
\eeq
demonstrates the existence of direct CP violation in K
decays.

When CP violation is turned on, the amplitudes $A_0$, $A_2$, $A_2^+$
acquire imaginary parts. To first order in CP violation, $\epsp$ is given by
\begin{equation}
\epsp  = - \frac{i}{\sqrt{2}} \, e^{i ( \chi_2 - \chi_0 )} \,
\frac{\real A_{2}}{ \real A_{0}} \,
\left[
\frac{\imag A_{0}}{ \real A_{0}} \, - \,
\frac{\imag A_{2}}{ \real A_{2}} \right] .
\label{eq:cp1}
\end{equation}
Since $\imag A_{I}$ is CP-odd the quantities $\real A_{I}$ and
$\chi_{I}$ are only needed in the CP limit ($I=0,2$).
$\epsp$ is suppressed by the small ratio $\real A_{2}/ \real A_{0}\approx 1/22$.
The phase $\phi_\epsp =\chi_2 - \chi_0 + \pi/2 = (42.5\pm 0.9)^\circ$
is very close to the so-called superweak phase \cite{Nakamura:2010zzb}
\begin{equation}
\phi_\eps \approx
\tan^{-1}
\left(
\frac{2 (M_{K_L}-M_{K_S})}{\Gamma_{K_S}-\Gamma_{K_L}}
\right)
= (43.51\pm 0.05)^\circ,
\end{equation}
implying that $\cos{(\phi_\epsp -\phi_\eps)}\approx 1$.

To obtain the theoretical SM prediction for $\epsp$, the CP-conserving
amplitudes $\real A_{I}$ are set to their experimentally determined values.
This procedure avoids the large uncertainties associated with the
hadronic matrix elements of the four-quark operators in $\cL_{\mathrm{eff}}^{\Delta S=1}$,
in particular $\real A_{0}$ that involves several octet
operators with a complicated
mixing under the renormalization group.
Thus, one only needs a first-principle calculation of the CP-odd amplitudes $\imag A_{0}$
and $\imag A_{2}$; the first one is completely dominated by the strong penguin operator
$Q_6$, while the leading contribution to the second one comes from the
electromagnetic penguin $Q_8$.
Fortunately, those are precisely the operators that are
expected to be better approximated
through our large-$N_C$ estimate of LECs.

Equation (\ref{eq:cp1}) involves a delicate balance between the two isospin contributions.
A naive estimate of $\imag A_{I}$ at lowest order in the chiral
expansion, i.e., using
the tree-level formulae in Eq.~(\ref{eq:LOamp}), results in a large
numerical cancellation leading
to unrealistically low values of $\epsp/\eps$ around $7\times 10^{-4}$
\cite{Buras:2003zz,Buras:2000qz,Bosch:1999wr,Buchalla:1995vs,
Ciuchini:1999xi,Ciuchini:1997bw,Ciuchini:1993vr,Ciuchini:1992tj}.
The true SM prediction is then very sensitive to the precise values of the two
contributing amplitudes \cite{Bertolini:1998vd,Hambye:1999yy}.
The one-loop CHPT corrections generate an important enhancement
($\sim 35\%$) of the isoscalar amplitude ($\pi\pi$ rescattering)
and a reduction of $A_{2}$, destroying the accidental
lowest-order cancellation and bringing the SM prediction
of $\epsp/\eps$ in good agreement with the experimental measurement
\cite{Pallante:2001he,Pallante:1999qf,Pallante:2000hk}.

Owing to the large ratio  $\real A_{0}/ \real A_{2}$, isospin violation
plays also an important role in $\epsp/\eps$. Small IV corrections
proportional to the large octet coupling $G_8$ feed into the small
amplitude $A_2$  generating relatively large contributions, which can
modify the predicted value of $\epsp$ in a sizable way.
A systematic analysis of isospin-breaking corrections in $\epsp$ was
undertaken in \textcite{Cirigliano:2003nn} where references to earlier
work can be found. To first order in isospin violation, one finds
\begin{equation}
\epsp \! = \! - \frac{i}{\sqrt{2}}  e^{i ( \chi_2 - \chi_0 )}
\omega_+ \!   \left[
\frac{\imag A_{0}^{(0)}}{ \real A_{0}^{(0)}}
(1 + \Delta_0 + f_{5/2}) \! - \! \frac{\imag A_{2}}{ \real
  A_{2}^{(0)}} \right] ,
\label{eq:cpiso}
\end{equation}
where
\begin{eqnarray}
\omega_+ &=& \real A_{2}^{+}/\real A_{0}, \nn
\Delta_0 &=& \frac{\imag A_0}{\imag A_0^{(0)}}
\frac{\real A_0^{(0)}}{\real A_0}  - 1
\label{eq:epspdefs}
\end{eqnarray}
and $f_{5/2}$ is defined in Eq.~(\ref{eq:f52}). The superscript $(0)$
on the amplitudes denotes the isospin limit.

$\imag A_2$ is itself first order in isospin breaking. One
usually separates the
electromagnetic penguin contribution to $\imag A_2$ from the
isospin-breaking effects generated by other four-quark operators:
\begin{equation}
\imag A_2 = \imag A_2^{\rm emp} \ + \ \imag  A_2^{\rm non-emp}.
\label{eq:empsep}
\end{equation}
A discussion of this scheme-dependent separation in the framework of
CHPT can be found in \textcite{Cirigliano:2003nn}. Splitting off the
electromagnetic penguin contribution to $\imag A_2$
in this way, one can write $\epsp$ in a more familiar form as
\begin{equation}
\epsp = - \displaystyle\frac{i}{\sqrt{2}}  e^{i ( \chi_2 - \chi_0 )}
\omega_+   \left[
\displaystyle\frac{\imag A_{0}^{(0)} }{ \real A_{0}^{(0)} }
(1 - \Omega_{\rm eff}) - \displaystyle\frac{\imag A_{2}^{\rm emp}}{ \real
  A_{2}^{(0)} } \right]  ,
\label{eq:cpeff}
\end{equation}
where
\begin{eqnarray}
\Omega_{\rm eff} &=& \Omega_{\rm IV} - \Delta_0 - f_{5/2},
\label{eq:omegaeff} \\
\Omega_{\rm IV} &=& \displaystyle\frac{\real A_0^{(0)} }
{ \real A_2^{(0)} } \times \displaystyle\frac{\imag A_2^{\rm non-emp} }
{ \imag A_0^{(0)} }. \label{eq:omegaIB}
\end{eqnarray}
The quantity $\Omega_{\rm eff}$ includes all effects to leading order
in isospin breaking and it generalizes the more traditional parameter
$\Omega_{\rm IV}$. Although $\Omega_{\rm IV}$ is in principle
enhanced by the large ratio $\real A_0^{(0)}/ \real A_2^{(0)}$, the
actual numerical analysis shows all three terms in
Eq.~(\ref{eq:omegaeff}) to be relevant when both
strong and electromagnetic isospin violation are included
\cite{Cirigliano:2003nn}.

The numerical analysis of \textcite{Cirigliano:2003nn} found large
cancellations among the different contributions to $\Omega_{\rm eff}$.
A well-known example are the
contributions of strong isospin violation via $\pi^0$--$\eta$ ~mixing where the sum
of $\eta$ and $\eta^\prime$ exchange generates an $\Omega_{\rm IV}$ of
the order of $25\%$. However, already at the level of
$\pi^0$--$\eta$ ~mixing alone, a complete NLO calculation
\cite{Ecker:1999kr} produces a destructive interference in
$\Omega_{\rm IV}$, with  $\Omega_{\rm IV}= (15.9 \pm 4.5)
\times 10^{-2}$.  Inclusion of electromagnetic effects slightly increases
$\Omega_{\rm IV}$ and generates sizable $\Delta_0$ and $f_{5/2}$, which
interfere destructively with $\Omega_{\rm IV}$ to produce the final
result \cite{Cirigliano:2003nn}:
\beq
\Omega_{\rm eff}\, =\, (6.0 \pm 7.7)\times 10^{-2} .
\eeq

The small value obtained for $\Omega_{\rm eff}$
reinforces the dominance of the gluonic penguin operator $Q_6$ in $\eps'$.
Taking this into account and updating all other inputs,
the SM prediction for $\epsilon'/\epsilon$ turns out to be
\cite{Pallante:2001he,Pich:2004ee}
\beq\label{eq:finalRes}
\mbox{Re}\left(\epsilon'/\epsilon\right) \; =\;
\left(19\pm 2\, {}_{-6}^{+9} \pm 6\right) \times 10^{-4}\, ,
\eeq
in excellent agreement with the experimental measurement
shown in Eq.~(\ref{eq:exp}).
The first error was estimated by varying the
renormalization scale $\mu$ between $M_\rho$ and $m_c$.
The uncertainty induced by $m_s$, which was taken in
the range 
$m_s(2\, \rm{GeV})=110\pm 20\, \rm{MeV}$,
is indicated by the second error.

The most critical step is the matching between the short- and long-distance
descriptions, which was done at leading order in $1/N_C$.
Since all next-to-leading ultraviolet (OPE) and infrared (CHPT)
logarithms have been
taken into account, our educated guess for the theoretical
uncertainty associated with subleading contributions is $\sim 30\% $
(third error).

The control of non-logarithmic corrections at NLO
in $1/N_C$ remains a challenge for future investigations.
Several dispersive analyses
\cite{Cirigliano:2002jy,Cirigliano:2001hs,Cirigliano:2001qw,Cirigliano:1999pv,Donoghue:1999ku,
Knecht:2001bc,Knecht:1998nn,Narison:2000ys,
Bijnens:2003hk,Bijnens:2001ps} 
and lattice calculations
\cite{Noaki:2001un,Blum:2001xb,Boucaud:2004aa}
of $\langle Q_8\rangle\equiv \langle 2\pi|Q_8|K\rangle$
already exist (most of them in the chiral limit). Taking the chiral corrections
into account, those results are compatible with the value used in
Eq.~(\ref{eq:finalRes}). Unfortunately, the hadronic matrix element
$\langle Q_6\rangle\equiv \langle 2\pi|Q_6|K\rangle$
is more difficult to compute.
Two analytical estimates in the chiral limit,
using the so-called minimal hadronic approximation
\cite{Hambye:2003cy} 
and the X-boson approach \cite{Bijnens:2000im,Bijnens:1998ee},
find large $1/N_C$ corrections to $\langle Q_6\rangle$.
It would be interesting to understand the physics behind those
contributions and to study whether corrections
of similar size are present for physical values of the quark masses.

Lattice calculations of $\langle Q_6\rangle$ are still not very
reliable
and give contradictory results, often with the wrong sign
\cite{Noaki:2001un,Blum:2001xb, 
Bhattacharya:2004qu,Pekurovsky:1998jd}.
The usual procedure in lattice simulations has been to calculate
$K\to\pi$ and $K\to\mathrm{vacuum}$ matrix elements and use lowest-order CHPT to recover
the physical $K\to 2\pi$ amplitudes. In addition to all usual lattice artifacts
(finite volume and lattice spacing, quenched approximation,
unphysical masses, etc.), this misses completely the crucial role of
final-state interactions in the isoscalar octet amplitude. Thus, a truly major effort is needed
to compute the $K\to 2\pi$ matrix elements directly. Some progress in this direction
has been achieved, relating the physical $K\to 2\pi$ amplitudes to the
corresponding matrix elements in a finite Euclidean volume, which are better suited
for lattice simulations \cite{Lellouch:2000pv,Lin:2001ek}.
So far,
this technique has been implemented in simulations of the
$\Delta I=3/2$ ~$K^+\to\pi^+\pi^0$ amplitude with promising results
\cite{Kim:2010sd,Boucaud:2004aa,Goode:2011kb,Liu:2010fb}.
A first estimate of the $A_0$ amplitude at
unphysical kinematics has been recently reported
\cite{Blum:2011pu}.

More work is needed to reduce the present uncertainty quoted
in Eq.~(\ref{eq:finalRes}). This is a difficult task, but progress in this
direction may be expected in the next few years.

\subsubsection{CP violation in $K^0$--$\overline{K^0}$ mixing}
\label{subsec:epsilon}

Since $\mathrm{Re}(\epsilon'/\epsilon)\ll 1$, the ratios $\eta_{_{+-}}$ and $\eta_{_{00}}$
provide a direct measurement of $|\epsilon|$ \cite{Nakamura:2010zzb}:
\beq\label{eq:eps_exp}
|\epsilon| \, =\, \frac{1}{3} \left( 2 |\eta_{_{+-}}| + |\eta_{_{00}}|\right) \, =\,
(2.228\pm 0.011)\times 10^{-3} ,
\eeq
in perfect agreement with the semileptonic asymmetry
\beqa
\lefteqn{\frac{\Gamma(K_L\to\pi^-\ell^+\nu) - \Gamma(K_L\to\pi^+\ell^-\bar\nu)}{\Gamma(K_L\to\pi^-\ell^+\nu) + \Gamma(K_L\to\pi^+\ell^-\bar\nu)}\, =\, \frac{2\mathrm{Re}(\epsilon)}{1+|\epsilon|^2}} &&
\nn
& \hskip 3cm  & =\; (3.32\pm 0.06)\times 10^{-3} .\hskip 1.5cm
\eeqa

The theoretical prediction
can be written in the form \cite{Buchalla:1995vs,Buras:2008nn}
\beqa\label{eq:epsil}
|\epsilon| & = & C_\epsilon k_\epsilon\hat B_K A^2\lambda^6\bar\eta
\times\left\{ A^2\lambda^4(1-\bar\rho)\eta_{tt}S_0(x_t)
\right. \\ &&\hskip 2.5cm\left.\mbox{}
+ \eta_{ct}S_0(x_c,x_t) - \eta_{cc}S_0(x_c)\right\} ,\quad\;\nonumber
\eeqa
where $x_i = m_i^2/M_W^2$, $S_0(x_i,x_j)$ and $S_0(x_i)$ are the
\textcite{Inami:1980fz} box functions,
\beq
C_\epsilon = \frac{G_F^2F_K^2M_{K^0}M_W^2}{3\sqrt{2}\pi^2(M_{K_L}-M_{K_S})} = 3.7\times 10^4 ,
\eeq
$\eta_{tt}^{\mbox{\tiny NLO}} = 0.5765\pm 0.0065$, $\eta_{cc}^{\mbox{\tiny NLO}} = 1.43\pm 0.23$ and $\eta_{ct}^{\mbox{\tiny NNLO}} = 0.496\pm 0.047$
are short-distance QCD corrections \cite{Buras:1990fn,Herrlich:1996vf,Brod:2010mj} and
$k_\epsilon = 0.94\pm 0.02$ accounts for small long-distance contributions \cite{Buras:2010pza}.

The renormalization-group-invariant  parameter $\hat B$ measures the hadronic matrix element
$\langle \bar K^0|Q_{\Delta S=2}|K^0\rangle$ in units of its vacuum saturation approximation.
In the large--$N_C$ limit, $\hat B = 3/4$ \cite{Gaiser:1980gx,Buras:1985yx}.
The most precise lattice
determinations, obtained with $2+1$ active flavors, quote
$\hat B = 0.724\pm 0.030$ \cite{Aubin:2009jh} and $\hat B = 0.749\pm
0.027$ \cite{Aoki:2010pe}.

Eq.~(\ref{eq:epsil}) provides a parabolic constraint in the plane $\bar\rho$--$\bar\eta$ that is included
in the SM unitarity triangle fits \cite{Bona:2006ah,Charles:2004jd}.
The recent precise lattice value for $\hat B$ introduces some tension in the fit. Determining
$\bar\eta$ and $\bar\rho$ from other observables, it implies $|\epsilon|=(1.90\pm 0.26)\times 10^{-3}$
\cite{Brod:2010mj}, slightly smaller than (\ref{eq:eps_exp}).


\subsection{$K \to 3\pi$ }
\label{subsec:3pi}
There are five CP-conserving decays to three pions (the $K^-$ decays
are not listed separately):
\begin{eqnarray}
A^L_{000}&=&A(K_L(k) \to \pi^0(p_1)\pi^0(p_2)\pi^0(p_3)),\nn
A^L_{+-0}&=&A(K_L(k) \to \pi^+(p_1)\pi^-(p_2)\pi^0(p_3)),\nn
A^S_{+-0}&=&A(K_S(k) \to \pi^+(p_1)\pi^-(p_2)\pi^0(p_3)),\nn
A_{00+}&=&A(K^+(k) \to \pi^0(p_1)\pi^0(p_2)\pi^+(p_3)),\nn
A_{++-}&=&A(K^+(k) \to \pi^+(p_1)\pi^+(p_2)\pi^-(p_3)).\qquad
\label{eq:K3pi}
\end{eqnarray}
For the kinematics one uses the variables
\begin{equation}
s_1 = \left(k-p_1\right)^2,~
s_2 = \left(k-p_2\right)^2,~
s_3 = \left(k-p_3\right)^2
\end{equation}
with
\begin{eqnarray}
s_0 &=&
\frac{1}{3}(s_1+s_2+s_3) \nn
&=& \frac{1}{3}\left(M_K^2+M_{\pi^1}^2+M_{\pi^2}^2+M_{\pi^3}^2\right),
\label{eq:s0}
\end{eqnarray}
where the masses are those of the particles
appearing in the decay under consideration.

In terms of the Dalitz plot variables
\begin{equation}
\label{eq:defxy}
x = \frac{s_2-s_1}{M_{\pi^+}^2},
\qquad
y = \frac{s_3-s_0}{M_{\pi^+}^2},
\end{equation}
the amplitudes are often expanded as \cite{Devlin:1978ye}
\begin{eqnarray}
\label{eq:ddexp}
A^L_{000}&=& 3(\alpha_1+\alpha_3)
         +3 (\zeta_1-2\zeta_3)\left(y^2+\frac{1}{3}x^2\right),
\nn
A^L_{+-0}&=& (\alpha_1+\alpha_3)-(\beta_1+\beta_3)y \nn
         && + (\zeta_1-2\zeta_3)\left(y^2+\frac{1}{3}x^2\right)
\nn
&& +         (\xi_1-2\xi_3)\left(y^2-\frac{1}{3}x^2\right),
\nn
A^S_{+-0}&=& \frac{2}{3}\sqrt{3}\,\gamma_3 x- \frac{4}{3}\xi_3^\prime
         xy,
 \\
A_{00+}&=& \left(-\alpha_1+\frac{1}{2}\alpha_3\right)
           +\left(\beta_1-\frac{1}{2}\beta_3-\sqrt{3}\gamma_3\right)y
\nn
&& -  (\zeta_1+\zeta_3)\left(y^2+\frac{1}{3}x^2\right) \nn
&& - (\xi_1+\xi_3+\xi_3^\prime)\left(y^2-\frac{1}{3}x^2\right),
\nn
A_{++-}&=& \left(-2\alpha_1+\alpha_3\right)
           +\left(-\beta_1+\frac{1}{2}\beta_3-\sqrt{3}\gamma_3\right)y
\nn
&& -  (2\zeta_1+2\zeta_3)\left(y^2+\frac{1}{3}x^2\right) \nn
&& +  (\xi_1+\xi_3-\xi_3^\prime)\left(y^2-\frac{1}{3}x^2\right).
\no
\end{eqnarray}

To NLO in the chiral expansion, the $K \to 3\pi$ amplitudes were first
calculated by \textcite{Kambor:1991ah} in the isospin limit. Their
analysis was repeated and updated by
\textcite{Bijnens:2002vr}, confirming the main observation of
\textcite{Kambor:1991ah,Kambor:1992he} that inclusion of NLO corrections
significantly improves the LO current algebra amplitudes. One main
reason for the much better agreement with experimental data is that
the quadratic slope parameters $\zeta_1$,\dots,$\xi_3^\prime$ vanish
at LO. On the other hand, NLO CHPT only provides the leading
contributions for these quadratic slope parameters, which moreover
depend on a number of LECs. Some assumptions about the combinations of
NLO LECs occurring in the amplitudes must be made for a comparison
with experiment.

Experiments on $K \to 3\pi$ decays provide information on the rates
and on the Dalitz plot distributions. The latter are conventionally
expanded up to second order in  $x,y$ (assuming again CP conservation):
\begin{equation}
\left|\frac{A(s_1,s_2,s_3)}{A(s_0,s_0,s_0)}\right|^2
= 1 + g y + h y^2 + k x^2,
\label{eq:dalitzexp}
\end{equation}
except for $K_S \to \pi^+\pi^-\pi^0$ [see Eq.~(\ref{eq:ddexp})]. The
experimental data available at the end of 2004 were confronted with
NLO CHPT by \textcite{Bijnens:2004ku,Bijnens:2004vz,Bijnens:2004ai},
including also isospin-violating and radiative corrections. Assuming
that the additional electromagnetic LECs at NLO in the Lagrangian
(\ref{eq:Lelweak}) all vanish at a certain scale, their conclusion was
that isospin breaking in $K \to 3\pi$ is in general small. That
analysis seems worth repeating, not only because of better
knowledge of LECs reviewed in Sec.~\ref{sec:lecs},
but also because of the more
precise experimental information on the Dalitz plot slopes in
Eq.~(\ref{eq:dalitzexp}), in particular for the $K^\pm$ modes from NA48/2
[see \textcite{Batley:2010fj} and references therein].

\subsubsection{CP violation in $K \to 3\pi$ decays}
\label{subsec:CP3pi}
The decay $K_S \to 3 \pi^0$ violates CP. In analogy to $K^0
\to 2 \pi$ in Eq.~(\ref{eq:epsilon_def}), one defines the amplitude
ratio
\begin{eqnarray}
\eta_{000} = \displaystyle
\left.\frac{A^S_{000}}{A^L_{000}}\right|_{x=y=0}
&=&   \epsilon + \epsilon'_{000}.
\end{eqnarray}
The parameter $\epsilon'_{000}$ is a measure of direct CP
violation. To lowest order in CHPT, there is a simple relation between
$\epsilon'_{000}$ and $\epsilon'$ \cite{Li:1979wa,D'Ambrosio:1996nm},
which implies
\begin{equation}
|\epsilon'_{000}| \ll |\epsilon|.
\end{equation}
Therefore, it will be very difficult to detect direct CP violation in
this decay. In fact, $K_S \to 3 \pi^0$ has not been observed at all so
far. The accurate theoretical prediction
\begin{equation}
\mathrm{BR}(K_S \to 3 \pi^0)= 1.9 \times 10^{-9}
\end{equation}
should be compared with
the best upper
bound \cite{Ambrosino:2005iw}
\begin{equation}
\mathrm{BR}(K_S \to 3 \pi^0) < 1.2 \times 10^{-7}
\quad
({\rm 90 \% \, \, C.L.}),
\end{equation}
corresponding to
\begin{equation}
|\eta_{000}| < 0.018.
\end{equation}

In the search for direct CP violation in $K$ decays, the three-pion
decays of charged kaons have played a prominent role. In addition to
rate asymmetries, both theory and experiment have paid special
attention to asymmetries in the linear Dalitz plot parameter $g$
defined in Eq.~(\ref{eq:dalitzexp}).

After a number of conflicting estimates [reviewed by
\textcite{D'Ambrosio:1996nm}], the theoretical state-of-the-art
Dalitz plot asymmetries come from a NLO CHPT calculation of
\textcite{Gamiz:2003pi} [see also
\textcite{Prades:2005gv,Prades:2007ud}]. In fact,
this calculation also involves an estimate of the dominant
contributions to the absorptive parts of the relevant amplitudes of
$\mathcal{O}(G_8 p^6)$.

The linear Dalitz plot slopes $g_{C,N}^\pm$ refer
to the decays of $K^\pm$ into three charged pions ($C$), and one
charged and two neutral pions ($N$), respectively. The CP-violating
quantities of interest are the slope asymmetries
\begin{equation}
A_g^{C,N} = \displaystyle\frac{g^+_{C,N} - g^-_{C,N}}{g^+_{C,N} +
  g^-_{C,N}}.
\end{equation}
These asymmetries depend at LO on the imaginary parts of the LECs
$G_8$ and $g_{\rm ewk}$ [Eqs.~(\ref{eq:Lweak}) and (\ref{eq:Lelweak})]
only. \textcite{Gamiz:2003pi} found that $A_g^C$ is relatively
insensitive to NLO LECs whereas $A_g^N$ is less stable.
The theoretical results are compared with the most precise
experimental data in Table \ref{tab:3picpv}.
\renewcommand{\arraystretch}{1.2}
\begin{table}[floatfix]
\centering

\caption{\label{tab:3picpv}
Dalitz slope asymmetries for $K^\pm \to 3\pi$ decays: comparison
between theory \cite{Gamiz:2003pi} and experiment \cite{Batley:2007yfa}.
\\  }

\begin{tabular}{|c|c|c|}
\hline
& & \\[-.4cm]
 $A_g^C$  &  $A_g^N$ &  Ref. \\[4pt]
\hline
& & \\[-.3cm]
$(-2.4 \pm 1.2)\times 10^{-5}$ & $(1.1 \pm 0.7)\times 10^{-5}$ &
\textcite{Gamiz:2003pi}   \\[.1cm]
$(-1.5 \pm 2.2)\times 10^{-4}$ & $(1.8 \pm 1.8)\times 10^{-4}$ &
\textcite{Batley:2007yfa}  \\[5pt]
\hline
\end{tabular}

\end{table}

To a good approximation, $\epsp/\eps$ also depends on the imaginary
parts of the LO LECs $G_8$ and $g_{\rm ewk}$ only
(see Sec.~\ref{subsec:epsprime}).
Therefore, one can
establish bounds for
$A_g^C$ within the SM, using the
experimental value for $\epsp/\eps$ \cite{Prades:2007ud}.

\subsubsection{$\pi\pi$ scattering lengths from $K \to 3\pi$ decays
  near threshold}
\label{subsec:cusp}

\begin{figure}[floatfix]
\leavevmode
\includegraphics[width=8cm]{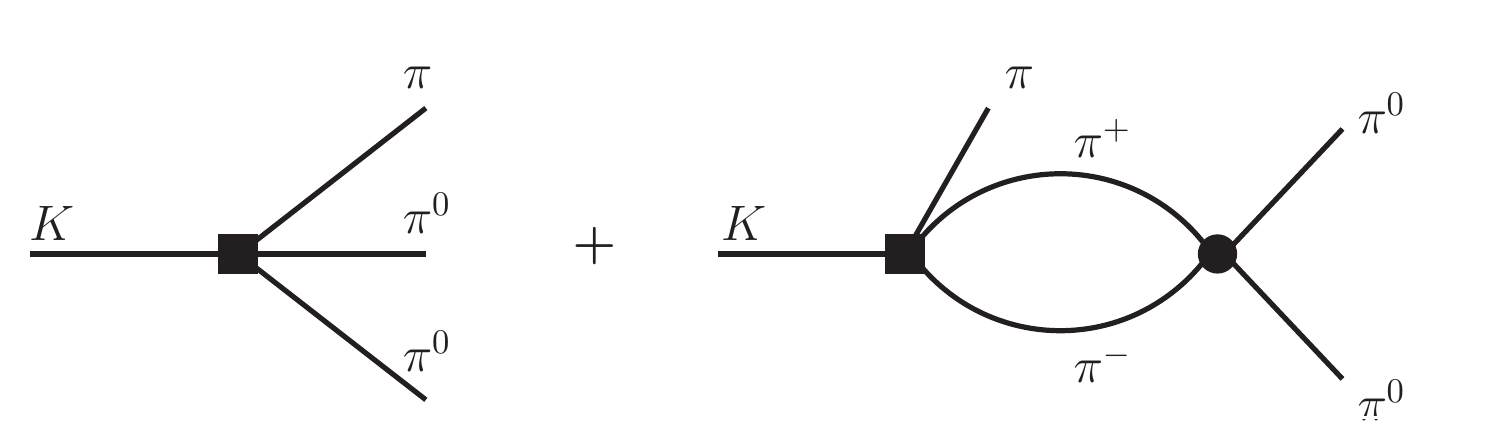}
\caption{\label{fig:cusp} Interference between tree and one-loop
  amplitudes generates a cusp at $M_{\pi^0\pi^0}^2 = 4 M^2_{\pi^+}$ in
   $K \to 3 \pi$ decays.}
\end{figure}

Because of the $\pi^+$--$\pi^0$ mass difference there is a cusp in the
$M_{\pi^0\pi^0}$ distribution at $M_{\pi^0\pi^0} = 2 M_{\pi^+}$ in
$K \to 3\pi$ decays with two $\pi^0$ in the final state.
It was first seen in $K^\pm \to \pi^\pm \pi^0 \pi^0$
\cite{Batley:2005ax,Batley:2009nv}, more recently also in
$K_L \to 3 \pi^0$ \cite{abouzaid:2008js}. It is due to the
charge exchange scattering of pions in the final state
\cite{Budini:1961bf,Cabibbo:2004gq}
\begin{equation}
K^\pm \to \pi^\pm (\pi^+ \pi^-)^* \to \pi^\pm \pi^0 \pi^0.
\label{eq:cusp}
\end{equation}
The interference between tree and one-loop
amplitudes in Fig.~\ref{fig:cusp} generates the cusp via
the square-root singularity at $M_{\pi^0\pi^0}^2 = 4 M^2_{\pi^+}$. From
Eq.~(\ref{eq:cusp}) the effect is seen to be mainly sensitive to the
combination of $\pi\pi$ scattering lengths
\begin{equation}
a_0 - a_2 \sim
A(\pi^+ \pi^- \to \pi^0 \pi^0)_{\mathrm{thresh}}.
\end{equation}
Various approaches have been pursued to extract the $\pi\pi$
scattering lengths from $K \to 3 \pi$ near threshold.
\begin{enumerate}
\item[i.] Following the original approach of \textcite{Cabibbo:2004gq},
  based on unitarity and analyticity,
a systematic expansion of the singular terms of the $M_{\pi^0\pi^0}$
distribution in powers of the scattering lengths was performed by
\textcite{Cabibbo:2005ez}.
\item[ii.] In a related method, unitarity and analyticity were combined
  with CHPT \cite{Gamiz:2006km}.
\item[iii.] A two-loop dispersive representation of $K \to 3 \pi$
  amplitudes in the presence of isospin breaking is under construction
  \cite{Kampf:2008ts}.
\item[iv.] In the most advanced approach, based on a nonrelativistic
  effective field theory (NRQFT), the $K \to 3 \pi$ amplitudes are expanded in
  powers of the scattering lengths and of the pion  momenta in the $K$
  rest frame \cite{Colangelo:2006va,Bissegger:2007yq,Gasser:2011ju}.
\end{enumerate}

\begin{figure}[floatfix]
\leavevmode
\includegraphics[width=7.7cm]{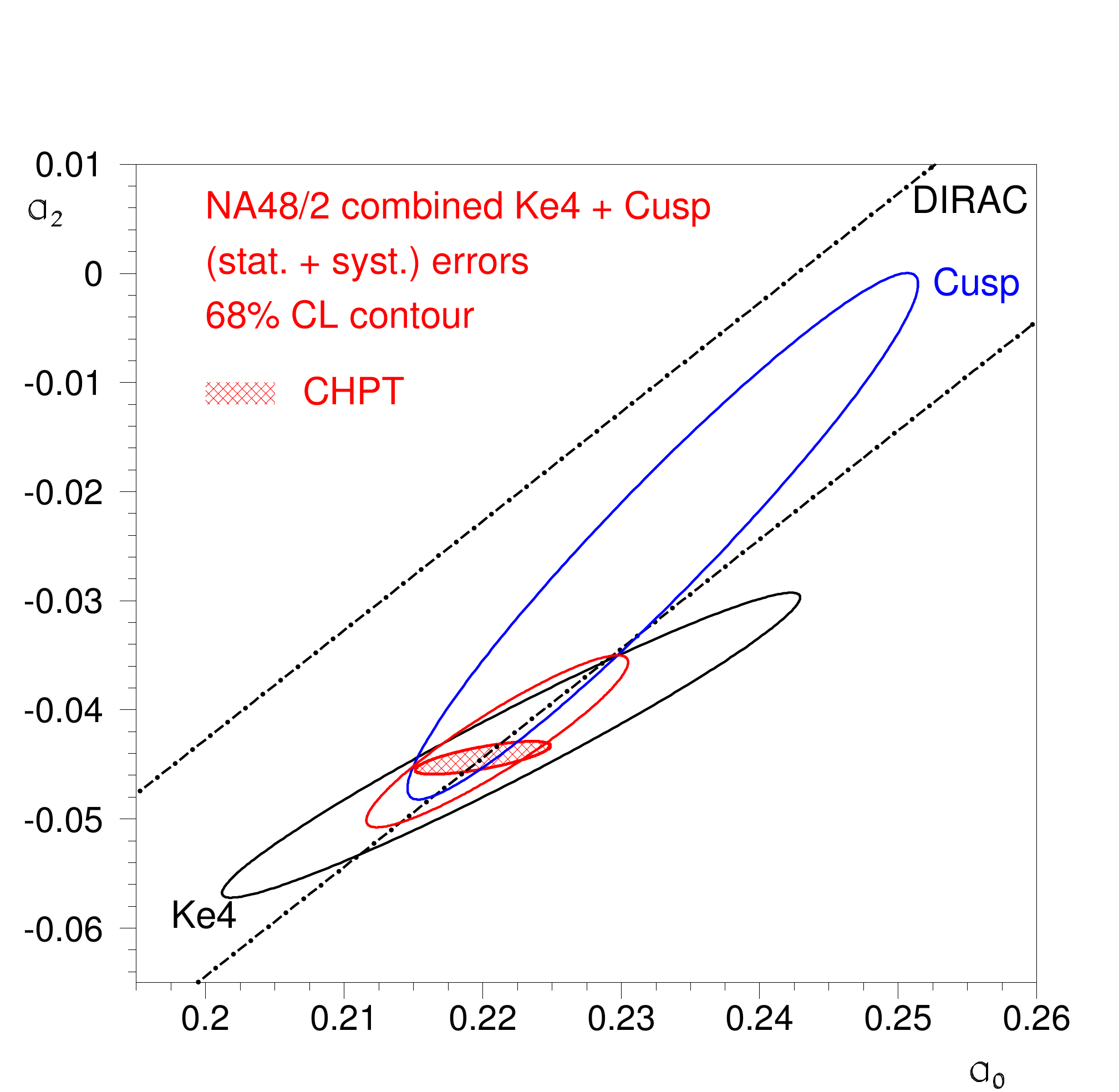}
\caption{\label{fig:combined}
NA48/2  $K_{e4}$ and cusp results from two-parameter fits in the
($a_0$, $a_2$) plane. The smallest contour corresponds to the combination
of NA48/2 results. The cross-hatched ellipse is the CHPT prediction
(\ref{eq:a0a2chpt}) of \textcite{Colangelo:2001sp,Colangelo:2001df}.
The dash-dotted lines correspond to the recent result from DIRAC
\cite{Adeva:2011}. We thank Brigitte Bloch-Devaux for updating the
original figure from \textcite{Batley:2010zza}.
}
\end{figure}

In the NRQFT approach of the Bern-Bonn group, the power counting
involves a small parameter $\ve$ that characterizes the size of pion
three-momenta: $|\vec p\,|/M_\pi=\mathcal{O}(\ve)$. In contrast to standard CHPT,
the scattering lengths are not predicted but extracted from the data.
The decay amplitudes are then given
by a two-fold expansion in $\ve$ and in the scattering lengths
(denoted below generically as $a$). At any given order in $a$ and
$\ve$, only a finite number of graphs contribute because each loop is
suppressed by one order in $\ve$.

NRQFT is manifestly Lorentz invariant and therefore frame
independent. Analyticity and unitarity are guaranteed as in standard
CHPT. However, unlike in standard CHPT, the amplitudes are valid to
all orders in the quark masses. They have been calculated up to
$\mathcal{O}(\ve^4,a\,\ve^5, a^2 \ve^2)$
\cite{Colangelo:2006va,Bissegger:2007yq}.

A major advantage of the NRQFT approach is that photons can be
incorporated in a straightforward manner, allowing for the systematic
inclusion of electromagnetic corrections \cite{Bissegger:2008ff}.
In the Coulomb gauge, only transverse photons appear as internal lines
in diagrams whereas the Coulomb photons give rise to a non-local
vertex via the equations of motion. Of course, also real photon
emission must be included, in particular to cancel infrared
divergences. However, the nonrelativistic power counting then shows
that finite bremsstrahlung effects are small near threshold.

The production of pionium, bound states of charged pions, upsets the
nonrelativistic counting. A certain region around the cusp is therefore
excluded from the data analysis such that one-photon exchange of
$\mathcal{O}(a\,e^2)$ is sufficient.

The decay spectra were calculated to $\mathcal{O}(e^2 \ve^4)$ for all $K \to 3
\pi$ channels and, in addition, to $\mathcal{O}(e^2 a\,\ve^2)$ for the channels of
main interest, $K^+ \to \pi^+ \pi^0 \pi^0$ and $K_L \to 3 \pi^0$
\cite{Bissegger:2008ff}. The radiatively corrected amplitudes were
used in the analysis of NA48/2 data by \textcite{Batley:2009nv} to
extract the scattering lengths.
Fig. \ref{fig:combined} shows a comparison of the cusp analysis of
\textcite{Batley:2009nv} and the $K_{e4}$ studies of
\textcite{Batley:2010zza}. The smallest contour corresponds to the
combination of the two methods. The result of the DIRAC experiment
\cite{Adeva:2011} and the CHPT
prediction of \textcite{Colangelo:2001sp,Colangelo:2001df}
are shown as well.
The corresponding numerical results are displayed in
Table \ref{tab:a0a2}. The agreement
between theory and experiment is impressive.

\begin{center}
\begin{table*}[floatfix]

\caption{\label{tab:a0a2} Experimental and theoretical results for
  s-wave $\pi\pi$ scattering lengths. First line: cusp analysis of NA48/2
  \cite{Batley:2009nv} using the Bern-Bonn framework
  \cite{Colangelo:2006va,Bissegger:2007yq}. The external error
    is due to the uncertainty in the ratio of the amplitudes for $K^+
    \to \pi^+ \pi^+ \pi^-$ and $K^+ \to \pi^+ \pi^0 \pi^0$.
  Second line: $K_{e4}$
  analysis of NA48/2 \cite{Batley:2010zza}. Third line: analysis of
  \textcite{Colangelo:2001df} on the basis of Roy equations and CHPT. \\}

\begin{tabular}{|c|c|c|c|}
\hline
& & & \\[-.4cm]
& $a_0 - a_2$ & $a_0$ & $a_2$
 \\[4pt]
\hline
 & & &    \\[-.3cm]
\textcite{Batley:2009nv} &
$0.2571(48)_{\mathrm{stat}}(25)_{\mathrm{syst}}(14)_{\mathrm{ext}}$   &
&  $- 0.0241(129)_{\mathrm{stat}}(94)_{\mathrm{syst}}(18)_{\mathrm{ext}}$   \\[4pt]
\textcite{Batley:2010zza} & &
$0.2220(128)_{\mathrm{stat}}(50)_{\mathrm{syst}}(37)_{\mathrm{th}}$
&
$- 0.0432(86)_{\mathrm{stat}}(34)_{\mathrm{syst}}(28)_{\mathrm{th}}$   \\[4pt]
\textcite{Colangelo:2001df} &  $0.264 \pm 0.004$   & $0.220 \pm 0.005$
 & $- 0.0444 \pm 0.0010$  \\[5pt]
\hline
\end{tabular}

\end{table*}
\end{center}

\section{Rare and radiative decays}
\label{sec:rare}
Kaon decays mediated by FCNC fall in the set of rare and
radiative decays. These modes  are suppressed in the SM and
their main interest, other than their own understanding, relies on the
possible observation
of New Physics effects.  Most of these processes are dominated by
long-distance contributions, such  as $K \rightarrow \gamma
\gamma$, $K \rightarrow \gamma \gamma^*$,
$K \rightarrow \pi \gamma^*$ and others. However, there are also
processes governed by short-distance amplitudes, such as $K \rightarrow
\pi \nu \bar{\nu}$.
\par
Long-distance dominated decays have been studied in the chiral framework.
The leading contributions, mostly
${\cal O}(p^4)$, have been evaluated within CHPT. In many of the
processes estimates of the dominant NLO corrections have also
been carried out.

\subsection{$K \rightarrow \pi \nu \overline{\nu},\, \pi \pi \nu
\overline{\nu}$ }

\begin{flushleft}
\begin{figure}[floatfix]
\leavevmode
\includegraphics[width=8cm]{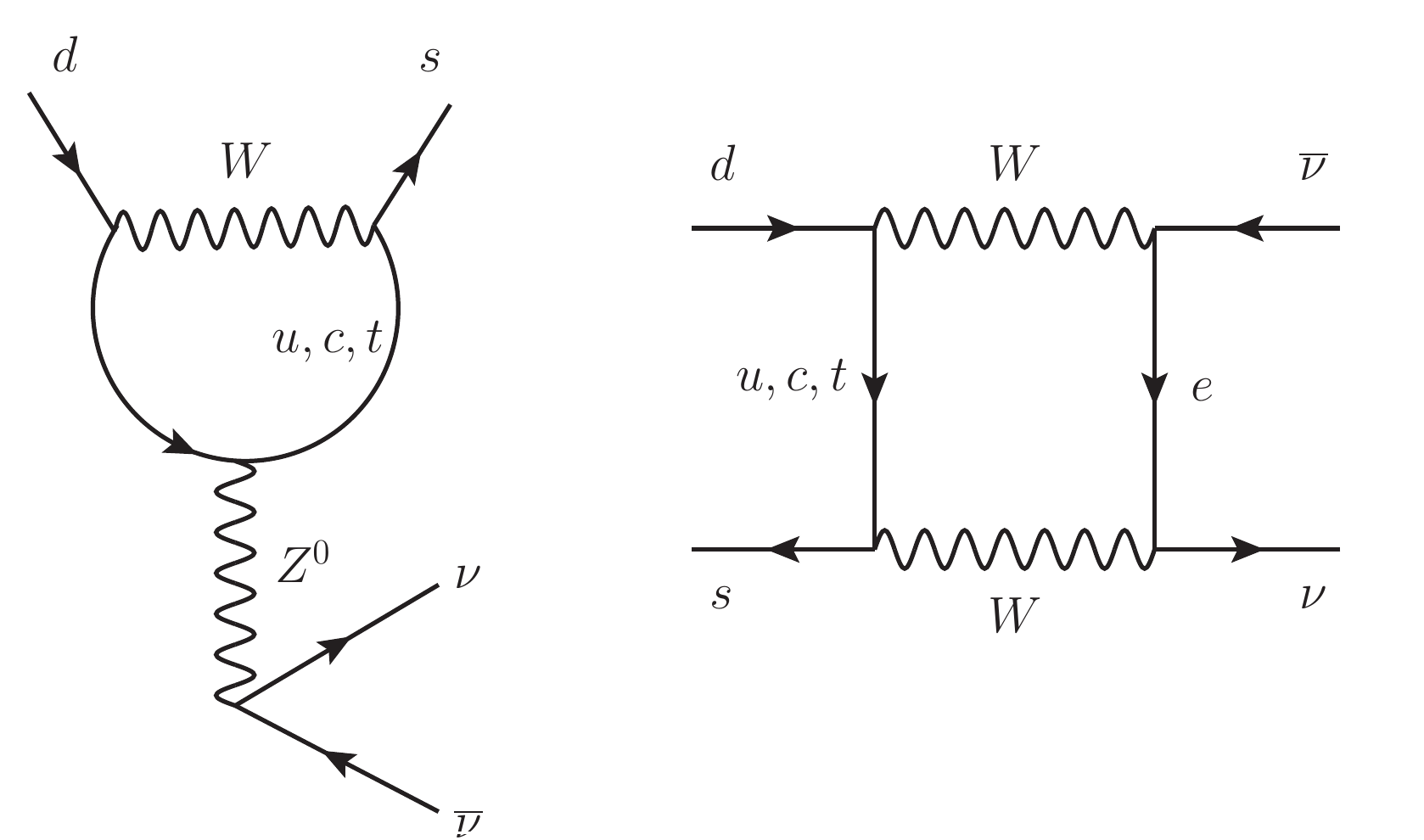}
\caption{\label{fig:fcnc1}
Z-penguin and box contributions to
$K \rightarrow \pi \nu \overline{\nu}$.}
\end{figure}
\end{flushleft}

The rare decays
$K_L \to  \pi^0 \nu \bar{\nu}$
and
$K^\pm  \to  \pi^\pm  \nu \bar{\nu}$
can be predicted with a  precision surpassing any other FCNC process
involving quarks, thus making them a clean place to look for non-standard signals.
The analysis of these processes makes use of the full arsenal of
effective field theory, from short-distance effective Hamiltonians to CHPT.
We outline below the main steps of the analysis and then quote the final
results for the branching ratios,  with a brief discussion of the uncertainties.

Let us begin with a qualitative discussion  that highlights the main features
of these modes.
In the SM, the $K \rightarrow \pi \nu \overline{\nu}$ decays
proceed  through FCNC amplitudes generated at the quark level  by
Z-penguin and box diagrams (see Fig.~\ref{fig:fcnc1}).
Separating the contributions according to the intermediate up-type
quark running inside the loops,
the quark-level amplitude has the structure
\beqa
\lefteqn{A (s \to d \nu \bar{\nu}) \sim
\sum_{q=u,c,t} \lambda_q\,  X_{\mathrm{SM}} (x_q)} &&
\nn
& &\qquad\sim \,
\frac{m_t^2}{M_W^2}   \lambda_t +
\frac{m_c^2}{M_W^2}  \ln \frac{M_W}{m_c}   \lambda_c +
\frac{\Lambda_{\rm QCD}^2}{M_W^2}   \lambda_u ,\qquad
\eeqa
where $\lambda_q = V_{qd}^{\phantom{*}}V_{qs}^*$ and $x_q =
m_q^2/M_W^2$.
The equation above reflects the quadratic (``hard") nature of the
GIM mechanism.
This in turn implies that the
 top-quark  contribution, which carries a large CP-violating phase, accounts for
$\sim 68\%$ of $A (s \to d \nu \bar{\nu})$
while the  charm- and up-quark contributions amount to
$\sim 29\%$ and $\sim 3\%$, respectively.
So we see that the power-like GIM mechanism implies a large
suppression of the $u$-quark (long-distance)
contribution and complete dominance of the CP-violating contribution
by the $t$-quark term.
These properties are often summarized by stating that the
$K \rightarrow \pi \nu \overline{\nu}$  decays are ``short-distance" dominated.
The short-distance domination also implies that to a good approximation
the interaction is described at low energy by
one single local operator, $Q_{13}$ in  Eq.~(\ref{eq:Q11-13}),
whose hadronic matrix element can be related to the form factors appearing in $K_{\ell 3}$
decays.  All these features make it possible to predict
$K \rightarrow \pi \nu \overline{\nu}$ rates very accurately.

The above qualitative discussion can be put on solid footing by systematically employing
effective-theory techniques.  As usual in the application to weak decays, three steps are required.
(i) Determine  the effective Lagrangian ${\cal L}_{\rm eff}$
at the weak scale $\mu \sim M_W$ by integrating out
the heavy gauge bosons and the top quark.
In this case, the diagrams of Fig.~\ref{fig:fcnc1}  generate the operator $Q_{13}$.
The Wilson coefficient $C_{13}$ is determined entirely in terms of
$\lambda_t$ and $x_t$,
and it now includes
NLO QCD  effects   \cite{Buchalla:1998ba,Misiak:1999yg}
and two-loop electroweak corrections \cite{Buchalla:1997kz,Brod:2010hi}.
(ii) Evolve ${\cal L}_{\rm eff}$ down to a low hadronic scale
using renormalization group techniques.
In this step charm-loop contributions (the $c$ quark is still an active
degree of freedom) generate
corrections to $C_{13}$ proportional to
$\lambda_c x_c \ln x_c$.
These are known up to and including NNLO QCD effects
 \cite{Buchalla:1993wq,Buras:2005gr,Buras:2006gb}
and NLO electroweak corrections \cite{Brod:2008ss}.
At the charm threshold $\mu \sim m_c$   one integrates out the charm quark.
At the current level of precision, one needs to keep  not only the
leading dimension-six operator,
but also dimension-eight operators \cite{Falk:2000nm},
whose relative importance is suppressed by $M_K^2/m_c^2 \sim 15\%$.
(iii)  Calculate hadronic matrix elements of the operators appearing in
${\cal L}_{\rm eff}$ at a low scale $\mu \sim 1$ GeV.
${\cal L}_{\rm eff}$ contains the purely semileptonic operator $Q_{13}$,
the dimension-eight operators generated at the charm threshold, and
the $|\Delta S| =1$ four-quark operators,
which generate long-distance contributions to $K \to \pi \nu \bar{\nu}$
(they correspond to the $u$-quark contributions to Z-penguin and box diagrams).
The matrix element of $Q_{13}$ can be related to $K_{\ell 3}$ form factors
\cite{Marciano:1996wy}.
This analysis was recently updated by \textcite{Mescia:2007kn}
to include first-order isospin-breaking effects
to NLO in CHPT as well as long-distance radiative corrections for the $K^\pm$ decay.
Concerning the long-distance contributions and dimension-eight
operators, the relevant matrix elements
were calculated in CHPT \cite{Lu:1994ww,Isidori:2005xm}.

Owing to the CP properties of the $K^0_2$ state, the operator $Q_{13}$ only
contributes to $K_L \to \pi^0 \nu \bar\nu$ through a violation of the CP symmetry.
The transition amplitude is completely dominated by direct CP
  violation, the contribution
from $K^0$--$\overline{K^0}$ mixing being only of the order of 1\%.
There exist tiny CP-conserving contributions through higher-order short-distance
operators and long-distance corrections, but their effect on the branching ratio is negligible \cite{Buchalla:1998ux}.

Putting all the ingredients together, the predicted SM rates for $K \to \pi \nu
\bar{\nu}$ decays can be written as
\beqa
\mathrm{BR} (K_L \to \pi^0 \nu \bar{\nu}) &=& \kappa_L \left(
\frac{{\rm Im}\, \lambda_t}{\lambda^5} \, X  \right)^2
\left( 1- \delta_\epsilon\right),
\\
\mathrm{BR} (K^+  \to \pi^+ \nu \bar{\nu}) &=& \kappa_+ \left(1 + \Delta_{\rm EM}\right)
 \Bigg[ \left( \frac{{\rm Im}\, \lambda_t}{\lambda^5} \, X  \right)^2
\quad\nonumber \\
+  \bigg(
\frac{{\rm Re} \lambda_t}{\lambda^5} \, X   &+&
\frac{{\rm Re} \lambda_c}{\lambda} \, (P_c + \delta P_{c,u})
\bigg)^2  \Bigg] ,
\eeqa
where
$\delta_\epsilon = \sqrt{2}|\epsilon|\, [1+P_c/(A^2\lambda)-\rho]/\eta$
contains the small $K^0$--$\overline{K^0}$ mixing contribution
\cite{Buchalla:1996fp}, with
$\lambda = V_{us}$, $A$, $\rho$ and $\eta$ the Wolfenstein CKM parameters.
The overall factors $\kappa_{L,+}$
encode the hadronic matrix element related to
$K_{\ell 3}$ data \cite{Mescia:2007kn}:
\begin{eqnarray}
\kappa_L &=& (2.231  \pm 0.013) \times 10^{-10} \; (\lambda/0.225)^8,
\nonumber \\
\kappa_+ &=& (5.173  \pm 0.025) \times 10^{-11} \; (\lambda/0.225)^8.
\end{eqnarray}
They are now dominated by experimental uncertainties in $K_{\ell 3}$.
Defining the infrared-safe photon-inclusive rate by the cut
$E_\gamma^{\rm cms} < 20$ MeV,
the electromagnetic correction takes the value $\Delta_{\rm EM} = -0.003$ \cite{Mescia:2007kn}.
The top-quark contribution is given by
$X = 1.469 \pm 0.017$
\cite{Buras:2005gr,Buras:2006gb,Brod:2010hi},
with an uncertainty due to the parametric error in $m_t$,  matching
scale $\mu_t$, and  higher-order electroweak effects.
The dimension-six  charm contribution has the value
$P_c =  0.38 \pm 0.04$, with error  dominated by the parametric uncertainty in $m_c$.
Finally, the long-distance ($u$ quark)  and dimension-eight charm
contributions can be lumped in $\delta P_{c,u} = 0.04 \pm 0.02$
\cite{Isidori:2005xm}.
Taking the CKM matrix elements from global fits, one arrives at
\cite{Buras:2005gr,Buras:2006gb,Brod:2010hi}:
\begin{eqnarray}
\mathrm{BR} (K_L \to \pi^0 \nu \bar{\nu}) &=&  (2.4 \pm 0.4) \times 10^{-11},
\\
\mathrm{BR} (K^+  \to \pi^+ \nu \bar{\nu}) &=& (0.78 \pm 0.08) \times 10^{-10}.
\end{eqnarray}
In both cases the uncertainty is largely parametrical ($\sim 80\%$ for
the $K_L$ mode and
$\sim 70 \%$ for the $K^+$ mode), due to CKM input, $m_{c}$, $m_{t}$  and
$\alpha_{s} (M_Z)$.
As the determination of CKM parameters improves in the next few years, we can expect
to reach
accuracies better than $10 \%$.
In summary, from a theoretical perspective $K \to \pi \nu \bar{\nu}$ decays
offer the  cleanest window on non-standard
contributions to the $s \to d$ transitions (complementary to
B  meson studies).   Moreover, within the general context of flavor physics,
$K \to \pi \nu \bar{\nu}$ decays are perhaps the most promising place to look
for non-standard signals, due to their SM-specific suppression induced
by CKM factors [$ A (s \to d \nu \bar{\nu}) \propto V_{us}^5$].
Even if  one considers SM extensions with the same CKM
suppression factor,
$K \to \pi \nu \bar{\nu}$ decays still provide an excellent probe,  because
the SM contributions can be predicted very accurately.

On the experimental side, the charged kaon mode was observed
\cite{Artamonov:2008qb}, while only an upper bound on the neutral mode
has been achieved \cite{Ahn:2007cd,Ahn:2009gb}:
\beqa
\mathrm{BR} (K^+  \to \pi^+ \nu \bar{\nu})
&=&  (1.73^{+1.15}_{-1.05}) \times 10^{-10},
\\
\mathrm{BR} (K_L \to \pi^0 \nu \bar{\nu})
&< &  2.6  \times 10^{-8}  \quad  (90 \% \, {\rm C.L.}). \quad
\eeqa
New experiments
are under development at CERN \cite{Spadaro:2011ue} and J-PARC \cite{Watanabe:2010zzf}
for charged and neutral modes, respectively.
These experiments aim to reach $\mathcal{O}(100)$ events (assuming SM rates), thus beginning to
seriously probe the new-physics potential of these rare $K$ decays.
Increased sensitivities could be obtained through the recent P996 proposal for a
$K^+  \to \pi^+ \nu \bar{\nu}$ experiment at Fermilab and the higher kaon fluxes available at Project-X
\cite{Tschirhart:2011zz}.

Finally, let us mention that  the decays
$K_{L,S} \to \pi \pi \nu \bar{\nu}$
and
$K^\pm \to \pi^\pm \pi^0 \nu \bar{\nu}$
share the same feature of short-distance
domination as the corresponding single-pion modes,
and could therefore provide another probe of the
underlying $s \to d \nu \bar{\nu}$ transition
within and beyond the SM.
The calculation of the decay amplitudes requires taking the
matrix element of the current $\bar{s} \gamma_\mu
(1 - \gamma_5) d$ between the kaon and two-pion states,
which can be extracted from the measured $K_{\ell 4}$
decays using isospin symmetry~\cite{Littenberg:1995zy,Chiang:2000bg},
or directly calculated in CHPT \cite{Geng:1994cw}.
Theoretical predictions can be summarized as follows
\cite{Littenberg:1995zy}:
\begin{eqnarray}
\mathrm{BR} (K_L \to \pi^+ \pi^- \nu \bar{\nu}) &\!\simeq &\!
\\ &&\hskip -2cm
1.8 \times\left[ (1.37 - \rho)^2  + 0.17\, \eta^2 \right] \times 10^{-13},
\nn
\mathrm{BR} (K_L \to \pi^0 \pi^0 \nu \bar{\nu}) &\!\simeq &\! (1.37 - \rho)^2 \times 10^{-13},
\nn
\mathrm{BR} (K^\pm \to \pi^\pm \pi^0 \nu \bar{\nu}) &\!\simeq &\!
7 \times \left[ (1.37 - \rho)^2  +  \eta^2 \right] \times 10^{-15}.
\nonumber
\end{eqnarray}
\noindent
Experimental searches are still far above the expected
SM rates and have reached the following $90 \%$ C.L. limits:
\begin{eqnarray}
\mathrm{BR} (K^+ \to \pi^+ \pi^0  \nu \bar{\nu}) &<& 4.3 \times 10^{-5}
\hspace*{.2cm} \mbox{\protect\cite{Adler:2000ic}},\nn
\mathrm{BR} (K_L \to \pi^0 \pi^0 \nu \bar{\nu}) &<&
8.1 \times 10^{-7}
\hspace*{.2cm} \mbox{\protect\cite{Ogata:2011ii}}. \nn
& &
\end{eqnarray}

\subsection{$K \rightarrow \gamma^{(*)} \gamma^{(*)}$}
The amplitude for a transition of the type
\begin{equation}
K (p) \longrightarrow \gamma^* (q_1) \; \gamma^*(q_2)
\end{equation}
is determined by a tensor amplitude $M^{\mu \nu}(q_1,q_2)$ that has
the most general form compatible with gauge invariance:
\begin{eqnarray} \label{eq:kgsgs}
 M^{\mu \nu} &\!\! = & \! \!  \left[ g^{\mu\nu} -
\frac{q_1^\mu q_1^\nu}{q_1^2} - \frac{q_2^\mu q_2^\nu}{q_2^2} +
\frac{q_1 \cdot q_2}{q_1^2 q_2^2}  q_1^\mu q_2^\nu \right]
M_K^2  a(q_1^2,q_2^2) \no \\
&\!\! + & \!\! \left[q_2^\mu q_1^\nu - q_1 \cdot q_2 \left(
\frac{q_1^\mu q_1^\nu}{q_1^2} + \frac{q_2^\mu q_2^\nu}{q_2^2} -
\frac{q_1 \cdot q_2}{q_1^2 q_2^2}  q_1^\mu q_2^\nu \right)\right]  \no \\
&& \, \times \; b(q_1^2,q_2^2) \no \\
&\!\! + & \!\! i \,  \ve^{\mu\nu\rho\sigma} q_{1\rho} q_{2\sigma} \,
c (q_1^2,q_2^2).
\end{eqnarray}
Thus, for instance, the amplitude into two real photons is given by  $A = M^{\mu \nu}(q_1,q_2) \, \varepsilon_{\mu}(q_1) \varepsilon_{\nu}(q_2)$.
Bose symmetry implies that the invariant amplitudes $a(q_1^2,q_2^2)$,
$b(q_1^2,q_2^2)$ and $c(q_1^2,q_2^2)$ are symmetric functions of their
arguments. If CP is conserved the amplitudes $a$ and $b$
contribute
to $K_1^0 (K_S) \rightarrow \gamma^* \gamma^*$ while the amplitude $c$
determines $K_2^0 (K_L) \rightarrow
\gamma^* \gamma^*$. When one of the photons is on-shell ($q_1^2=0$ for
instance), $M^{\mu \nu}$ is described by two invariant amplitudes,
\begin{eqnarray} \label{eq:kggs}
 M^{\mu \nu} & = & \left( q_2^{\mu} q_{1}^{\nu} -  q_1 \cdot q_2 g^{\mu \nu} \right)\,  b(0,q_2^2) \no \\
&& + \, i \,
\varepsilon^{\mu \nu \rho \sigma} q_{1\rho} q_{2 \sigma} \, c(0,q_2^2)  ,
\end{eqnarray}
which also remains valid for both photons on-shell.

These processes are dominated by
long-distance dynamics, in particular
those with both photons on-shell or when off-shell photons produce
lepton pairs. Short-distance amplitudes give only tiny contributions.

Since the photon does not couple directly to neutral particles,
the only possible local contributions should come from field strength tensors, which are absent at
${\cal O}(p^2)$. Moreover, the ${\cal O}(p^4)$ Lagrangians  ${\cal L}_{G_8 p^4}^{\Delta S=1}$
and ${\cal L}_{G_{27} p^4}^{\Delta S=1}$ in Eq.~(\ref{eq:Lweak})
  only contain two operators with
$F^{\mu\nu}$ terms, but they are coupled to at least two charged particles.
Therefore, the leading $K \rightarrow \gamma^* \gamma^*$
contribution is generated by a non-local
${\cal O}(p^4)$ loop amplitude that is necessarily finite.
In the following and unless explicitly stated we assume CP invariance.

\subsubsection{$K_S \rightarrow \gamma \gamma$}

\begin{figure}[floatfix]
\leavevmode
\includegraphics[width=7cm]{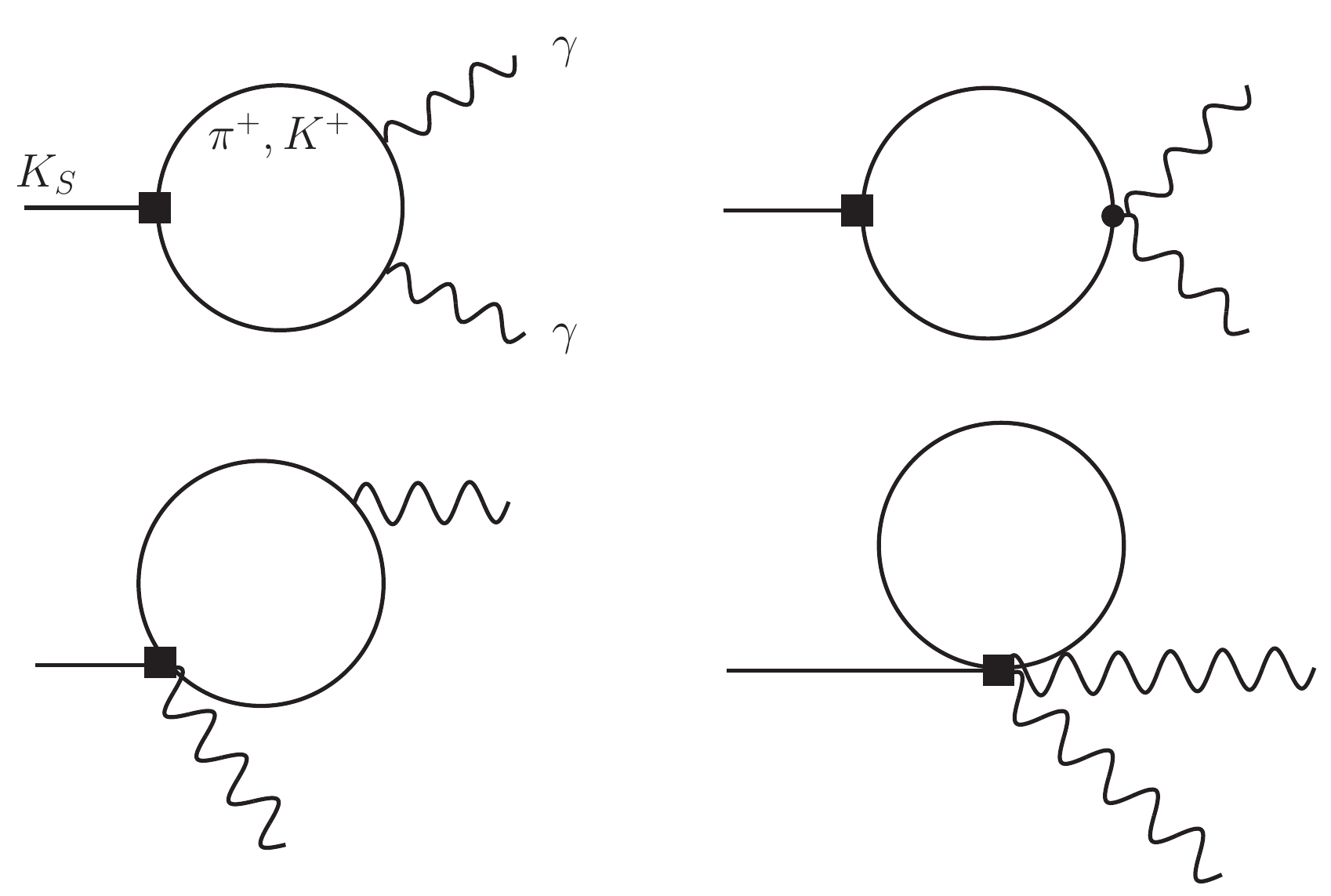}
\caption{\label{fig:ksgg}
Lowest-order [${\cal O}(p^4)$] contributions to $K_S \rightarrow \gamma \gamma$.}
\end{figure}

The photons produced in this decay
have parallel polarizations ($F_{\mu \nu} F^{\mu \nu}$) and,
up to one loop, there is no short-distance contribution due to Furry's
theorem \cite{Gaillard:1974hs}.
This decay arises, at leading order, from a finite two-pion one-loop
amplitude with one vertex from
${\cal L}_{G_8 p^2}^{\Delta S=1}$ or
${\cal L}_{G_{27} p^2}^{\Delta S=1}$
\cite{D'Ambrosio:1986ze,Goity:1986sr}, as shown in Fig.~\ref{fig:ksgg}.
Its contribution is
\begin{equation} \label{eq:ksgg4}
 b^{(4)}(0,0)  =   \frac{2}{\pi} \, \alpha \, F_0 \left( G_8 +
 \frac{2}{3} G_{27} \right) \left(1-r_{\pi}^2 \right)
F \left(\frac{1}{r_{\pi}^2}\right) ,
\end{equation}
where
$r_{\mbox{\tiny P}} = M_{P}/M_K$
and $F(x)$ is given in Eq.~(\ref{eq:1lfz}).
The charged-kaon loop contribution vanishes because of
  the factor $(1-r_K^2)=0$.

The decay width is given by
\begin{equation} \label{eq:wksgg}
 \Gamma \left( K_S \rightarrow \gamma \gamma \right) \, = \, \frac{M_K^3}{64 \, \pi}\,  \big|b(0,0)\big|^2  .
\end{equation}
Taking the LO values of the couplings $G_8$ and $G_{27}$,
Eq.~(\ref{eq:ksgg4}) results in
$\mathrm{BR}(K_S \rightarrow \gamma \gamma) = 2.0 \times 10^{-6}$,
which compares rather well with the present experimental world average
\cite{Nakamura:2010zzb}
\begin{equation}\label{eq:KSgg_exp}
\mathrm{BR}(K_S \rightarrow \gamma \gamma) = (2.63\pm 0.17) \times 10^{-6} .
\end{equation}

Although the full ${\cal O}(p^6)$ amplitude has not been calculated,
the dominant effects were implemented through
(i) the inclusion of unitarity corrections from
$K_S\!\to\!\pi \pi\!\to\!\pi^+ \pi^-\!\to\!\gamma \gamma$ \cite{Kambor:1993tv}, and (ii) a
local contribution introducing an unknown coupling constant
\cite{Buchalla:2003sj},
\begin{equation} \label{eq:ksgg6}
 b^{(4+6)}(0,0) = \frac{2 \, \alpha \, F_0}{M_K^2} B(M_K^2) + \frac{4 \, \alpha \, G_8}{\pi \, F_0}M_K^2 \left( 1 - r_{\pi}^2
\right) a_1  ,
\end{equation}
where $B(s)$ can be found in \textcite{Kambor:1993tv}.
Including the 27-plet contribution in $B(s)$, the
  experimental rate (\ref{eq:KSgg_exp})
implies $a_1 = (-1.2 \pm 1.3) \times 10^{-3}$, showing that unitarity corrections are enough to reproduce
the measured branching fraction.

\subsubsection{$K_L \rightarrow \gamma \gamma$}
This decay produces photons with perpendicular polarizations
($\varepsilon_{\mu \nu \rho \sigma} F^{\mu \nu} F^{\rho \sigma}$) and
then it is the amplitude $c(0,0)$
in Eq.~(\ref{eq:kggs}) that contributes.
Owing to its GIM suppression, the short-distance amplitude only gives a few-percent contribution
to the full width
\cite{Gaillard:1974hs,Ma:1981eg,Pramudita:1988jv,Herrlich:1991bq}.
Accordingly this decay is also dominated by long-distance dynamics.

\begin{figure}[floatfix]
\leavevmode
\includegraphics[width=7cm]{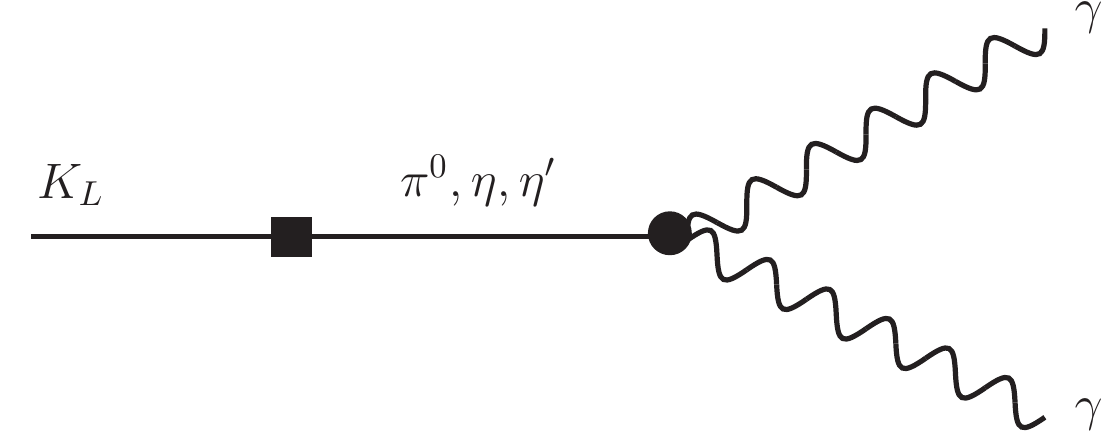}
\caption{\label{fig:klgg}
Dominant contribution to $K_L \rightarrow \gamma \gamma$.}
\end{figure}

The dominant contribution, shown in Fig.~\ref{fig:klgg}, is given by the weak transition of the kaon into
a non-flavored pseudoscalar meson and its corresponding decay into two
photons \cite{Ma:1981eg}, the latter being determined by the
anomalous Lagrangian ${\cal L}_{\rm WZW}$ in Eq.~(\ref{eq:Lstrong}).
At lowest
order in the chiral $\rm SU(3)$ expansion [${\cO}(p^4)$],
only $\pi^0$ and $\eta_8$ propagate. However, the amplitude vanishes
exactly due to the Gell-Mann-Okubo mass relation
\cite{GM:1961,Okubo:1961jc}. Therefore, the decay
starts at ${\cal O}(p^6)$ where the singlet $\eta_1$ state is
  also included:
\begin{equation} \label{eq:klgg}
 c^{(6)}(0,0) \, = \, - \frac{2}{\pi} \, \alpha \,  F_0 \left( G_8 -
 G_{27} \right) F_2(\hat\rho, \xi, \theta)  ,
\end{equation}
where
\begin{eqnarray} \label{eq:f2}
 F_2 & = & \frac{1}{1-r_{\pi}^2}
 + \frac{1}{3 \left( 1-r_{\eta}^2 \right)}\left[ \left(1+\xi \right) \cos \theta +
2 \sqrt{2}\hat\rho \sin \theta \right]  \nonumber \\
&& \; \; \; \; \; \; \; \; \;\; \; \; \; \; \; \; \; \; \; \; \; \;
 \;\; \; \; \; \;   \times\,  \left[ \frac{F_{\pi}}{F_{\eta_8}}
   \cos \theta - 2 \sqrt{2} \frac{F_{\pi}}{F_{\eta_0}} \sin \theta \right]
 \nonumber \\
&& - \frac{1}{3 ( 1-r_{\eta'}^2)} \left[ 2 \sqrt{2} \hat\rho \cos \theta - \left( 1 + \xi \right)
\sin \theta \right] \nonumber \\
&& \; \; \; \; \; \;\; \; \; \; \; \; \; \; \; \; \; \;  \times \left[
   \frac{F_{\pi}}{F_{\eta_8}} \sin \theta + 2 \sqrt{2} \frac{F_{\pi}}{F_{\eta_0}}
   \cos \theta \right]  .
\end{eqnarray}
Here $\theta$ is the mixing angle between  $\eta_8$ and
$\eta_1$ states:
\begin{equation}\label{eq:theta}
 \left( \begin{array}{c} \eta \\ \eta' \end{array} \right) =
\left( \begin{array}{cc} \cos \theta & -\sin \theta \\
                         \sin \theta & \cos \theta
       \end{array} \right) \left( \begin{array}{c} \eta_8 \\ \eta_1 \end{array} \right)  .
\end{equation}
Although there is still some discussion on the value of $\theta$, we
take $\theta = -20^\circ$ arising in the large-$N_C$ analyses
  \cite{HerreraSiklody:1997kd,Kaiser:1998ds}.
$\xi$ parametrizes the amount of $\rm SU(3)$ breaking
\cite{Donoghue:1986ti}:
\begin{equation} \label{eq:xi}
 \xi \, = \, \sqrt{3} \, \frac{\langle \eta_8 | {\cal L}^{|\Delta S|=1} | K_2^0 \rangle}{\langle \pi^0 | {\cal L}^{|\Delta S|=1} | K_2^0 \rangle} \,- \, 1  ,
\end{equation}
while $\hat\rho$ carries the information of the breaking of nonet symmetry
through the weak interactions of the
singlet at ${\cal O}(p^2)$ \cite{Donoghue:1983hi}:
\begin{equation} \label{eq:rho}
 \hat\rho \, = \, - \sqrt{\frac{3}{8}} \, \frac{\langle \eta_1 | {\cal L}^{|\Delta S|=1} | K_2^0 \rangle}{\langle \pi^0 | {\cal L}^{|\Delta S|=1} | K_2^0 \rangle}  .
\end{equation}
If $\rm SU(3)$ and nonet symmetries are exact we have $\xi = 0 $ and $\hat\rho
= 1$. Finally,
$F_{\pi}$, $F_{\eta_8}$ and $F_{\eta_1}$
are the decay constants of the pion, $\eta_8$ and $\eta_1$, respectively.
The values of the symmetry-breaking parameters
are still not precisely known. We take
$F_{\eta_8}/F_{\pi}=1.34$ and $F_{\eta_1}/F_{\pi}=1.0$ from
\textcite{Kaiser:1998ds} [see also
\textcite{Feldmann:2002kz}]. The $\rm SU(3)$ breaking parameter
was estimated to be $\xi \simeq 0.17$ \cite{Donoghue:1986ti},
but this value was challenged with the claim that it cancels
with an additional
$s\rightarrow d gg$ contribution \cite{He:2002as}.
Hence we will consider, conservatively, $\xi \sim 0.0 - 0.2$.
Finally,
dominance of the pion pole seems to require a small
breaking of nonet symmetry, $\hat\rho \simeq 0.8$
\cite{Cheng:1990mw,D'Ambrosio:1997ta}.

The decay width is given by Eq.~(\ref{eq:wksgg}) with the
function $c(0,0)$ instead of $b(0,0)$.
If we consider the experimental determination
\cite{Nakamura:2010zzb}
\begin{equation}
\mathrm{BR}(K_L \rightarrow
\gamma \gamma) = (5.47 \pm 0.04) \times 10^{-4},
\end{equation}
a value of $\theta= -20^\circ$
accommodates $\xi = 0$ with a
value of $\hat\rho \simeq 0.7$ while $\xi = 0.2$ requires $\hat\rho \simeq 0.8$.

\subsubsection{$K_S \rightarrow \gamma \ell^+ \ell^-$}
The amplitude for processes with one off-shell photon
decaying into a lepton pair is given by
\begin{equation} \label{eq:gammall}
 A = \frac{e}{q_2^2} \, M^{\mu \nu}(q_1,q_2)\, \varepsilon_{\mu}^*(q_1)\, \bar{u}(k)\gamma_{\nu}
v(k')  ,
\end{equation}
with $q_2 = k+k'$.
At ${\cal O}(p^4)$ the amplitude $b(0,q_2^2)$ in Eq.~(\ref{eq:kggs}) is uniquely determined by a
one-loop calculation of the  $K_1^0 \rightarrow \gamma \gamma^*$
transition \cite{Sehgal:1973we,Ecker:1987hd}. The dominant octet
contribution is
\begin{equation} \label{eq:b4q2}
 b^{(4)}(0,q_2^2) \, = \, \frac{4}{\pi} \, G_8 \,  \alpha \, F_0 \left( 1 - r_{\pi}^2 \right)
H(z)  ,
\end{equation}
where $z=q_2^2/M_K^2$ and $H(z)$ is given in
Eq.~(\ref{eq:1lhz}). The result for the spectrum in $q_2^2$ is
usually normalized to the two-photon width as
\begin{equation} \label{eq:specksgll}
 \frac{1}{\Gamma_{\gamma\gamma}} \frac{d \Gamma}{d z} \, = \,
 \frac{2}{z} \left(1-z\right)^3
\Bigg| \frac{H(z)}{H(0)} \Bigg|^2 \,\frac{1}{\pi} \mbox{Im}\, \Pi(z)  ,
\end{equation}
with $\mbox{Im}\, \Pi(z)$ the electromagnetic spectral function associated to the lepton pair:
\begin{equation} \label{eq:speclep}
 \frac{1}{\pi} \mbox{Im}\, \Pi(z) \, = \, \frac{\alpha}{3 \pi} \left( 1+2\frac{r_{\ell}^2}{z} \right)
\sqrt{1-4\frac{r_{\ell}^2}{z}} \; \theta \left( z-4r_{\ell}^2 \right) ,
\end{equation}
where $r_{\ell} = m_{\ell}/M_K$.
$\Gamma_{\gamma \gamma}\equiv \Gamma(K_S\to\gamma\gamma)$ is given by $b^{(4)}(0,0)$ in Eq.~(\ref{eq:ksgg4}) [notice that $H(0) = -\frac{1}{2}\, F(1/r_{\pi}^2)$]. 

Integrating the spectrum, one predicts the ratios
\beq
\frac{\Gamma(K_S \rightarrow \gamma \ell^+ \ell^-)}{\Gamma(K_S \rightarrow \gamma\gamma)}\, =\,\left\{
\ba{cc}  1.2 \times 10^{-2}\qquad & (\ell =e),\\ 2.8 \times 10^{-4}\qquad & (\ell =\mu).\ea\right.
\eeq
These processes have not been measured yet.

\subsubsection{$K_L \rightarrow \gamma \ell^+ \ell^-$}
As in the $K_S$ decay, the normalized spectrum is given by Eq.~(\ref{eq:specksgll})
with the ratio
$H(z)/H(0)$ substituted by
\beq
f(z) \equiv \frac{c(0,q_2^2)}{c(0,0)}  = 1 + b \, z + {\cal O}(z^2) .
\eeq
Being of higher order in CHPT, our knowledge of the form factor
$f(z)$ is unfortunately rather limited.
The decay mechanism in Fig.~\ref{fig:klgg} generates an unambiguous
contribution from the
electromagnetic form factor of the $\pi^0$ ($\eta$, $\eta'$), which
amounts to a
slope $b_V = M_K^2/M_{\rho}^2 \simeq 0.41$. However, there are other
possible contributions
that have been analyzed only within explicit models
\cite{D'Ambrosio:1996sw,Ecker:1990in,Bergstrom:1983rj,Bergstrom:1990uh,Sehgal:1973we},
giving slopes in the range $b\simeq 0.8-1.0$.

The experimental data can be fitted with a simple parametrization such as \cite{D'Ambrosio:1997jp}
\begin{equation} \label{eq:dipff}
 f(z)\big|_{\mbox{\tiny DIP}} = 1 +
 \alpha_{\mbox{\tiny DIP}} \frac{z}{z-M_{\rho}^2/M_K^2}  ,
\end{equation}
which provides the slope
$b^{\mbox{\tiny DIP}} = - \alpha_{\mbox{\tiny DIP}} \, M_K^2/M_{\rho}^2$.
A recent determination by the
KTeV collaboration \cite{Abouzaid:2007cm} for the electron case finds
$\alpha_{\mbox{\tiny DIP}}=-1.73 \pm 0.05$
($b^{\mbox{\tiny DIP}}=0.71 \pm 0.02$), while
$\alpha_{\mbox{\tiny DIP}}=-1.54 \pm 0.10$ ($b_{\mbox{\tiny DIP}}=0.63 \pm 0.04$) was previously extracted
from the muon mode \cite{AlaviHarati:2001wd}.

\subsubsection{$K_L \rightarrow  \ell^+_1 \ell^-_1 \ell^+_2 \ell^-_2$}
The dominant long-distance amplitude for this process is driven by the
creation of two lepton pairs:
\begin{equation} \label{eq:kl4l}
 A \, = \,\frac{4\pi\alpha}{q_1^2q_2^2} \,  M^{\mu \nu}(q_1,q_2)\, \bar{u}(k_1)\gamma_{\mu} v(k_1') \, \bar{u}(k_2)\gamma_{\nu} v(k_2') ,
\end{equation}
with $q_i = k_i + k_i'$. If CP is conserved only the amplitude
$c(q_1,q_2)$ in Eq.~(\ref{eq:kgsgs}) contributes. The invariant
spectrum in the variables $z_i = q_i^2/M_K^2$, normalized to
$\Gamma_{\gamma \gamma}\equiv \Gamma(K_L\to\gamma\gamma)$,
is given by
\begin{eqnarray} \label{eq:kllll}
\frac{1}{\Gamma_{\gamma \gamma}} \frac{d^2 \Gamma}{dz_1 dz_2} & = &
\frac{2}{z_1z_2} \left(
\frac{\alpha}{3 \pi} \right)^2 |f(z_1,z_2)|^2 \lambda^{3/2}(1,z_1,z_2)   \nonumber \\
&& \times \prod_{i=1,2} \left( 1+2\frac{r_{\ell_i}^2}{z_i} \right) \sqrt{1-4\frac{r_{\ell_i}^2}{z_i}} ,
\end{eqnarray}
where
$f(z_1,z_2) = c(q_1^2,q_2^2) / c(0,0)$.
A parametrization suitable to analyze the data is provided by \cite{D'Ambrosio:1997jp}
\begin{eqnarray}\label{eq:DIPff}
 f(z_1,z_2)\big|_{\mbox{\tiny DIP}}& = & 1 + \alpha_{\mbox{\tiny DIP}} \sum_{i=1,2}
\frac{z_i}{z_i-M_{\rho}^2/M_K^2} \nonumber \\
&&  + \,\beta_{\mbox{\tiny DIP}} \prod_{i=1,2} \frac{z_i}{z_i-M_{\rho}^2/M_K^2}  .\qquad
\end{eqnarray}

The channels $\ell_1=\ell_2=e$  and $\ell_1 = e$, $\ell_2=\mu$
have already been observed experimentally.
The first one is only sensitive to the parameter
  $\alpha_{\mbox{\tiny DIP}}$ due to the low invariant mass of
the $e^+e^-$ pairs
\cite{AlaviHarati:2001ab}. Although the second
channel should be more suitable to extract the parameter
$\beta_{\mbox{\tiny DIP}}$, in
practice that is still not feasible due to poor statistics
\cite{AlaviHarati:2002eh}. Consequently, $\beta_{\mbox{\tiny
      DIP}}=0$ is assumed and
$\alpha_{\mbox{\tiny DIP}}$ is extracted.
The most precise determination is obtained for the channel with muons
giving $\alpha_{\mbox{\tiny DIP}}= -1.59 \pm 0.37$, in good
agreement with the results from
$K_L \rightarrow \gamma \mu^+ \mu^-$. The world average is dominated by the $K_L \rightarrow \gamma e^+ e^-$ mode and is given by $\alpha_{\mbox{\tiny DIP}}= -1.69 \pm 0.08$.
A complete lowest-order calculation of
QED radiative corrections was also performed
\cite{Barker:2002ib}. Its impact
on the slope $\alpha_{\mbox{\tiny DIP}}$ could amount to a 15$\%$
correction, well within the present error.

The high-$q^2$ behaviour of the function $f(z_1,z_2)$ in the
  model of \textcite{D'Ambrosio:1997jp}
enforces a constraint on its parameters that is
violated only in a very mild way,
namely the sum rule
$1+2 \alpha_{\mbox{\tiny DIP}}+ \beta_{\mbox{\tiny DIP}}=0$. Hence,
from the world average value of $\alpha_{\mbox{\tiny DIP}}$, we
can estimate the  phenomenologically elusive second
parameter: $\beta_{\mbox{\tiny DIP}} \simeq~2.4$.

\subsection{$K \rightarrow \ell^+ \ell^-$ }
\label{sec:kll}
The most general amplitude for this process is given by
\begin{equation}\label{eq:kllgen}
 A(K \rightarrow \ell^+ \ell^-)\, =\,  \overline{u}(k) \left( i B + C \gamma_5 \right) v(k').
\end{equation}
If CP is conserved the amplitude $B$ (p-wave $^3P_0$)
determines $K^0_1 \rightarrow \ell^+ \ell^-$
while $K^0_2 \rightarrow \ell^+ \ell^-$ proceeds via the s-wave $^1S_0$ amplitude $C$. The associated width is
\begin{equation}\label{eq:gakllgen}
\Gamma(K \rightarrow \ell^+ \ell^-) = \frac{M_K}{8 \pi} \beta_{\ell} \left( \beta_{\ell}^2 |B|^2 + |C|^2 \right),
\end{equation}
with $\beta_{\ell} = \sqrt{1-4m_{\ell}^2/M_K^2}$.

\subsubsection{$K_S \rightarrow \ell^+ \ell^-$}

The main contribution comes from the amplitude $B_{\gamma \gamma}$ that gives
the transition $K_S \rightarrow \gamma^* \gamma^* \rightarrow \ell^+
\ell^-$ \cite{Ecker:1991ru}. This two-loop amplitude is finite
because chiral symmetry forbids any CP-invariant local contribution at this order.
The result can be written as
\begin{eqnarray}\label{eq:epksll}
R_S^{\ell} &=& \frac{\Gamma(K_S \rightarrow \ell^+
  \ell^-)}{\Gamma(K_S \rightarrow \gamma \gamma)} \\
&=& \frac{\alpha^2 \beta_{\ell}^2 m_{\ell}^2}{2 \pi^2 |H(0)|^2 M_K^2}
 \left| I_{\ell,\mbox{\footnotesize disp}} + i
 I_{\ell,\mbox{\footnotesize abs}}\right|^2 , \no
\end{eqnarray}
where $H(0)$ follows from Eq.~(\ref{eq:1lhz}) and $I_{\ell,
  \mbox{\footnotesize disp}}$ ($I_{\ell, \mbox{\footnotesize abs}}$)
indicate the dispersive (absorptive) parts of the two-loop
diagrams. The $\ell=e$ case is dominated
by the absorptive contribution:
$I_{e,\mbox{\footnotesize disp}}\simeq 1.4$,
$I_{e, \mbox{\footnotesize abs}}\simeq -35$. This gives
$R_S^e = 7.8 \times 10^{-9}$
corresponding to
\begin{equation}
\mathrm{BR}(K_S \rightarrow e^+ e^-) = 2.1\times 10^{-14} ,
\end{equation}
to be compared
with the recent bound
by \textcite{Ambrosino:2008zi}:
\begin{equation}
\mathrm{BR}(K_S \rightarrow e^+ e^-) < 9\times 10^{-9}
\quad  (90\% \, \, \mbox{C.L.}).
\end{equation}
For $\ell=\mu$ there
is a slight dominance of the
dispersive contribution,
$
I_{\mu,\mbox{\footnotesize disp}}\simeq
-2.8$,
$I_{\mu, \mbox{\footnotesize abs}}\simeq 1.2$,
which implies
$R_S^{\mu}=1.9 \times 10^{-6}$ and
\begin{equation}
\mathrm{BR}(K_S \rightarrow \mu^+ \mu^-) = 5.1\times 10^{-12},
\end{equation}
to be
compared with the (almost 30-year-old) bound \cite{Gjesdal:1973zn}
\begin{equation}
\mathrm{BR}(K_S \rightarrow
\mu^+ \mu^-) < 3.2\times 10^{-7} \quad (90 \% \, \, {\rm C.L.}) .
\end{equation}

\begin{figure}[floatfix]
\leavevmode
\includegraphics[width=7cm]{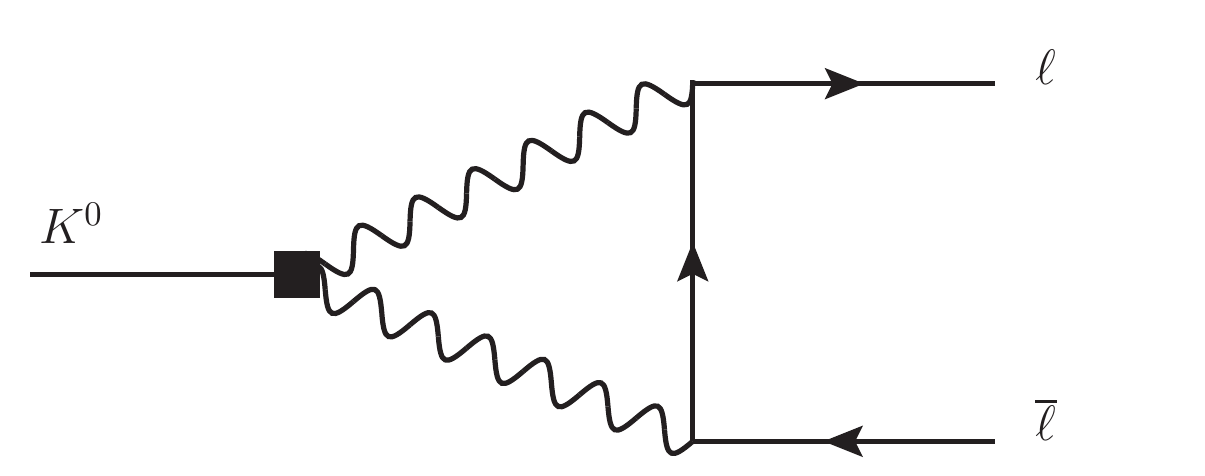}
\caption{\label{fig:kll}
Two-photon contribution to $K^0 \rightarrow \ell^+ \ell^-$.}
\end{figure}

There exists a small short-distance contribution to this decay
through the CP-violating component of the s-wave amplitude $C$ in
Eq.~(\ref{eq:gakllgen}),
which in the SM was estimated for the muon case as
\cite{Isidori:2003ts,Buchalla:1995vs}
\begin{equation}\label{eq:ksmumusd}
C_{\mbox{\footnotesize sd}}
= - \frac{G_F \, \alpha(M_Z)}{\pi \, \sin^2 \theta_W}\, \sqrt{2}\,
m_{\mu}\, F_K \, \mbox{Im} \, (V_{ts}^* V^{\phantom{*}}_{td}) \, Y(x_t),
\end{equation}
where $Y(x_t)$ is given in \textcite{Buchalla:1995vs}. Using the
Wolfenstein parametrization,
the short-distance contribution to the branching ratio is given by
\beqa\label{eq:ksuuwolf}
 \mathrm{BR}(K_S \rightarrow \mu^+ \mu^-)_{\mbox{\footnotesize sd }}  
&\!\! =&\!\! 1.4 \times 10^{-12} \left| \frac{V_{cb}}{0.041}
\right|^4  \left| \frac{\lambda}{0.225}\right|^2  \overline{\eta}^2
\nn
&\!\! \simeq &\!\! 1.7 \times 10^{-13},
\eeqa
an order of magnitude short of the CP-invariant contribution.

\subsubsection{$K_L \rightarrow \ell^+ \ell^-$}

If CP invariance is preserved this decay is given by the
amplitude $C$ in Eq.~(\ref{eq:kllgen}).
It is convenient to normalize the rate to the $K_L\to\gamma\gamma$ mode:
\beqa \label{eq:klllsim}
R_L^{\ell} & =&
\frac{ \Gamma (K_L \rightarrow \ell^+ \ell^-)}{\Gamma (K_L \rightarrow \gamma
  \gamma)}
\\
&=& 2 \beta_{\ell}\left( \frac{\alpha}{\pi} \, r_{\ell}\right)^2
\,\left( |F_{\ell ,\mbox{\footnotesize disp}}|^2 + |F_{\ell ,\mbox{\footnotesize abs}}|^2 \right)  .
\nonumber\eeqa
The absorptive amplitude gets contributions from several available
on-shell states \cite{Martin:1970ai}, but the
$\gamma \gamma$ intermediate state dominates completely (Fig.~\ref{fig:kll}):
\begin{equation} \label{eq:imcld}
F_{ \ell,\mbox{\footnotesize abs}} = \frac{\pi}{2 \, \beta_{\ell}} \,
\ln{\left( \frac{1-\beta_{\ell}}{1+\beta_{\ell}} \right) } .
\end{equation}
The dispersive part arises from the one-loop diagram
$K_L \rightarrow \gamma^* \gamma^* \rightarrow \mu^+ \mu^-$
and local CHPT terms to absorb the loop divergence
\cite{GomezDumm:1998gw,Knecht:1999gb,Isidori:2003ts}:
\begin{eqnarray} \label{eq:gammagammasd}
F_{\ell,\mbox{\footnotesize disp}} & = &
 \frac{1}{4\beta_\ell} \ln^2{\left(\frac{1-\beta_\ell}{1+\beta_\ell}\right)}
+  \frac{1}{\beta_\ell} \mathrm{Li}_2\left(\frac{\beta_\ell-1}{\beta_\ell+1}\right)
+ \frac{\pi^2}{12\beta_\ell}
\nn &&\mbox{} + 3\ln{\frac{m_\ell}{\mu}} + \chi(\mu) ,
\end{eqnarray}
where the local contribution can be split into its long-distance
and short-distance components:
\begin{equation}\label{eq:chiss}
 \chi(\mu) = \chi_{\gamma \gamma}(\mu) + \chi_{sd}  .
\end{equation}
$\chi_{\gamma \gamma}(\mu)$ compensates
the scale dependence of the one-loop amplitude, while $\chi_{sd}$
accounts for the short-distance contribution $\bar s \, d
  \rightarrow \mu^+ \mu^-$.

The $K_L\to\mu^+\mu^-$ decay is well established experimentally. The
measured rate \cite{Nakamura:2010zzb}
\begin{equation}
R_L^{\mu} = (1.25 \pm 0.02) \times 10^{-5}
\end{equation}
appears to be nearly saturated by the absorptive contribution (\ref{eq:imcld}),
$R_L^{\mu}|_{\mbox{\footnotesize abs}} = 1.195 \times 10^{-5}$. The difference between these two numbers
provides an experimental measurement of the modulus of the dispersive amplitude and,
therefore, up to a two-fold ambiguity, of the local term:
\beq\label{eq:chi_KL}
\chi(M_\rho) = \left\{ \ba{c} 3.75\pm 0.20 \\  1.52\pm 0.20\ea\right. .
\eeq
The $K_L\to e^+e^-$ rate is then predicted to be
$R_L^e = (1.552\pm 0.014)\times 10^{-8}$ or $(1.406\pm 0.013)\times 10^{-8}$, respectively.
Both values are in agreement with the present experimental result
$R_L^e|_{\mbox{\footnotesize exp}} = (1.65\pm 0.91)\times 10^{-8}$
\cite{Ambrose:1998cc}, obtained with only four events.
Incidentally, this is at present
the tiniest branching ratio ever measured: ${\rm BR}(K_L
\rightarrow e^+ e^-) = 9^{+6}_{-4} \times 10^{-12}$.
The corresponding branching ratio for the muon case is
${\rm BR}(K_L \rightarrow \mu^+ \mu^-) = (6.84 \pm  0.11) \times
10^{-9}$
\cite{Ambrose:2000gj}.

At $\cO(p^6)$ in the chiral expansion, the $K_L\to\gamma\gamma$ amplitude is
given by the $\pi^0,\eta,\eta'$ exchange mechanism shown in Fig.~\ref{fig:klgg}.
The details of the weak transition vertex cancel out in the ratio $R_L^{\ell}$.
Therefore, removing the short-distance contribution $\chi_{sd}$, Eqs.~(\ref{eq:klllsim}) to
(\ref{eq:gammagammasd}) can be directly applied to the electromagnetic $2\gamma$ decays
of the neutral unflavored pseudoscalar mesons. Moreover, the local CHPT contributions
are also the same. Thus, from the measured $\pi^0\to e^+ e^-$
and $\eta\to\mu^+\mu^-$ rates one can determine $\chi_{\gamma \gamma}(\mu)$
\cite{GomezDumm:1998gw}, again up to a two-fold ambiguity. One gets
$\chi_{\gamma \gamma}(M_\rho)= 0.8\pm 1.0$ or $-20.7\pm 1.0$ from
$\pi^0\to e^+ e^-$,
and $\chi_{\gamma \gamma}(M_\rho)= 5.5\pm 0.9$ or $-0.8\pm 0.9$
  from $\eta\to\mu^+\mu^-$.
The second $\pi^0\to e^+ e^-$ solution is clearly excluded. The first one agrees very well with the
negative $\eta\to\mu^+\mu^-$ solution, but it is only $3.2\sigma$ away from the positive one.
Keeping the two alternative possibilities and following the PDG
average prescription, we get
\beq\label{eq:chi2g}
\chi_{\gamma \gamma}(M_\rho)\, = \, \left\{ \ba{c} -0.1\pm 0.7 \\ 3.4\pm 2.3 \ea\right. ,
\eeq
to be compared with the theoretical estimate
$\chi_{\gamma \gamma}(M_\rho)\, = \, -0.3\pm 0.9$ \cite{Knecht:1999gb},
evaluated within lowest-meson dominance in the large-$N_C$ framework.
An alternative theoretical estimate was obtained
\cite{Isidori:2003ts,D'Ambrosio:1997jp}
performing the $2\gamma$ loop
integration with the form factor (\ref{eq:DIPff}), with
$\alpha_{\mbox{\tiny DIP}}= -(1+\beta_{\mbox{\tiny DIP}})/2 = -1.69 \pm 0.08$
as determined from the radiative decays. One gets in this way
$\chi_{\gamma \gamma}(M_\rho)\, = \, 3.3\pm 1.3$, also compatible with (\ref{eq:chi2g}).

The short-distance contribution to $K_L\to\mu^+\mu^-$ is well known at
NLO \cite{Buchalla:1995vs} and
a NNLO evaluation of the charm-quark contribution $Y_{\mathrm{NL}}$
was carried out more recently \cite{Gorbahn:2006bm}.
The relative sign with respect to the long-distance amplitude can be
fixed unambiguously in the
large-$N_C$ limit \cite{GomezDumm:1998gw,Isidori:2003ts}:
\beqa
 \chi_{sd}^{\mathrm{SM}} & = &
4.965 \times 10^3 \; \left[\mbox{Re}(\lambda_t)\, Y(x_t) + \mbox{Re} (\lambda_c)\, Y_{\mathrm{NL}} \right]
 \nn & = & -1.82\pm 0.04 .
\eeqa
Adding the long-distance component (\ref{eq:chi2g}), one gets
$\chi(M_\rho) = -1.9\pm 0.7$ or $1.6\pm 2.3$, to be compared
with Eq.~(\ref{eq:chi_KL}).
Although the present uncertainties are still too large to make a meaningful test of $\chi_{sd}$,
a better understanding of the long-distance amplitude could well uncover new-physics
contributions.

The longitudinal polarization $P_L$ of either muon in the decay
$K_L \rightarrow \mu^+ \mu^-$ is a measure of \,CP violation
\cite{Pais:1969bi,Sehgal:1969zk}.
Within the SM, the main source for $P_L$ is indirect CP violation due to
$K^0 - \overline{K^0}$ mixing.
The longitudinal polarization arises from the interference of both
amplitudes in
Eq.~(\ref{eq:kllgen}):
\begin{equation}
 \label{eq:muonlonpol}
P_L = \frac{N_R - N_L}{N_R+N_L} = - \frac{M_K \beta_{\mu}^2}{4 \pi
  \Gamma} \; \imag  (B \, C^*) ,
\end{equation}
where $N_R, N_L$ are the numbers of outgoing $\mu^-$ with positive or
negative helicity, respectively
and $\Gamma$ is the full decay width.

Within the SM, one expects \cite{Ecker:1991ru}
\beq
|P_L| = (2.6\pm 0.4)\times 10^{-3} .
\eeq
This polarization has not been measured yet.

\subsection{$K \rightarrow \pi \gamma \gamma^{(*)}$}
\label{sec:kpgg}
The general amplitude for $K(k) \rightarrow \pi(p) \gamma(q_1)\gamma(q_2)$,
\begin{equation}
\label{eq:kpgg}
 A(K\to\pi\gamma\gamma)\, =\, \varepsilon_{\mu}^*(q_1)\, \varepsilon_{\nu}^*(q_2) \, M^{\mu \nu}(k,q_1,q_2)  ,
\end{equation}
contains four Lorentz structures:
\begin{eqnarray}
\label{eq:kpggm}
 M^{\mu \nu} & = & \frac{A(z,y)}{M_K^2}\, \left( q_2^{\mu} q_1^{\nu} - q_1 \cdot q_2\, g^{\mu \nu} \right) \nonumber \\
&+& \frac{2 B(z,y)}{M_K^4}\, \left( -k\cdot q_1\, k \cdot q_2\, g^{\mu \nu} - q_1 \cdot q_2\, k^{\mu} k^{\nu}  \right. \nonumber \\
&& \hskip 1.3cm\left.\mbox{}  + k \cdot q_1\, q_2^{\mu} k^{\nu}
+ k \cdot q_2\, k^{\mu} q_1^{\nu} \right)  \nonumber \\
&+&  \frac{C(z,y)}{M_K^2}\; \varepsilon^{\mu \nu \rho \sigma} q_{1 \rho} q_{2 \sigma} \,  \\
&+&   \frac{D(z,y)}{M_K^4}\, \left[\phantom{l^\lambda\!\!\!\!} \varepsilon^{\mu \nu \rho \sigma} \left( k \cdot q_2\, q_{1 \rho} + k \cdot q_1\, q_{2 \rho} \right)
k_{\sigma} \right. \nonumber \\
&& \hskip 1.2cm \left.\mbox{} + \left( k^{\mu} \varepsilon^{\nu \alpha \beta  \gamma} +
k^{\nu} \varepsilon^{\mu \alpha \beta \gamma} \right) k_{\alpha} q_{1 \beta} q_{2 \gamma} \right]  , \quad\;
\nonumber
\end{eqnarray}
where $z=(q_1 + q_2)^2/M_K^2$ and $y = k \cdot (q_1 - q_2)/M_K^2$.
Bose symmetry requires the invariant amplitudes
$A(z,y)$, $B(z,y)$ and $C(z,y)$ to be even in $y$, while $D(z,y)$ is odd.
In the limit where CP is conserved, $A$ and $B$
contribute only to $K_L \rightarrow \pi^0 \gamma \gamma$, while
$C$ and $D$ contribute to $K_S \rightarrow \pi^0 \gamma \gamma$.
All of them are involved in
$K^+ \rightarrow \pi^+ \gamma \gamma$.

The double differential rate for unpolarized photons is
\begin{eqnarray}
\label{eq:kpggsp}
 \frac{d^2 \Gamma}{d y \,d z} &=& \frac{M_K}{2^9
   \pi^3}\; \bigg\{ \, z^2 \, \left( |A+B|^2 + |C|^2 \right)
\\
&& \hskip 1cm  +\, \left[ y^2- \frac{1}{4}\, \lambda(1,r_{\pi}^2,z) \right]^2 \left( |B|^2 + |D|^2 \right) \bigg\} .
\nonumber
\end{eqnarray}
The physical region is given by
$0  \leq |y| \leq \lambda^{1/2}(1,r_{\pi}^2,z)/2$ and $0 \leq z \leq (1-r_{\pi})^2$.

The processes $K \rightarrow \pi \gamma \gamma$ have no tree-level
contributions of ${\cal O}(p^2)$. At ${\cal O}(p^4)$  the
amplitudes $B$ and $D$ are still zero, since there are not enough
powers of momenta to generate the gauge structure. Therefore, $B$ and $D$
arise only at ${\cal O}(p^6)$. Notice that both $B$ and $D$ lead
to contributions also for small $z$.
\par
The amplitudes for $K \rightarrow \pi \gamma \gamma^*$ have
a related but much more involved structure than with
both photons on-shell. 
We comment briefly on these processes.

\subsubsection{$K^+ \rightarrow \pi^+ \gamma \gamma$}
The leading contribution of ${\cal O}(p^4)$ to the dominant octet
amplitude for $K^+ \rightarrow \pi^+ \gamma \gamma$ was determined in
\textcite{Ecker:1987hd}:
\begin{eqnarray}
 \label{eq:k3pgga}
A^{(4)}(z) &=& \frac{G_8 M_K^2 \alpha}{2 \pi z} \left[ (z+1-r_{\pi}^2) \, F\left( z/r_{\pi}^2 \right)  \right. \nonumber  \\
&& \; \; \; \; \; \; \;   \; \;  \; \;  \; \; \;  + (z-1+r_{\pi}^2) \,  F(z) - \hat{c} z \Big]  ,\qquad
\end{eqnarray}
where $F(z)$ is defined in Eq.~(\ref{eq:1lfz}) and
\begin{equation}
\hat{c} = 128 \pi^2
\left[ 3 (L_9 + L_{10}) + N_{14} - N_{15} - 2 N_{18} \right] /3
\end{equation}
in terms of the LECs in Eqs.~(\ref{eq:Lstrong}),
(\ref{eq:Lweak}). Note that the loop contribution is finite and,
consequently, the counterterm combination is scale independent. The
subdominant 27-plet contribution to the amplitude $A^{(4)}(z)$
was also determined \cite{Gerard:2005yk}. In addition, starting
at ${\cal O}(p^4)$ but going beyond it,
\begin{eqnarray}
 \label{eq:k3pggc1}
C(z) & = & \frac{G_8 M_K^2 \alpha}{\pi}  \left[ \frac{z-r_{\pi}^2}{z-r_{\pi}^2+i r_{\pi} \frac{\Gamma_{\pi^0}}{M_K}}
- \frac{3 z -2 -r_{\pi}^2}{3 (z-r_{\eta}^2)}  \right] , \nonumber \\
&& \;
\end{eqnarray}
where $r_{\eta}= M_{\eta}/M_K$ and $\Gamma_{\pi^0} \equiv \Gamma ( \pi^0 \rightarrow \gamma \gamma )$. Notice that the imaginary
part of the $\eta$ pole has not been included.  This amplitude is generated
by the Wess-Zumino-Witten functional in Eq.~(\ref{eq:Lstrong}) through
the sequence $K^+ \rightarrow \pi^+ (\pi^0, \eta) \rightarrow
\pi^+ \gamma \gamma$.
The $\eta'$ contribution, though fairly suppressed by its mass,
was also considered within $\rm U(3)$ CHPT \cite{Gerard:2005yk}.

\begin{figure}[floatfix]
\leavevmode
\includegraphics[width=8.5cm]{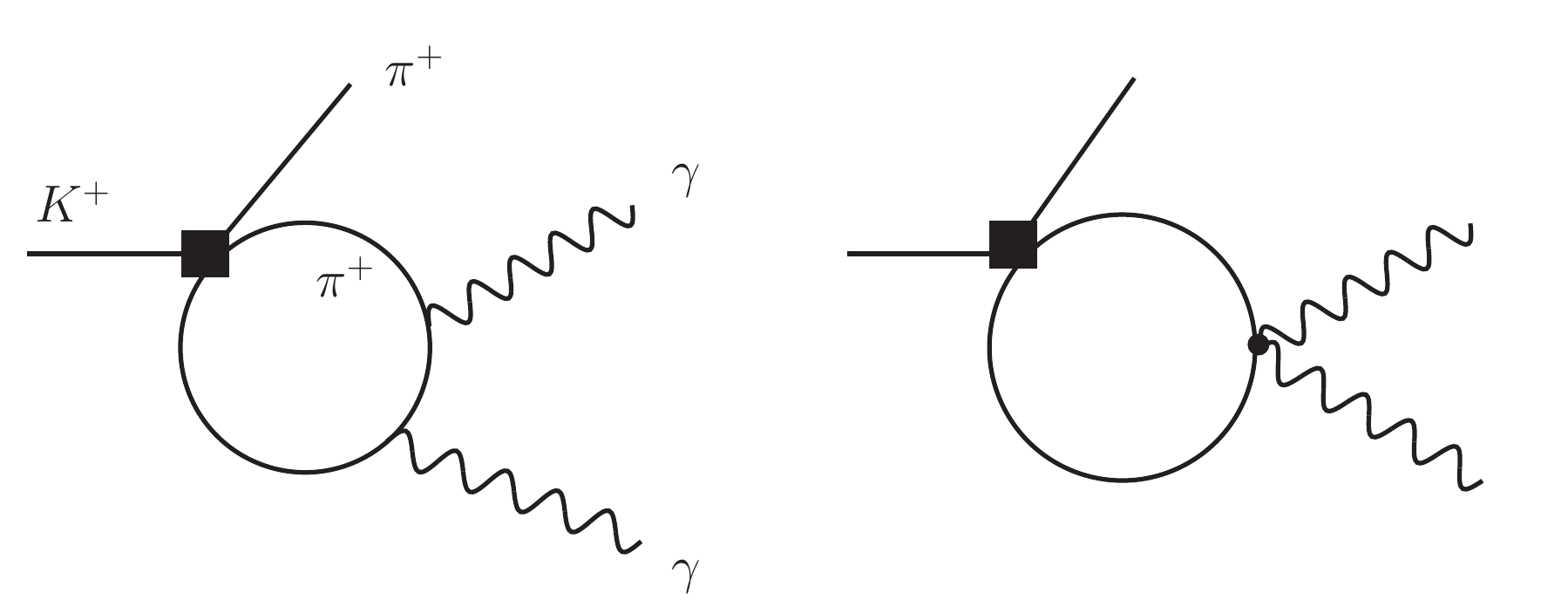}
\caption{\label{fig:kpgg}
Unitarity contribution from $K^+ \rightarrow \pi^+ \pi^+ \pi^-$ to $K^+ \rightarrow \pi^+ \gamma \gamma$.}
\end{figure}

Although there is no complete evaluation to  ${\cal O}(p^6)$,
the most important contributions have been estimated.
The unitarity corrections from $K^+ \rightarrow \pi^+ \pi^+ \pi^-$, shown in Fig.~\ref{fig:kpgg},
were determined by \textcite{D'Ambrosio:1996zx}. They
contribute to both amplitudes $A(z,y)$ and $B(z)$, where $A$
becomes $y$-dependent. Local non-resonant
contributions were also studied and a naive chiral dimensional estimate
indicates that they are small.
Vector-resonance contributions to the ${\cal O}(p^6)$ LECs are
introduced through their effect on the
amplitude $B$ \cite{Cohen:1993ta}:
\begin{equation}
 \label{eq:avkpgg}
a_{V}^+ = - \frac{\pi}{2 G_8 M_K^2 \alpha} \lim_{z \rightarrow 0} B_V(z)  .
\end{equation}
The vector-exchange contribution to $A$ can also be written in terms
of the parameter $a_{V}^+$, if we assume that
these local amplitudes are generated through strong resonance exchange
supplemented with a weak transition in the external
legs \cite{Ecker:1990in}:
\begin{equation}
A_V = G_8 M_K^2 \alpha \, a_{V}^+ (3+r_{\pi}^2-z)/\pi.
\end{equation}
These vector-exchange contributions were estimated by
  \textcite{D'Ambrosio:1996sw}
and found to be very small: $a_{V}^+ = -0.2 \pm 0.3$.

\begin{figure}[floatfix]
\leavevmode
\includegraphics[width=8.2cm]{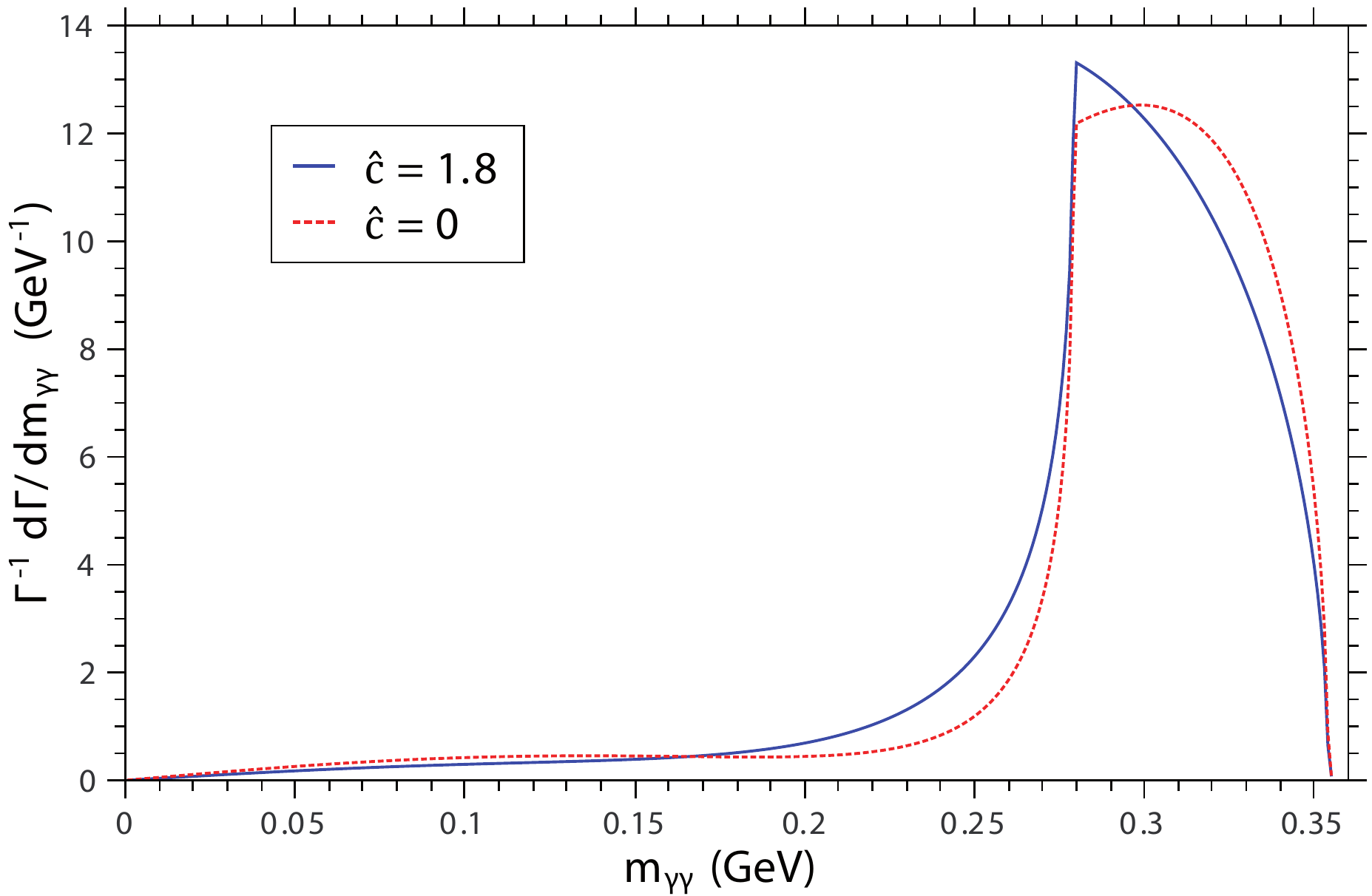}
\caption{\label{fig:kppggspectrum}
Predicted spectra in the diphoton invariant mass at different
values of $\hat{c}$ for $K^+ \rightarrow \pi^+ \gamma \gamma$.}
\end{figure}

\begin{figure}[floatfix]   
\leavevmode
\includegraphics[width=8.2cm]{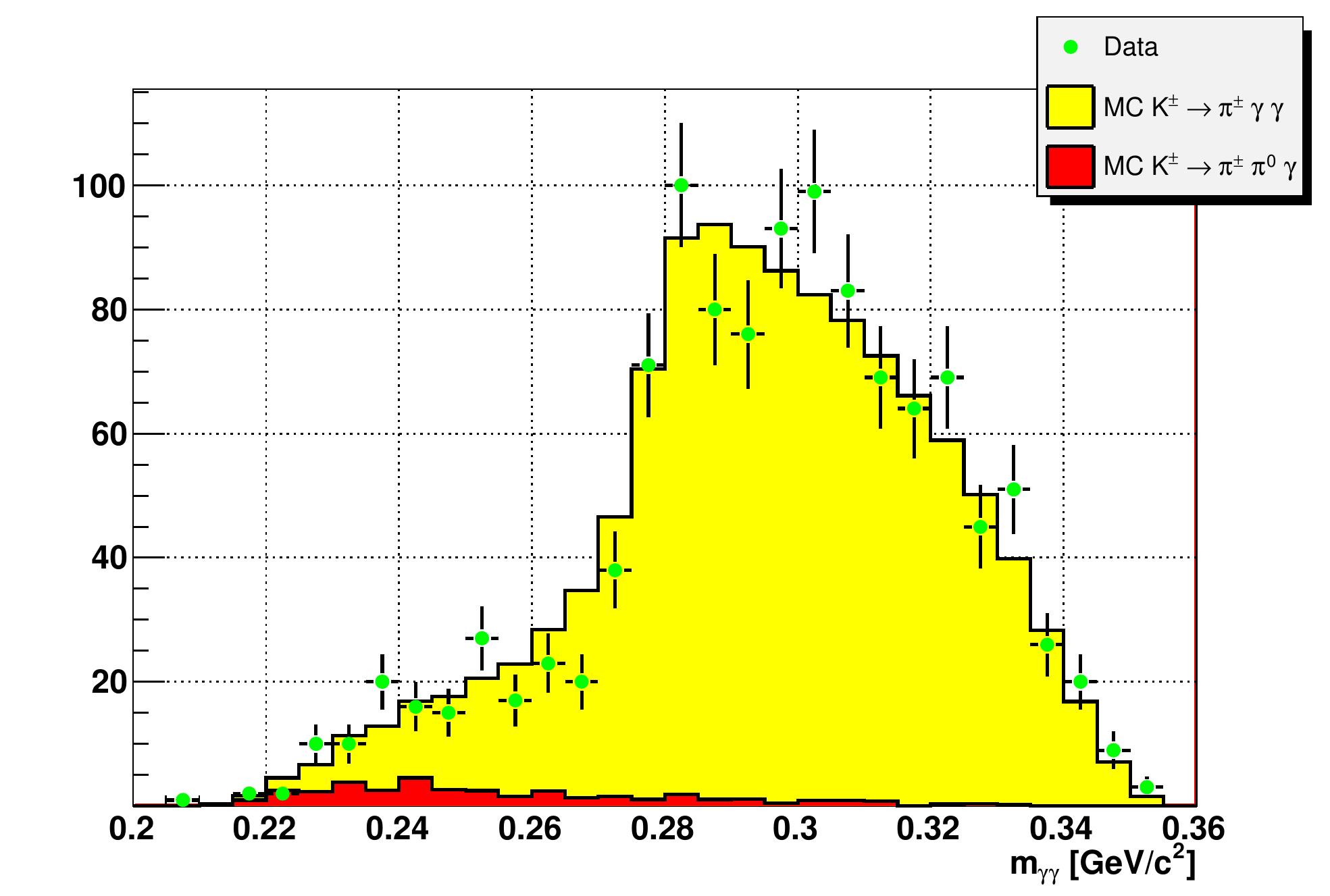}
\caption{\label{fig:kppggna48}
Spectrum for $K^+ \rightarrow \pi^+ \gamma \gamma$ in the
  diphoton invariant mass by NA48/2. From
\textcite{Morales:2008cz}.}
\end{figure}

The branching ratio of this decay and its spectrum were
measured by \textcite{Kitching:1997zj}:
\beq
\mathrm{BR}(K^+ \rightarrow \pi^+ \gamma \gamma)\, =\, (1.1  \pm 0.3) \times 10^{-6} .
\eeq
Hence it is possible to determine the value of
$\hat{c}$ in Eq.~(\ref{eq:k3pgga}). From the analysis of
the normalized spectrum at ${\cal O}(p^4)$ the value $\hat{c} = 1.6 \pm
0.6$ ($\chi^2/\mathrm{dof} = 0.9$) is
obtained. Including the unitarity corrections and neglecting local
${\cal O}(p^6)$ contributions, the fit improves
($\chi^2/\mathrm{dof} = 0.7$) yielding $\hat{c}=1.8  \pm 0.6$.
The diphoton invariant-mass distribution is shown in Fig.~\ref{fig:kppggspectrum}
for different values of $\hat{c}$.
The spectrum observed by NA48/2 is shown in Fig.~\ref{fig:kppggna48}.

The electromagnetic penguin operators induce
a CP-violating charge asymmetry in $K^{\pm} \rightarrow \pi^{\pm}
\gamma \gamma$.
This asymmetry is estimated
to be tiny within the SM \cite{Ecker:1987hd,Gao:2002ub}.

\subsubsection{$K_S \rightarrow \pi^0 \gamma \gamma$}
The leading contribution to this decay is given by
$K^0
\rightarrow \pi^0 (\pi^0, \eta) \rightarrow \pi^0 \gamma \gamma$,
where the first part is dominated by the octet weak transition in
the Lagrangian (\ref{eq:Lweak}) and the second part arises from the
Wess-Zumino-Witten anomalous term in Eq.~(\ref{eq:Lstrong}). The
amplitude starts at ${\cal O}(p^4)$ but includes also higher orders
\cite{Ecker:1987fm}:
\begin{eqnarray}
 \label{eq:k3pggc2}
C(z) &\!\! = &\!\! \frac{G_8 M_K^2 \alpha}{\pi}  \left[ \frac{2-z-r_{\pi}^2}{z-r_{\pi}^2+i r_{\pi} \frac{\Gamma_{\pi^0}}{M_K}}
- \frac{F_{\pi}(2-3 z+r_{\pi}^2)}{3 F_{\eta}\left(z-r_{\eta}^2\right)}  \right] , \nonumber \\
&& \,
\end{eqnarray}
keeping $F_{\pi} \neq F_{\eta}$ to account for chiral corrections.
An amplitude $D(z,y)$ would arise
at higher chiral orders. As in the $K^\pm$ decay,
the $\eta'$ contribution was also included
by \textcite{Gerard:2005yk}. It is seen, however,
that the amplitude is dominated by the pion pole.

In order to eliminate the overwhelming background from $K_S
\rightarrow \pi^0 \pi^0$, one restricts the kinematical region to
$z > 0.2$. The chiral prediction gives
\begin{equation}
\mathrm{BR}(K_S \rightarrow \pi^0
\gamma \gamma)_{z>0.2} = 3.8  \times 10^{-8}
\end{equation}
that compares
well with the experimental measurement \cite{Lai:2003vc}
\begin{equation}
\mathrm{BR}(K_S \rightarrow \pi^0 \gamma \gamma)_{z>0.2} = (4.9  \pm 1.8) \times
10^{-8}.
\end{equation}

\subsubsection{$K_L \rightarrow \pi^0 \gamma \gamma$}
The relevance of this channel extends beyond its own interest as it
provides a CP-conserving contribution via the
two-photon cut to the decay $K_L \rightarrow \pi^0 \ell^+ \ell^-$
\cite{Donoghue:1987awa,Ecker:1987hd,Sehgal:1988ej,Heiliger:1992uh}
that competes with direct and indirect CP-violating contributions.

The absence of ${\cal O}(p^4)$ counterterms for this process indicates
that the only contribution is a finite one-loop result at this order
\cite{Ecker:1987fm,Sehgal:1989pw,Cappiello:1988yg}. The octet part of
the amplitude $A(z)$ is
\begin{eqnarray}
\label{eq:klp0gga}
 A^{(4)}(z) &=& \frac{G_8 M_K^2 \alpha}{\pi \, z} \left[ \left( z-r_{\pi}^2 \right)  F\left(z/r_{\pi}^2
 \right) \right. \nonumber \\
&& \; \; \; \; \; \; \; \; \; \; \; \; \; \; \;  -
\left(z-1-r_{\pi}^2\right) F(z) \Big], \qquad
\end{eqnarray}
where $F(z)$ is given in Eq.~(\ref{eq:1lfz}). This result gives rise
to  $\mathrm{BR}(K_L \rightarrow \pi^0 \gamma \gamma) = 6.8 \times 10^{-7}$,
significantly smaller than the present PDG average
\cite{Nakamura:2010zzb}
\beq
{\rm BR}(K_L \rightarrow \pi^0 \gamma \gamma)\, =\, (1.27 \pm 0.03) \times 10^{-6} .
\eeq
This indicates that higher-order chiral corrections should be considered.

\begin{figure}[floatfix]
\leavevmode
\includegraphics[width=8.2cm]{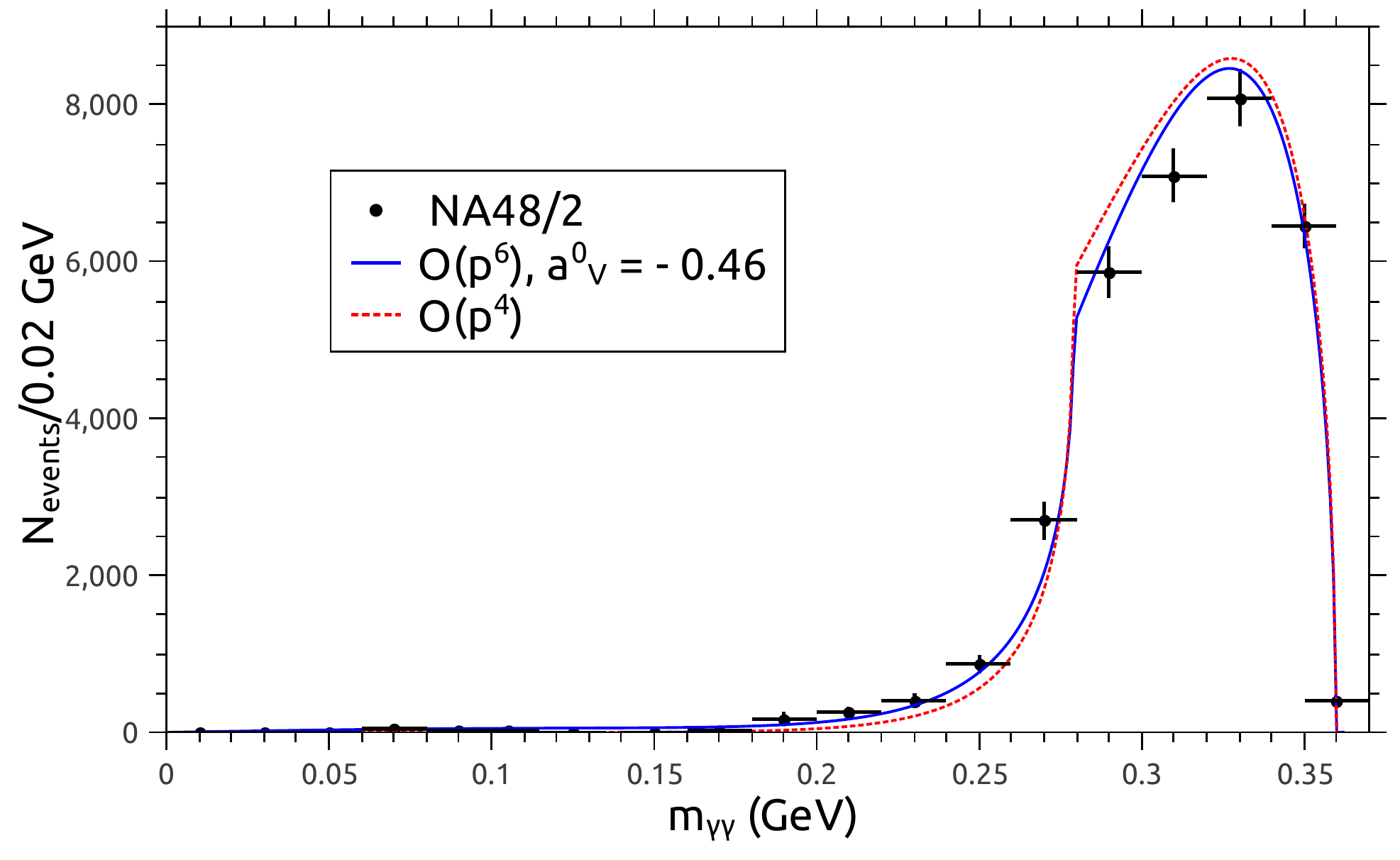}
\caption{\label{fig:klpggspectra}
Comparison of the normalized spectra in the diphoton invariant mass
for  $K_L \rightarrow \pi^0 \gamma \gamma$ at
${\cal O}(p^4)$ and  ${\cal O}(p^6)$ in CHPT. The data are from
\textcite{Lai:2002kf}.
 }
\end{figure}

The unitarity corrections from $K_L \rightarrow \pi^+ \pi^- \pi^0$
contributing at ${\cal O}(p^6)$
\cite{Cappiello:1992kk,Cohen:1993ta} give rise to both $A$ and $B$
  amplitudes. The
inclusion of the 27-plet contribution was carried out by
\textcite{Cappiello:1992kk}. As in the $K^\pm$ case,
those amplitudes also get local contributions
from both vector exchange and
non-resonant parts. From dimensional analysis
the latter seem to be negligible, but the situation of vector resonance
contributions is still not settled. This is parametrized by
$a_V^0$, defined as in Eq.~(\ref{eq:avkpgg}), that
was estimated theoretically to take the value $a_V^0 =
-0.7 \pm 0.3$ \cite{D'Ambrosio:1996sw}.
The measured spectrum
clearly favors a non-zero value for this parameter:
NA48 \cite{Lai:2002kf} gets $a_V^0 = -0.46  \pm 0.05$ while KTeV
\cite{Abouzaid:2008xm} obtains $a_V^0 = -0.31 \pm 0.09$.
The predicted spectrum is compared with the measured distribution
in Fig.~\ref{fig:klpggspectra}.

On the basis of the experimental situation at the time,
  \textcite{Gabbiani:2002bk} argued that the complete set of three
  $\cO(p^6)$ counterterms is needed to describe both branching ratio
  and spectrum. However, the latest experimental analyses
  \cite{Lai:2002kf,Abouzaid:2008xm} find a satisfactory representation
  in terms of $a_V^0$ only, supporting the dominance of vector exchange.

\subsubsection{$K \rightarrow \pi \gamma \ell^+ \ell^-$}
These decays are obviously mediated by
$K \rightarrow \pi \gamma\gamma^*$ and are
closely related to the case with
both photons on-shell. The study of $K^+ \rightarrow \pi^+ \gamma
\ell^+ \ell^-$ and $K_L \rightarrow \pi^0 \gamma \ell^+ \ell^-$
was carried out by \textcite{Donoghue:1997rr,Donoghue:1998ur}
and \textcite{Gabbiani:1998tj}.

The decay amplitude has a more complex Lorentz structure
than in Eq.~(\ref{eq:kpggm}).
Nevertheless, at ${\cal O}(p^4)$ these decays have the same properties as $K^+
\rightarrow \pi^+ \gamma \gamma$ and
$K_L \rightarrow \pi^0 \gamma \gamma$, i.e., both have finite one-loop
contributions, the charged channel has the same
scale-independent counterterm $\hat{c}$ and the neutral one
has no local contributions.
Taking $\hat{c}= 1.8$, one obtains
at this order:
\beqa
\mathrm{BR}(K^+ \rightarrow \pi^+ \gamma e^+ e^-) & = & 1.4 \times 10^{-8} ,
\nn
\mathrm{BR}(K_L \rightarrow\pi^0 \gamma e^+ e^-)& =& 1.0 \times 10^{-8}.
\eeqa
These predictions have to be compared with the recent
experimental measurements \cite{Batley:2007uh}
\begin{equation}
\mathrm{BR}(K^+ \rightarrow \pi^+ \gamma e^+ e^-)
= (1.29 \pm 0.13) \times 10^{-8}
\end{equation}
and
\cite{Abouzaid:2007xe}
\begin{equation}
\mathrm{BR}(K_L \rightarrow \pi^0 \gamma e^+ e^-) = (1.62 \pm 0.17) \times 10^{-8}.
\end{equation}
As can be seen, the agreement
is rather good in the charged-kaon channel.
\par
Unitarity corrections from $K \rightarrow \pi \pi \pi$, extending beyond
${\cal O}(p^4)$, were performed for both decays. We comment on
them in turn.
\begin{itemize}
 \item $K^+ \rightarrow \pi^+ \gamma e^+ e^-$: \ Local ${\cal O}(p^6)$
   counterterms were found to be negligible in $K^+ \rightarrow
\pi^+ \gamma \gamma$. However, neglecting them also in this case
and using $\hat{c}=1.8$, one finds
$
\mathrm{BR}(K^+ \rightarrow \pi^+ \gamma e^+ e^-) = 1.7 \times 10^{-8},
$
which disagrees with the experimental result.
Indeed, the analysis of the spectrum (for $m_{ee\gamma} > 260
\, \mbox{MeV}$) carried out in \textcite{Batley:2007uh}
gets $\hat{c}=0.90 \pm 0.45$. Hence local contributions of ${\cal O}(p^6)$
and beyond might be non-negligible after all.
\item  $K_L \rightarrow \pi^0 \gamma e^+ e^-$: \ Together with unitarity
  corrections,
\textcite{Donoghue:1997rr}
included explicit contributions from vector resonances.
Their prediction
$
\mathrm{BR}(K_L \rightarrow \pi^0 \gamma e^+ e^-) = 2.3 \times 10^{-8}
$
is in slight disagreement with the experimental
measurement quoted above.
\textcite{Abouzaid:2007xe}
concluded that employing a value of $a_V^0 = -0.46$ (in accordance with
$K_L \rightarrow \pi^0 \gamma \gamma)$, the branching ratio goes down
to
$
\mathrm{BR}(K_L \rightarrow \pi^0 \gamma e^+ e^-) = 1.51 \times 10^{-8},
$
in better agreement with the measured width.
\end{itemize}


\subsection{$K \rightarrow \pi \ell^+ \ell^-$ }
\label{sec:kpill}
The FCNC transitions $K \rightarrow \pi \ell^+ \ell^-$ ($\ell = e,
\mu$) are dominated
by single virtual-photon exchange ($K \rightarrow \pi \gamma^*$) if
allowed by CP invariance as in $K_S$ and $K^{\pm}$ decays.
This contribution is CP violating for $K_L \rightarrow \pi^0 \ell^+
\ell^-$ and consequently this process has become a point
of reference for studying the CP-violating sector of the SM.

\subsubsection{$K_S, \, K^{\pm} \rightarrow \pi \ell^+ \ell^-$}
The amplitude for $K(k) \rightarrow \pi(p) \ell^+(p_+)
\ell^-(p_-)$ is determined by an
electromagnetic transition form factor in the presence of the
nonleptonic weak interactions \cite{Ecker:1987qi,D'Ambrosio:1998yj}:
\begin{equation}
\label{eq:AVz}
 A  = -
\frac{G_F \, \alpha}{4 \pi}\, V(z) \, (k+p)^{\mu} \,
\overline{u}_{\ell}(p_-) \gamma_{\mu} v_{\ell}(p_+)  ,
\end{equation}
where $z = q^2/M_K^2$ and $q = k-p$. The spectrum in the
dilepton invariant mass is given by
\begin{equation}
\label{eq:spectdilep}
\frac{d \Gamma}{dz} = \frac{G_F^2 \alpha^2 M_K^5}{12 \pi (4 \pi)^4} \;  \bar\lambda^{3/2} \,
\sqrt{1-4\frac{r_{\ell}^2}{z}} \, \left( 1 + 2 \frac{r_{\ell}^2}{z}\right) \,  |V(z)|^2  ,
\end{equation}
where $\bar\lambda \equiv \lambda(1,z,r_{\pi}^2)$ and $4 r_{\ell}^2 \leq z \leq (1-r_{\pi})^2$.
\par
Because of gauge invariance $V(z) = 0$ at ${\cal O}(p^2)$.
The first evaluation at ${\cal O}(p^4)$ of the dominant octet amplitude was
carried out by \textcite{Ecker:1987qi}:
\begin{equation}
\label{eq:vpmz}
V_{+}(z) = - \frac{G_8}{G_F} \, \left[ \, \Phi(z) + \Phi(z/r_\pi^2) +
  w_{+} \, \right]
\end{equation}
for $K^{\pm} \rightarrow \pi^{\pm} \gamma^*$, and
\begin{equation}
\label{eq:vsz}
 V_S(z) = \frac{G_8}{G_F} \, \left[ \, 2 \,\Phi(z) + w_S \, \right]
\end{equation}
for $K_S \rightarrow \pi^0 \gamma^*$. Here
\begin{equation}
 \label{eq:phipz}
\Phi(z) = \frac{5}{18} - \frac{4}{3z} + \frac{4}{3z}\left(1-\frac{z}{4}\right) G(z)  ,
\end{equation}
where
$G(z)$ is given in Eq.~(\ref{eq:1lgz}).
In terms of the LECs in Eqs.~(\ref{eq:Lstrong}) and (\ref{eq:Lweak}),
the local ${\cal O}(p^4)$ contributions are
\begin{eqnarray}
\label{eq:ww}
 w_+ & = & \frac{64 \pi^2}{3} \left( N_{14}^r - N_{15}^r + 3 L_9^r \right) + \frac{1}{3} \ln \frac{\mu^2}{M_K M_{\pi}} \, , \nonumber \\
w_S & = & \frac{32 \pi^2}{3} \left( 2 N_{14}^r + N_{15}^r \right) + \frac{1}{3} \ln \frac{\mu^2}{M_K^2}  .
\end{eqnarray}
\begin{figure}[floatfix]
\leavevmode
\includegraphics[width=6.5cm]{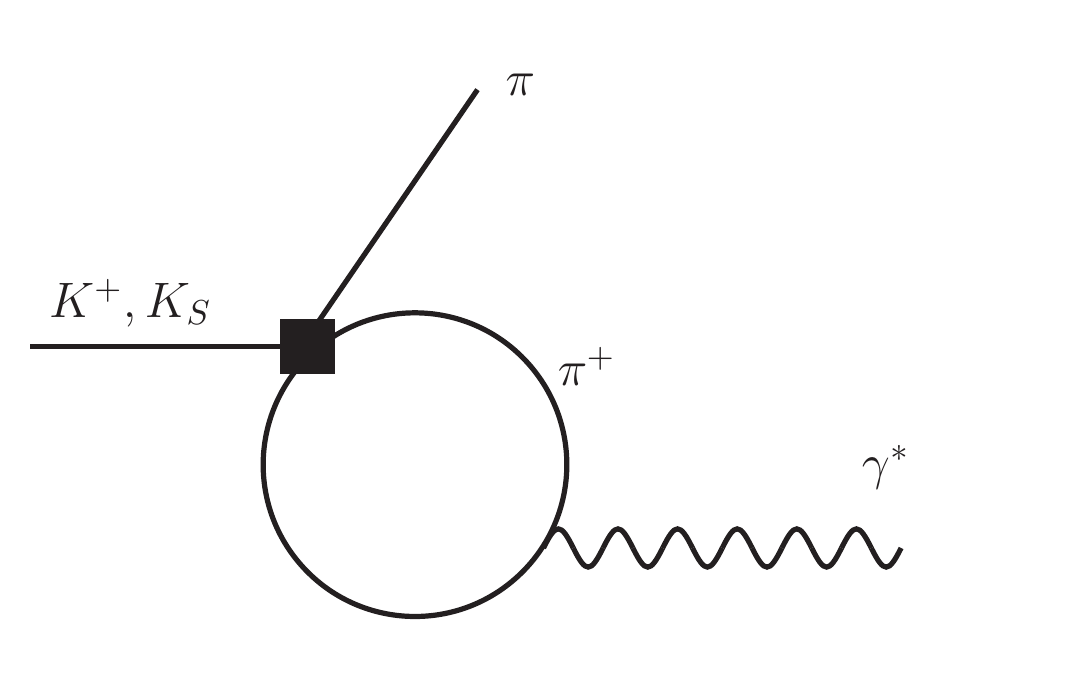}
\caption{\label{fig:kpgs}
Unitarity $K \rightarrow \pi \pi \pi$ contribution to $K^+, K_S \rightarrow \pi \gamma^*$.}
\end{figure}

Although a complete study to ${\cal O}(p^6)$ has
not been performed yet, the unitarity
corrections from $K \rightarrow \pi \pi \pi$, shown in  Fig.~\ref{fig:kpgs},
were determined by \textcite{D'Ambrosio:1998yj}.
The full result can be decomposed in a
polynomial contribution (linear in $z$ up to this order)
plus the unitarity loop corrections,
\begin{equation}
\label{eq:viz}
 V_i(z) = a_i + b_i z + V_i^{\pi \pi}(z)  \qquad\quad (i = +, S)  ,
\end{equation}
the latter being given by
\begin{equation}
\label{eq:vippz}
 V_i^{\pi \pi}(z) = \frac{\alpha_i + \beta_i (z-z_0)/r_{\pi}^2}{G_F M_K^2 r_{\pi}^2} \,
 \left[1+\frac{z}{r_V^2}\right] \, \left[\Phi(z/r_{\pi}^2) + \frac{1}{6}\right] ,
\end{equation}
with $z_0 = r_{\pi}^2 + 1/3$, $r_V = M_\rho/M_K$
and
\begin{eqnarray}
\label{eq:alphasbetas}
 \alpha_+ = \beta_1 - \frac{1}{2} \, \beta_3 + \sqrt{3} \, \gamma_3 , &\quad &
\beta_+ = 2 \, (\xi_1+\xi_3-\xi_3')  ,
\nonumber \\
\alpha_S = - \frac{4}{\sqrt{3}} \,  \gamma_3 , \hskip 1.54cm &&
\beta_S = \frac{8}{3} \, \xi_3'  ,
\end{eqnarray}
in terms of the parameters in Eq.~(\ref{eq:ddexp}). The
  parametrization (\ref{eq:viz}) includes the $\cO(p^4)$
  contributions.
The polynomial part incorporates the local counterterms in Eq.~(\ref{eq:ww})
  through $a_+^{(4)} = G_8/G_F (1/3 - w_+)$ and $a_S^{(4)} =
  -G_8/G_F (1/3 - w_S)$, and tiny contributions to the slopes $b_i^{(4)}$
  from the ${\cal O}(p^4)$ kaon loop.

$V_i(z)$ in Eq.~(\ref{eq:viz}) is expected to be an excellent approximation to the complete form factor of $\cO(p^6)$. It only assumes that all contributions except the two-pion intermediate state can be well approximated by a linear polynomial for small values of $z$. The
predicted rates can then be expressed in terms of $a_i$ and $b_i$ in
Eq.~(\ref{eq:viz}):
\begin{eqnarray}
\label{eq:Kbrnum}
\mathrm{BR}_{K^{\pm}}^e &=& \left[ \,  0.15 - 3.31 \, a_+ - 0.90 \,
  b_+  + 60.51 \, a_+^2 \right. \nonumber \\
&& \left. \mbox{}  + 16.36 \, a_+ b_+ + 1.77 \, b_+^2 \, \right] \times
10^{-8} \, , \nonumber \\
\mathrm{BR}_{K^{\pm}}^{\mu} &=& \left[ \,  1.19 - 19.97 \, a_+ - 6.56
  \, b_+  + 120.16 \, a_+^2 \right. \nonumber \\
&& \left. \mbox{}  + 69.42 \, a_+ b_+ + 10.59 \, b_+^2 \, \right] \times
10^{-9} \, , \nonumber \\
\mathrm{BR}_{K_S}^e &=& \left[ \,  0.01 - 0.55 \, a_S - 0.17 \, b_S  +
  43.76 \, a_S^2 \right. \nonumber \\
&& \left. \mbox{}  + 11.83 \, a_S b_S + 1.28 \, b_S^2 \, \right] \times
10^{-10} \, , \nonumber \\
\mathrm{BR}_{K_S}^{\mu} &=& \left[ \,  0.07 - 3.96 \, a_S - 1.34 \,
  b_S  + 86.90 \, a_S^2 \right. \nonumber \\
&& \left. \mbox{}  + 50.21 \, a_S b_S + 7.66 \, b_S^2 \, \right] \times 10^{-11}  .
\end{eqnarray}
If $a_i, b_i \sim 1$ as expected, the polynomial contribution
dominates over the unitarity-cut loop corrections coming from $K
\rightarrow \pi \pi \pi$. In particular, due to the strong suppression
of $K_S \rightarrow \pi^+ \pi^- \pi^0$, the latter are tiny for
the $K_S$ decay.
Thus, these decays are very sensitive to the chiral LECs.

\begin{figure}[floatfix]
\leavevmode
\includegraphics[width=7.5cm]{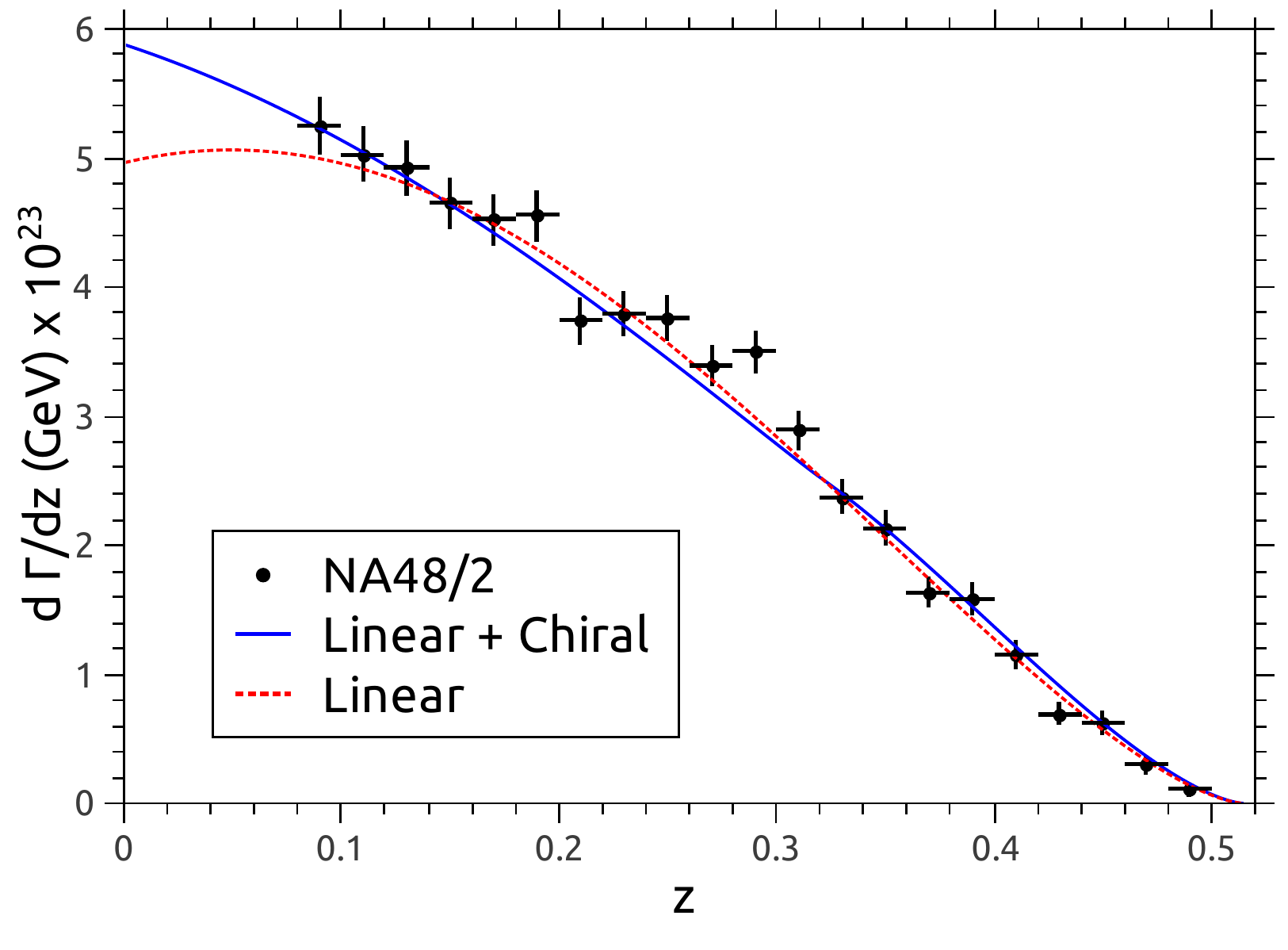}
\caption{\label{fig:kpgspectra1}
Spectra for $K^+ \rightarrow \pi^+ e^+ e^-$ in the dilepton invariant mass. The data are from
  \textcite{Batley:2009pv}.
 }
\end{figure}

\begin{figure}[floatfix]
\leavevmode
\includegraphics[width=8.2cm]{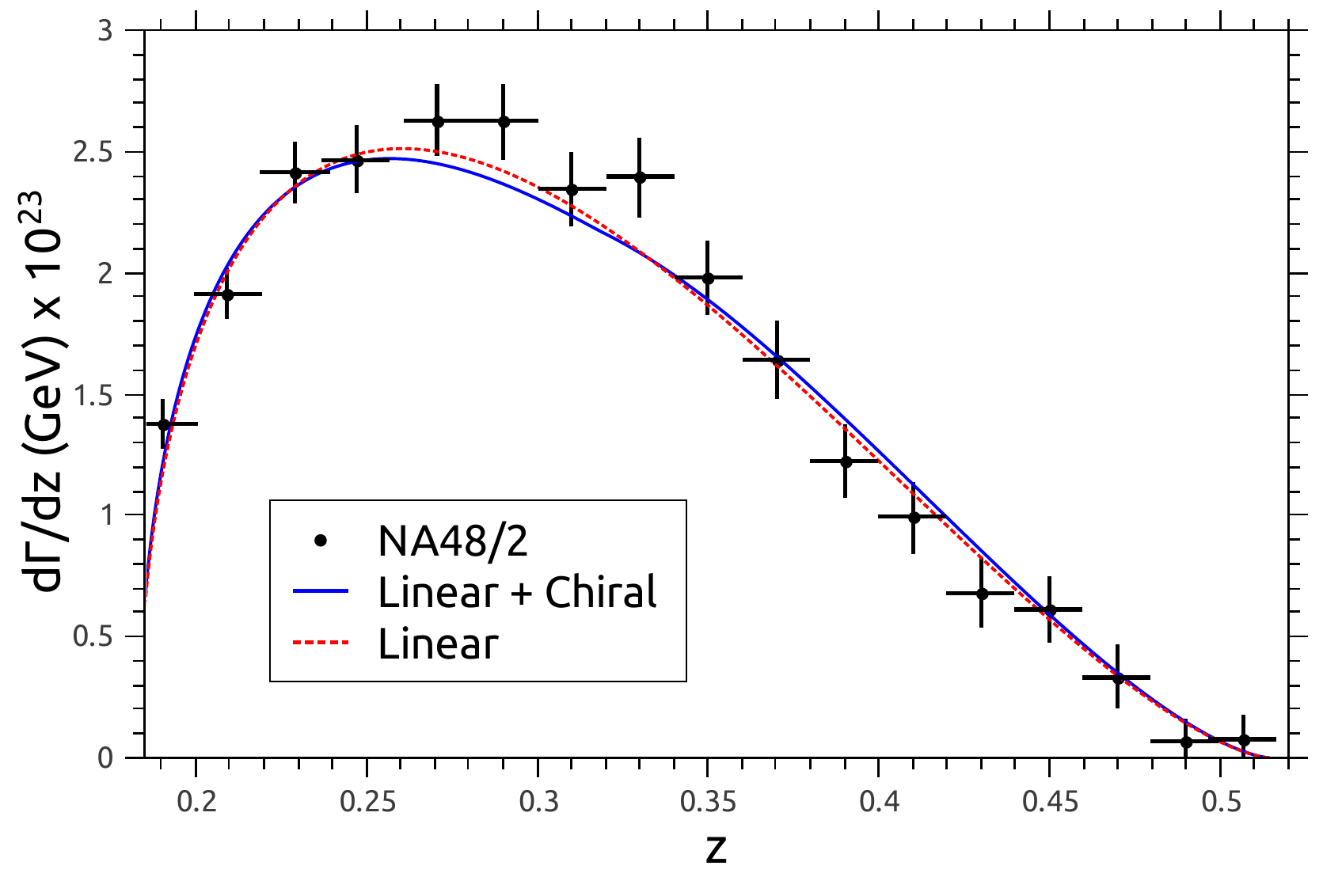}
\caption{\label{fig:kpgspectra2}
Spectra for $K^+ \rightarrow \pi^+ \mu^+ \mu^-$ in the dilepton
invariant mass. The data are from  \textcite{Batley:2011zz}.}
\end{figure}

The values of $a_+$ and $b_+$ have been fitted from the $K^+
\rightarrow \pi^+ \ell^+ \ell^-$ spectra.
One finds $a_+ = -0.578 \pm 0.016$ and $b_+ = -0.779 \pm 0.066$ for
$\ell=e$ \cite{Batley:2009pv},
while the muonic mode gives $a_+ = -0.575 \pm 0.039$ and $b_+ = -0.813
\pm 0.145$  \cite{Batley:2011zz}.
The resulting experimental branching ratios are
\beqa
\mathrm{BR}(K^\pm\to\pi^\pm e^+e^-)\, &=& (3.14\pm 0.10)\times 10^{-7} ,
\nn
\mathrm{BR}(K^\pm\to\pi^\pm\mu^+\mu^-) &=& (9.62\pm 0.25)\times 10^{-8} .\qquad
\eeqa
In Figs.~\ref{fig:kpgspectra1} and \ref{fig:kpgspectra2} we show the
spectra of $K^+ \rightarrow \pi^+ \ell^+ \ell^-$
($\ell = e, \mu$) and the comparison with the theoretical predictions
for the chiral form factors (\ref{eq:viz}) and for
a linear form factor [without $V_+^{\pi \pi}(z)$ in
  Eq.~(\ref{eq:viz})].

For the $K_S$ decay only branching ratios are available. By using the
ratio $b_i/a_i = 1/r_V^2$ given by vector meson dominance
\cite{D'Ambrosio:1998yj}, it is found that
$|a_S| = 1.06^{+0.26}_{-0.21}$ for the electron \cite{Batley:2003mu} and
$|a_S| = 1.54^{+0.40}_{-0.32}$ for the muon case \cite{Batley:2004wg}.
Notice though that the vector meson dominance
ratio $b_i/a_i$ quoted above fails in the charged kaon decay.
The measured rates are
\beqa
\mathrm{BR}(K_S\to\pi^0 e^+e^-)\, &=& (5.8\, {}^{+2.9}_{-2.4})\times 10^{-9} ,
\nn
\mathrm{BR}(K_S\to\pi^0\mu^+\mu^-) &=& (2.9\, {}^{+1.5}_{-1.2})\times 10^{-9} ,\qquad
\eeqa
where the large errors reflect the low statistics.

Other parametrizations going beyond ${\cal O}(p^4)$ have also been put
forward, focusing on the polynomial part of $V(z)$.
\textcite{Friot:2004yr} assume a minimal narrow-resonance structure, while
a related vector meson dominance approach is introduced in \textcite{Dubnickova:2006mk}.
Unfortunately, the size of present data samples is insufficient to distinguish between models.

In addition to the dominant $K \rightarrow \pi\gamma^*$ amplitude, the form factor $V(z)$
receives a short-distance contribution from the
operator $Q_{7V}$ in Eq.~(\ref{eq:Q11-13}).
Although negligible in the branching ratios and spectra, its
interference with the long-distance amplitude
leads to a CP-violating charge asymmetry in $K^{\pm} \rightarrow \pi^{\pm} \ell^+ \ell^-$
\cite{Ecker:1987hd}.
This asymmetry is tiny within the SM
\cite{D'Ambrosio:1998yj,D'Ambrosio:2002fa}.

\subsubsection{$K_L \rightarrow \pi^0 \ell^+ \ell^-$}
\begin{figure}[floatfix]
\leavevmode
\includegraphics[width=7cm]{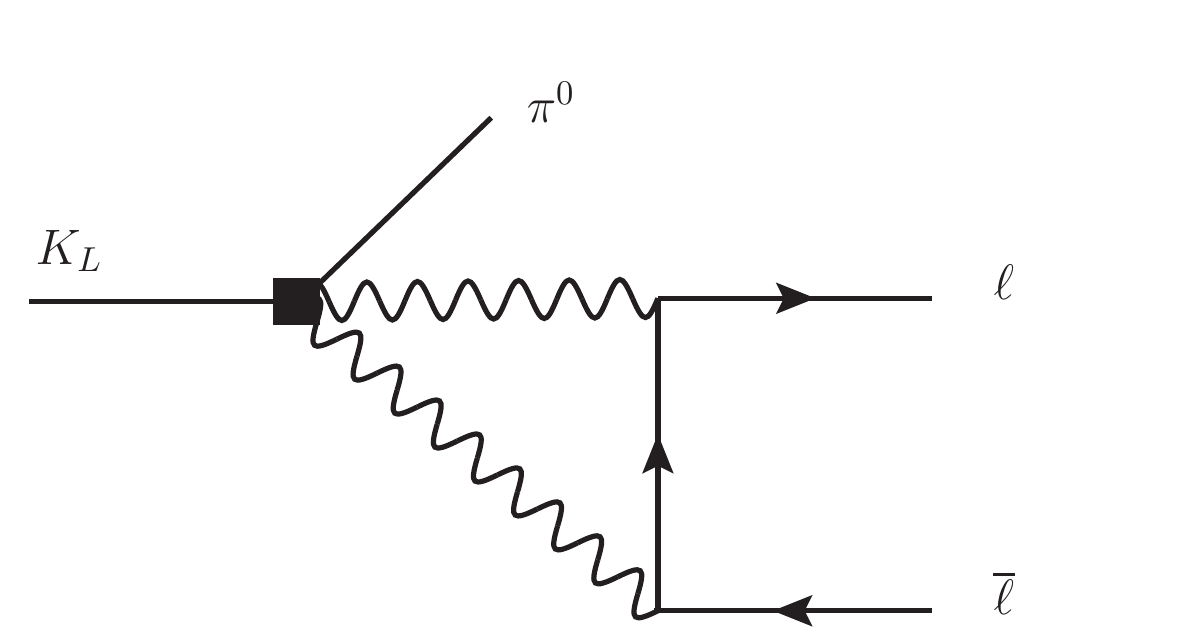}
\caption{\label{fig:klpgg}
CP-conserving contribution
to $K_L \rightarrow \pi^0 \ell^+ \ell^-$.}
\end{figure}

$K_L \rightarrow \pi^0 \left( \ell^+ \ell^- \right)_{J=1}$ with the
lepton pair in a vector or axial-vector state is
CP violating and, accordingly, also $K_L \rightarrow \pi^0
\gamma^*$. There are three main contributions to the
$K_L \rightarrow \pi^0 \ell^+ \ell^-$ decay that, in principle, could be
of the same order of magnitude
\cite{Ecker:1987hd,Donoghue:1994yt}:

\begin{enumerate}
\item[i.] A direct CP-violating transition induced by the
short-distance Lagrangian (\ref{eq:Leff}).
The relevant contributions come from the operators $Q_{7V}$ and
$Q_{7A}$  in Eq.~(\ref{eq:Q11-13})
and are proportional to $\mathrm{Im}\,\lambda_t$.

\item[ii.] An indirect CP-violating amplitude due to the
   $K^0 - \overline{K^0}$ oscillation, proportional to
   the CP-violating parameter $\epsilon$:
\beq
A(K_L \rightarrow \pi^0 \ell^+ \ell^-)|_{\mbox{\tiny CPV-ind}} =
 \epsilon\, A (K_S \rightarrow \pi^0  \ell^+ \ell^-).
\eeq

\item[iii.] A CP-conserving contribution from $K_L \rightarrow
  \pi^0 \gamma\gamma$
  through $\gamma \gamma \rightarrow \ell^+ \ell^-$ rescattering,
  as indicated in Fig.~\ref{fig:klpgg}.
\end{enumerate}

Let us first consider the CP-violating contributions
\cite{Heiliger:1992uh,D'Ambrosio:1998yj,Isidori:2004rb,Buchalla:2003sj,Flynn:1988ve}.
  Together with a vector contribution  $V(z)$, as in Eq.~(\ref{eq:AVz}), from
 indirect CP violation and the $Q_{7V}$ amplitude, the $Q_{7A}$ operator generates
 axial-vector and pseudoscalar terms:
\begin{eqnarray}
\label{eq:vvap}
 V^{\mbox{\tiny ind}}(z)
&\simeq &  \, \pm\, \epsilon
\, \left[ \,  a_S + b_S z \, \right]  ,  \nonumber \\
 V^{\mbox{\tiny dir}}(z) &= & i \,\frac{4 \pi \, y_{7V}}{\sqrt{2} \alpha} \, \imag \lambda_t\; f_{+}^{K \pi}(z)  , \nonumber \\
A(z) &= &  i \,\frac{4 \pi \, y_{7A}}{\sqrt{2} \alpha} \, \imag \lambda_t \; f_{+}^{K \pi}(z)  , \nonumber \\
P(z) &= &  -i \,\frac{8 \pi \, y_{7A}}{\sqrt{2} \alpha}\, \imag \lambda_t \; f_{-}^{K \pi}(z)   ,
\end{eqnarray}
where
$f_{\pm}^{K \pi}(z)$ are the $K_{\ell 3}$ form factors.
In $V^{\mbox{\tiny ind}}(z)$ we have neglected the tiny unitarity corrections from
$K_S \rightarrow \pi^+ \pi^- \pi^0$ and have made explicit a possible ambiguity
in the relative sign with respect to the short-distance contributions.

The pseudoscalar amplitude is helicity suppressed and can be neglected in the electron case.
Using $b_S/a_S = 1/r_V^2$,
the total CP-violating contribution to the rate is \cite{Buras:1994qa,Buchalla:2003sj}
\begin{eqnarray}
\label{eq:ecase}
&& \! \! \! \! \! \! \!
\mathrm{BR}(K_L \rightarrow \pi^0 e^+ e^-)_{\mbox{\tiny CPV}} =
   \\
&& \! \! \! \! \! \! \!
10^{-12} \times \left[ \, 15.7 \, |a_S|^2 \pm 6.2 \, |a_S| \left(
\frac{\imag
\lambda_t}{10^{-4}}
\right)
  + 2.4 \left( \frac{\imag \lambda_t}{10^{-4}} \right)^2 \right].
 \nonumber
\end{eqnarray}
For the muon case (including the dependence on $m_\mu$),
one obtains \cite{Isidori:2004rb}
\begin{eqnarray}
\label{eq:mucase}
&& \! \! \! \! \! \! \!
\mathrm{BR}(K_L \rightarrow \pi^0 \mu^+ \mu^-)_{\mbox{\tiny CPV}} =
\\
&& \! \! \! \! \! \! \!
10^{-12} \times \left[ \, 3.7 \, |a_S|^2 \pm 1.6 \, |a_S| \left(
\frac{\imag
\lambda_t}{10^{-4}}
\right)
  + 1.0 \left( \frac{\imag \lambda_t}{10^{-4}} \right)^2 \right] . \nonumber
\end{eqnarray}
In both cases the error in the numerical coefficients is estimated to be $10\% - 20\%$.
For an expected value of $a_S \sim 1$, the largest contribution comes from the indirect
CP-violating component, particularly in the case  $\ell=e$.
Based on several assumptions on the matching between short-
and long-distance dynamics, \textcite{Buchalla:2003sj} concluded that
a positive interference  between the
indirect and direct CP-violating components is
the most natural setting. Using
the values of $|a_S|$ extracted from the analyses of $K_S \rightarrow \pi^0
\ell^+ \ell^-$ and considering a positive interference, one finds
\begin{eqnarray}
\label{eq:klcpvbr}
\mathrm{BR}(K_L \rightarrow \pi^0 e^+ e^-)_{\mbox{\tiny CPV}} & = & (3.1  \pm 0.9)
\times 10^{-11} \, , \nonumber \\
\mathrm{BR}(K_L \rightarrow \pi^0 \mu^+ \mu^-)_{\mbox{\tiny CPV}} & = & (1.4 \pm
0.5) \times 10^{-11}  . \qquad\;
\end{eqnarray}

Now we consider the CP-conserving contribution from $K_L \rightarrow \pi^0\gamma \gamma$.
Neglecting the $y$ dependence of the $2\gamma$ amplitudes $A$, $B$ in Eq.~(\ref{eq:kpggm}),
the combination $A+B$ gives
  $|\gamma \gamma \rangle_{J=0}$ and the amplitude $B$ provides
  $|\gamma \gamma \rangle_{J=2}$. However, the transition to $|\ell^+
  \ell^-\rangle$ is helicity suppressed ($\propto m_{\ell}$) for the
  initial $J=0$ state \cite{Donoghue:1987awa}.
  Accordingly, for $\ell = e$ only the
  amplitude $B$ [arising beyond ${\cal O}(p^4)$ in the chiral counting]
  could give a non-negligible contribution, while for $\ell=\mu$ both amplitudes
can be significant \cite{Ecker:1987hd,Isidori:2004rb,Flynn:1988gy,Morozumi:1988vy}.

Through a naive dimensional analysis of the amplitude $B(z)$, \textcite{Ecker:1987hd}
concluded that the CP-conserving contribution to $K_L \rightarrow\pi^0 e^+ e^-$ was much
smaller than the CP-violating one, making this decay a good testing ground of the SM.
This was later confirmed by a two-loop calculation, including
unitarity corrections from
$K_L \rightarrow \pi^+ \pi^- \pi^0$ and ignoring ${\cal O}(p^6)$ local
contributions
\cite{Isidori:2004rb}, which found values below $10^{-12}$ for the
CP-conserving $K_L \rightarrow\pi^0 e^+ e^-$ rate. A local
contribution to $B(z)$, parametrized through the coupling $a_V^0$
[see Eq.~(\ref{eq:avkpgg})] determined from $K_L \rightarrow
\pi^0\gamma \gamma$ data, results
in smaller values around $10^{-13}$. The contribution from
  $B(z)$ turns out to be very small because of phase space and
angular momentum suppression. Therefore,
the $K_L \rightarrow\pi^0 e^+ e^-$ decay is dominantly CP violating.

The CP-conserving contribution to the muon channel,
\begin{equation}
 \mathrm{BR}(K_L \rightarrow \pi^0 \mu^+ \mu^-)_{\mbox{\tiny CPC}}  =  (5.2
  \pm 1.6) \times 10^{-12}  ,
\end{equation}
is of the same order but slightly smaller than the CP-violating component
(\ref{eq:klcpvbr}). The interference between CP-conserving and
CP-violating amplitudes
can then generate a sizable transverse polarization of the
muons that could be within reach
of the next generation of experiments \cite{Ecker:1987hd}.

The experimental upper bounds on these decays are getting close to the
SM predictions. At $90 \%$ C.L., we have \cite{AlaviHarati:2003mr}
\begin{equation}
\mathrm{BR}(K_L \rightarrow \pi^0 e^+ e^-) < 2.8 \times
10^{-10}
\end{equation}
and \cite{AlaviHarati:2000hs}
\begin{equation}
\mathrm{BR}(K_L \rightarrow \pi^0 \mu^+ \mu^-) < 3.8 \times 10^{-10} .
\end{equation}
\subsection{$K \rightarrow \pi \pi \gamma^{(*)}$}
\label{sec:kppg}
The amplitude for $K(p_K) \rightarrow \pi_1(p_1) \pi_2(p_2) \gamma(q)$
\begin{equation}
 \label{eq:kppga}
A(K\to\pi\pi\gamma)\, =\, \varepsilon_{\mu}^{*}(q) \, M^{\mu}(q,p_1,p_2)
\end{equation}
is decomposed into dimensionless electric ($E$) and magnetic ($M$)
components defined by
\begin{equation}
 \label{eq:kppgem}
M_{\mu} = \frac{E(z_i)}{M_K} \left[ \,  z_1 \, p_{2 \mu} -
  z_2 \, p_{1 \mu} \, \right]+ \frac{M(z_i)}{M_K^3} \,
\varepsilon_{\mu \nu \rho \sigma} p_1^{\nu} p_2^{\rho} q^{\sigma}  ,
\end{equation}
with $z_i = q \cdot p_i / M_K^2$  $(i=1,2)$ and $z_3 = z_1+z_2 = E_{\gamma}^*/M_K$,
where $E_{\gamma}^*$ is the photon energy in the kaon rest frame. Summing
over photon helicities, there is no interference between both amplitudes
and the double differential rate for an unpolarized photon is given by
\begin{eqnarray}
 \label{eq:kppgddr}
\lefteqn{\frac{d^2 \Gamma}{d z_1 \,d z_2} = \frac{M_K}{(4
  \pi)^3} \, \left( \, |E(z_i)|^2 + |M(z_i)|^2 \,  \right)}&&
\\
&&  \quad\times \left[ z_1 z_2 \left( 1-2(z_1+z_2)-r_1^2-r_2^2 \right)
-r_1^2 z_2^2 -r_2^2 z_1^2 \right] , \nonumber
\end{eqnarray}
with $r_i = M_{\pi_i}/M_K$. The electric amplitude can be decomposed
as the sum of
inner-bremsstrahlung (IB) and direct-emission (DE) components, $E =
E_{\mbox{\tiny IB}} + E_{\mbox{\tiny DE}}$, while the magnetic
amplitude contains DE only.
$E_{\mbox{\tiny IB}}$ is determined by the Low  theorem
  \cite{Low:1958sn}:
\begin{equation}
 \label{eq:lowkppg}
E_{\mbox{\tiny IB}}(z_i) = \frac{e}{M_K \, z_1 z_2} \,  A(K
\rightarrow \pi_1 \pi_2).
\end{equation}
In general, $E_{\mbox{\tiny IB}}$ dominates the photon spectrum for low
values of the photon energy (in the kaon rest
frame), but it is obviously absent for $K_S, K_L \rightarrow
\pi^0 \pi^0 \gamma$.

The DE amplitudes explore the electromagnetic structure
of the hadron interaction.
They are customarily decomposed in a multipole expansion \cite{D'Ambrosio:1992bf}:
\begin{eqnarray}
 \label{eq:multipolekppg}
E_{\mbox{\tiny DE}} &=& E_1(z_3) + E_2(z_3) \cdot z_{12} + {\cal O}
\left( z_{12}^2 \right) , \nonumber \\
M &=& M_1(z_3) + M_2(z_3) \cdot z_{12} +  {\cal O} \left( z_{12}^2 \right)  ,
\end{eqnarray}
with $z_{12} = z_1-z_2$.
This decomposition is particularly relevant for initial states with
definite CP (neutral channels), as even electric and odd magnetic
multipoles are CP odd while the others are CP even. Hence CP
invariance would require the amplitudes $E_1, M_2, E_3, \ldots$ to
vanish for $K_L \rightarrow \pi^+ \pi^- \gamma$, whereas $M_1, E_2,
M_3, \ldots$  vanish for $K_S \rightarrow \pi^+ \pi^- \gamma$. For
$K^{\pm} \rightarrow \pi^{\pm} \pi^0 \gamma$ all of them are
allowed. Moreover, the limited phase space ($|z_{12}| \lsim 0.17$)
indicates that higher-order multipoles are suppressed.

The decays $K \rightarrow \pi(p_1) \pi(p_2) \ell^+(k_+) \ell^-(k_-)$
where the lepton pair is produced by an off-shell photon
have amplitudes
\begin{equation}
 \label{eq:kpplla}
A(K\to\pi\pi\ell^+\ell^-) = \frac{e}{q^2} \, V_{\mu} \,
\overline{u}_{\ell}(k_-) \gamma^{\mu} v_{\ell}(k_+) ,
\end{equation}
where $q= k_+ + k_-$ and
\begin{equation}
\label{eq:kppllv}
 V_{\mu} = \frac{{\cal E}_1(z_i)}{M_K} p_{1 \mu} + \frac{{\cal
 E}_2(z_i)}{M_K} p_{2 \mu}  + \frac{M(z_i)}{M_K^3} \, \varepsilon_{\mu
 \nu \rho \sigma} p_1^{\nu} p_2^{\rho} q^{\sigma}  .
\end{equation}

\subsubsection{$K^+ \rightarrow \pi^+ \pi^0 \gamma$}
The IB contribution to this decay is suppressed by the $\Delta I =
1/2$ rule because it is proportional to the
$K^+ \rightarrow \pi^+ \pi^0$ amplitude, a $\Delta I = 3/2$
transition. However, this is still the largest
contribution to the rate. The dominant DE contribution is ${\cal O}(p^4)$
in CHPT \cite{Bijnens:1992ky,Ecker:1991bf}.
At this order, the electric
amplitude has a finite one-loop contribution of which
the octet part turns out to be tiny
\cite{Ecker:1993cq,D'Ambrosio:1994du}:
\begin{equation}
 \label{eq:e4loopkp}
E_{\mbox{\tiny loop}}^{(4)} = -ie \frac{G_8 M_K}{8 \pi^2 F_0} (M_K^2-M_{\pi}^2) \left[ h_{\pi K}(z_+)+h_{K \eta}(z_+) \right],
\end{equation}
where the functions $h_{ij}(z)$ are given in Eq.~(\ref{eq:hijz}).
A sizeable 27-plet contribution to the DE
electric amplitude, from an enhanced two-pion loop, has been recently identified \cite{Mertens:2011ts}.
There is also a local scale-independent contribution
\begin{equation}
 \label{eq:e4ctkp}
E^{(4)}_{\mathrm{ct}} = 2 i e \frac{G_8 \, M_K^3}{F_0} \, \left(
N_{14} - N_{15} - N_{16} -N_{17} \right)
\end{equation}
in terms of the LECs in Eq.~(\ref{eq:Lweak}). Model-dependent
estimates \cite{Ecker:1992de,D'Ambrosio:1997tb}
indicate that the vector meson contributions to $E^{(4)}_{\mathrm{ct}}$ are
suppressed  whereas an axial-vector meson exchange piece remains.
This amplitude could be measured from the interference with the IB
contribution. The ${\cal O}(p^4)$ magnetic amplitude has a
reducible component from the Wess-Zumino-Witten anomalous
Lagrangian in Eq.~(\ref{eq:Lstrong}) and a direct
  non-anomalous
contribution from odd-intrinsic-parity operators in
Eq.~(\ref{eq:Lweak}):
\begin{equation}
 \label{eq:m4ctkp}
M^{(4)} = - \,\frac{e}{2 \pi^2} \frac{G_8 M_K^3}{F_0} \left[ 1 - 16
\pi^2 \left( 3 N_{29} - N_{30} \right) \right] .
\end{equation}
A factorization estimate of the couplings \cite{D'Ambrosio:1997tb}
indicates that the different contributions are of the same order
and interfere constructively.
Some ${\cal O}(p^6)$ analyses
were also performed \cite{Ecker:1993cq,D'Ambrosio:1994du}
but the lack of knowledge on the LECs at this order makes it difficult
to assess their relevance. The role of
form factors induced by vector meson exchange was also explored
\cite{Cappiello:2007rs,D'Ambrosio:2000yc}.
\par
The differential cross section is usually written in terms of $T^*_C$,
the kinetic energy of the charged pion in the kaon
rest frame, and of $W^2=(q \cdot p_K)(q \cdot p_+)/(M_{\pi^+}^2 M_K^2)$,
factorizing the IB contribution \cite{D'Ambrosio:1992bf},
\begin{eqnarray}
\label{eq:kppgtw}
\frac{d^2 \Gamma^{\pm}}{d T_c^* \,d W^2} &=&
\frac{d^2 \Gamma^{\pm}_{IB}}{d T_c^* \,d W^2}
\left[ 1 +  2 \cos \left( \delta_1^1-\delta_0^2 \pm \phi\right) Y_E
  W^2 \right.  \nonumber \\
&& \; \; \; \; \; \; \; \; \; \; \; \; \; \; \;  \left.  + \left(
  Y_E^2+Y_M^2 \right) W^4 \right]  ,
\end{eqnarray}
where $\Gamma^+$ ($\Gamma^-$) corresponds to the decay of $K^+$ ($K^-$).
Here $\phi$ is a possible CP-violating phase, $\delta_{\ell}^I$ are
the strong rescattering phases for a final $\pi \pi$
  state of isospin
$I$ and orbital angular momentum ${\ell}$. $Y_E$,
$Y_M$ are the DE electric and magnetic amplitudes, respectively,
normalized to $K^{\pm} \rightarrow \pi^{\pm} \pi^0$. It is usually
assumed that $Y_E$ and $Y_M$ are nearly independent of $T_c^*$ and
are therefore constants over the Dalitz plot.  The linear term in $W^2$
corresponds to the interference between the amplitudes
$E_{\mathrm{DE}}$ and $E_{\mathrm{IB}}$.
A recent measurement by NA48/2 \cite{Batley:2010uja} establishes that
$Y_M/Y_E = -11 \pm 3$, showing that among
the DE amplitudes the magnetic contribution dominates.

Charge asymmetries in this channel are interesting observables for
CP-violating effects \cite{D'Ambrosio:1996nm}.
A bound on the fully integrated asymmetry $\delta \Gamma = (\Gamma^+-
\Gamma^-)/(\Gamma^+ + \Gamma^-)$ has recently been published
\cite{Batley:2010uja}: $|\delta \Gamma| < 1.5 \times 10^{-3}$ at
90$\%$ C.L. translates into $|\sin \phi| < 0.56$.
The asymmetry constructed with the partially integrated decay widths
\cite{Colangelo:1999kr} is also of interest:
\begin{equation}
 \label{eq:omegakppg}
\frac{d A_{W}}{d W^2} \, = \,
\frac{d \Gamma^+ / d W^2 - d \Gamma^- / d W^2}
{d \Gamma^+ /  d W^2 + d \Gamma^- / d W^2}.
\end{equation}
\textcite{Batley:2010uja} obtain $A_W = (-0.6 \pm 1.0) \times 10^{-3}$.

\subsubsection{$K_L \rightarrow \pi^+ \pi^- \gamma$}
The IB component is proportional to the amplitude for
$K_L \rightarrow \pi^+ \pi^-$. Consequently, it violates CP
and it is rather small. Assuming CP invariance, the dominant
contribution is given by the $M_1$ amplitude because $E_2$ is
suppressed by phase space. These features make this decay very suitable
to study the DE magnetic amplitude
\cite{Lin:1987de,Picciotto:1991ae,Ecker:1993cq,Ecker:1991bf,D'Ambrosio:1997ta,D'Ambrosio:2000yc,D'Ambrosio:1994du}.

The leading $E_2$ amplitude arises at ${\cal O}(p^4)$ in CHPT. There
is no local contribution and therefore $E_2$ is given
by a finite one-loop result \cite{Ecker:1993cq,D'Ambrosio:1994du}:
\begin{eqnarray}
 \label{eq:e2kppgp4}
E_{\mbox{\tiny loop}}^{(4)} &=& ie \, \frac{G_8 M_K}{8 \pi^2 F_0} (M_K^2 - M_{\pi}^2) \,
\left[ h_{\pi K}(z_{-}) + h_{K \eta}(z_{-}) \right.
\nonumber \\
&& \left. \mbox{} - h_{\pi K}(z_{+}) - h_{K \eta}(z_{+}) \right] ,
\end{eqnarray}
with $h_{ij}(z)$ given in Eq.~(\ref{eq:hijz}).
However, as commented above, this contribution is
even smaller than the CP-violating IB component: $| E_{\mbox{\tiny
    loop}}^{(4)} / E_{\mbox{\tiny IB}} | \sim 10^{-2}$.
The 27-plet contribution to the $E_2$ amplitude has also been
determined \cite{Mertens:2011ts}.

The $M_1$ multipole at leading ${\cal O}(p^4)$ is generated by a
constant local contribution \cite{Ecker:1993cq} in terms of the
LECs in Eq.~(\ref{eq:Lweak}):
\begin{equation}
 \label{eq:m1kppgp4}
M^{(4)} = - 16 \, e \, \frac{G_8 M_K^3}{F_0} \, \left( N_{29} + N_{31} \right) .
\end{equation}
\par
The magnetic amplitude was also studied at ${\cal O}(p^6)$,
including both local \cite{D'Ambrosio:1997ta,Ecker:1993cq,Lin:1987de}
and one-loop contributions \cite{D'Ambrosio:1997ta}.
A piece of the local term is completely fixed by the anomalous
WZW Lagrangian in Eq.~(\ref{eq:Lstrong}), as shown in
  Fig.~\ref{fig:kppg}a:
\begin{figure}[floatfix]
\leavevmode
\includegraphics[width=8.5cm]{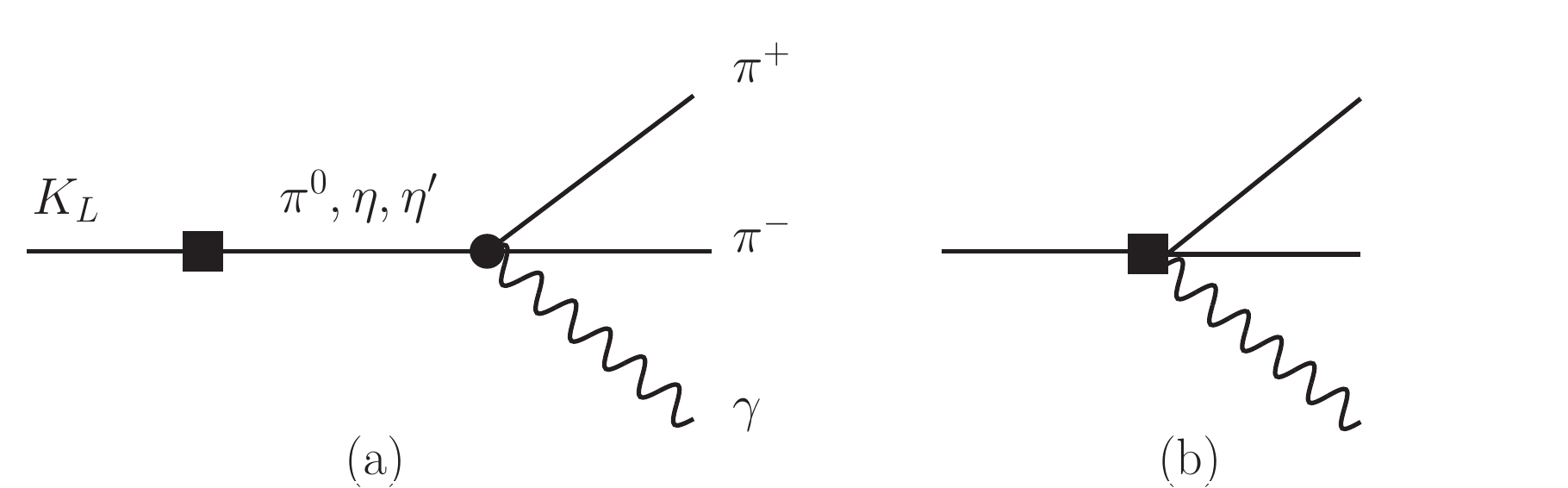}
\caption{\label{fig:kppg}
Diagram (a) has anomalous and resonance-dominated contributions to $K_L \rightarrow \pi^+ \pi^- \gamma$.
Diagram (b) gives a direct weak transition in terms of the couplings
in Eq.~(\ref{eq:Lweak}).}
\end{figure}
\begin{equation}
 \label{eq:m6kppgwzw}
M_{\mbox{\tiny anom}}^{(6)} = e \, \frac{G_8 M_K^3}{2 \pi^2 F_0} \,
F_1(\hat\rho, \xi, \theta) \, ,
\end{equation}
in terms of $\theta$, $\xi$ and $\hat\rho$ defined in
Eqs.~(\ref{eq:theta}), (\ref{eq:xi}) and (\ref{eq:rho}), where
\begin{eqnarray} \label{eq:f1}
 F_1 & = & \frac{1}{1-r_{\pi}^2}
 + \frac{1}{3 \left( 1-r_{\eta}^2 \right)}\left[ \left(1+\xi \right) \cos \theta +
2 \sqrt{2}\hat\rho \sin \theta \right] \nonumber \\
&& \; \; \; \; \; \; \; \; \;\; \; \; \;
 \;\; \times  \left[ \left(\frac{F_{\pi}}{F_{\eta_8}} \right)^3
   \cos \theta -  \sqrt{2} \left(\frac{F_{\pi}}{F_{\eta_0}}\right)^3 \sin \theta \right]\;
 \nonumber \\
&-&  \frac{1}{3 ( 1-r_{\eta'}^2)} \left[ 2 \sqrt{2} \hat\rho \cos \theta - \left( 1 + \xi \right)
\sin \theta \right] \nonumber \\
&& \; \; \;   \times \left[
  \left( \frac{F_{\pi}}{F_{\eta_8}}\right)^3 \sin \theta +  \sqrt{2} \left(\frac{F_{\pi}}{F_{\eta_0}}\right)^3
   \cos \theta \right].
\end{eqnarray}
This expression includes higher orders beyond ${\cal O}(p^6)$.

The first $z_3$ dependence comes with resonance dominated
local counterterms that depend both on strong vertices (with a weak
transition in the external kaon leg) and pure weak vertices
\cite{D'Ambrosio:1997ta}. A similar situation arises in the one-loop
amplitude $M_{\mbox{\tiny loop}}^{(6)} = M_{\mbox{\tiny WZW}} +
M_{N_i}$ where the first term originates from the WZW
Lagrangian with a weak transition in the external kaon leg, and
the second  is determined by vertices of the weak Lagrangian
(\ref{eq:Lweak}). Unfortunately, due to the presence of
weak LECs, it is not possible to provide a model-independent
theoretical prediction.
\par
The phenomenological analysis was originally
carried out with only the magnetic amplitude and with a $\rho$-pole
dominated form factor \cite{Lin:1987de}:
\begin{equation}
\label{eq:ga1a2kppg}
 M = e \, \frac{G_8 M_K^3}{2 \pi^2 F_0} \, \tilde{g}_{M_1} \, \left( 1
 + \frac{a_1/a_2}{(M_{\rho}^2-M_K^2) + 2 M_K^2 z_3} \right).
\end{equation}
Both parameters were determined by \textcite{Abouzaid:2006hy}:
$|\tilde{g}_{M_1}| = 1.20  \pm 0.09$ and $a_1/a_2 = (-0.738  \pm 0.019) \,
\mbox{GeV}^2$. A  model-independent parametrization is provided by a
power series expansion in the variable $z_3$:
\begin{equation}
 \label{eq:phenokppg}
M =  e \, \frac{G_8 M_K^3}{2 \pi^2 F_0} \, g_{M_1}  \, ( 1 + c \, z_3) \, .
\end{equation}
The values obtained from the expression (\ref{eq:ga1a2kppg})
correspond to $|g_{M_1}| =
1.30  \pm 0.12$ and $c = -1.74  \pm 0.08$. It was noticed \cite{Ecker:1993cq}
that the slope is rather large compared with naive expectations based
on resonance saturation. Indeed, a quadratic approximation also shows
rather large values for both linear and quadratic slopes
\cite{AlaviHarati:2000pt}. Although the value of $c$ can be easily
accommodated for a reasonable range of the parameters involved in the
theoretical analyses \cite{D'Ambrosio:1997ta,Ecker:1993cq},
the predicted photon spectrum  is slightly
at variance with the experimental determination. The ratio
between the width due to the DE magnetic amplitude and the total width
was also measured: $\Gamma_{\mbox{\tiny DE}} /
(\Gamma_{\mbox{\tiny DE}}+ \Gamma_{\mbox{\tiny IB}}) = 0.689  \pm 0.021$ for
$E_{\gamma} \geq 20 \,\mbox{MeV}$ \cite{Abouzaid:2006hy}.
The full branching ratio is
\begin{equation}
\mathrm{BR}(K_L \rightarrow
\pi^+ \pi^- \gamma) = (4.15 \pm 0.15) \times 10^{-5}.
\end{equation}

Both IB and $E_1$ amplitudes are CP violating.
It is usual to parametrize this violation through
\begin{eqnarray}
 \label{eq:cpvkppg}
|\eta_{_{+ - \gamma}}|\,\mathrm{e}^{i\phi_{_{+-\gamma}}} &
= & \frac{A(K_L \rightarrow \pi^+ \pi^-
  \gamma)|_{\mathrm{IB}+E_1}}{A(K_S \rightarrow \pi^+ \pi^- \gamma)}  ,
\nonumber \\
\epsilon_{_{+ - \gamma}}'& = & \eta_{+-\gamma} - \eta_{+-} ,
\end{eqnarray}
where $\eta_{_{+-}}$ is defined in Eq.~(\ref{eq:epsilon_def}). The
experimental determination of these quantities is already rather old
\cite{Matthews:1995bm,Ramberg:1992yt}: $|\eta_{_{+-\gamma}}| =
(2.36  \pm 0.06) \times 10^{-3}$,
$\phi_{_{+-\gamma}} = (44  \pm 0.4)^{\circ}$ and
$|\epsilon_{_{+-\gamma}}'|/\epsilon < 0.3$ at $90 \%$
C.L., in accordance with SM predictions
\cite{Tandean:2000qk}. More recent is a measurement of the
equivalent of the parameter $g_{M_1}$ in Eq.~(\ref{eq:phenokppg})
for the amplitude $E_1$:
\textcite{Abouzaid:2006hy} find
$|g_{E_1}| \leq 0.21$ ($90 \%$ C.L.).

\subsubsection{$K_S \rightarrow \pi^+ \pi^- \gamma$}
Contrary to the $K_L$ decay, this channel is
dominated by the IB amplitude
because there
is no suppression in the $K_S \rightarrow \pi^+ \pi^-$ decay. From the
remaining contributions, $M_1$ and $E_2$
are CP violating.
Therefore, the dominant DE amplitude is $E_1$ with a leading
contribution of ${\cal O}(p^4)$. The octet piece
\cite{D'Ambrosio:1993et,D'Ambrosio:1994du,Mertens:2011ts}
contains a scale-independent local amplitude
\begin{equation}
 \label{eq:ksppglocal}
E_{\mathrm{ct}}^{(4)} = 4ie \,  \frac{G_8 M_K^3}{F_0} \left[
  N_{14}-N_{15}-N_{16}-N_{17} \right] ,
\end{equation}
with the same combination of LECs that appears in $K^{\pm} \rightarrow
\pi^{\pm} \pi^0 \gamma$ in Eq.~(\ref{eq:e4ctkp}). In addition, there is a
finite one-loop contribution
\begin{eqnarray}
\! \! \! \! \! \! \! \! \!  \! \! \! \! \! \! E_{\mbox{\tiny
    loop}}^{(4)} &=& -ie \, \frac{G_8 M_K}{8 \pi^2 F_0} \, (M_K^2 -
M_{\pi}^2) \left[ \, - \, 4 \,  h_{\pi \pi}(-z_3)
\right. \nonumber \\
&&   \left. + \, h_{\pi K}(z_{+})  + h_{K \eta}(z_{+}) + ( z_{+}
\leftrightarrow z_{-} ) \right] .\quad
\end{eqnarray}
The 27-plet amplitude has been determined by
\textcite{Mertens:2011ts}. Depending on the unknown combination of
LECs
in Eq.~(\ref{eq:ksppglocal}), the ratio of
IB and DE contributions amounts to
$\Gamma_{\mathrm{IB}}/\Gamma_{\mathrm{DE}} \sim 10^3$ and IB
therefore dominates the experimental branching ratio
\cite{Nakamura:2010zzb}
\begin{equation}
\mathrm{BR}(K_S \rightarrow \pi^+ \pi^- \gamma,\, E_{\gamma}^* > 50 \, \mbox{MeV})
= (1.79 \pm 0.05) \times 10^{-3} .
\end{equation}

\subsubsection{$K \rightarrow \pi  \pi \ell^+ \ell^-$}
The lepton pair in this decay is produced by an off-shell
photon and in consequence the dynamics is that of the
$K \rightarrow \pi \pi \gamma^*$ process. The widths are smaller than
those for a real photon and only the decays
$K_L, K_S \rightarrow \pi^+ \pi^- e^+ e^-$ have been observed
experimentally \cite{Lai:2003ad,Abouzaid:2005te}.
This decay provides a useful tool to look for the CP-violating
interference between electric and magnetic amplitudes because the
lepton plane furnishes a measurement  of the photon polarization
vector.
Bounds on the decays $K_L \rightarrow \pi^0 \pi^0 \ell^+ \ell^-$
have also been determined for $\ell = e$ \cite{AlaviHarati:2002nf} and
$\ell = \mu$ \cite{Abouzaid:2011mi}.
As the $\mu^+\mu^-$ case is strongly
suppressed by the available phase space, we only consider
$e^+ e^-$ production.

The decay amplitude is given by
Eqs.~(\ref{eq:kpplla}) and (\ref{eq:kppllv}). A first study  of the decay
$K_L \rightarrow \pi^+ \pi^- e^+ e^-$ with constant electric and
magnetic amplitudes was carried out by \textcite{Sehgal:1992wm}, while
the calculation at leading ${\cal O}(p^4)$ of the electric amplitudes
(keeping a constant magnetic contribution) was performed in
\textcite{Elwood:1995xv}. The complete results up to
${\cal O}(p^4)$ for both  $K_L \rightarrow \pi^+ \pi^- e^+ e^-$ and
$K^+ \rightarrow \pi^+ \pi^0 e^+ e^-$  can be found in
\textcite{Pichl:2000ab}.
\begin{enumerate}
\item[i.] $K_L \rightarrow \pi^+ \pi^- e^+ e^-$
\vspace*{0.1cm}\\
For this decay ${\cal E}_2 = -{\cal E}_1 (p_1 \leftrightarrow p_2)$
and we will only quote the amplitude ${\cal E}_1$.
The leading tree-level contribution to the electric amplitudes is
provided by $K^0 - \overline{K^0}$ mixing with an off-shell photon
radiating off the pion legs. Therefore, it is proportional to the
indirect CP-violation parameter $\epsilon$:
\begin{equation}
 \label{eq:e1kppee}
{\cal E}_1^{(2)} = -4 i \,  e \, \epsilon \, \frac{G_8 M_K F_0}{2
  M_K^2 z_1 + q^2} ( M_K^2 - M_{\pi}^2) .
\end{equation}
At ${\cal O}(p^4)$ a local scale-dependent contribution arises,
\begin{equation}
 \label{eq:e1kppee4l}
{\cal E}_{1 \, \mbox{\tiny ct}}^{(4)} = 2 i \, e \, \frac{G_8 M_K}{3
  F_0} q^2 \left[ N_{14}^r - N_{15}^r - 3
( N_{16}^r -N_{17}) \right] ,
\end{equation}
together with a divergent one-loop amplitude \cite{Pichl:2000ab}. At
this order, the magnetic amplitude is constant and
coincides with the case of a real photon in Eq.~(\ref{eq:m1kppgp4}).
In \textcite{Pichl:2000ab} the magnetic amplitude was modified by
including a $\rho$-pole parametrization analogous to
(\ref{eq:ga1a2kppg}).
\item[ii.] $K^+ \rightarrow \pi^+ \pi^0 e^+ e^-$
\vspace*{0.1cm}\\
For this channel there is no relation between the two
electric amplitudes as in the $K_L$ decay.
It is interesting to observe that at leading ${\cal O}(p^2)$
the octet contribution vanishes in the isospin
limit and only the 27-plet representation gives a non-zero
contribution \cite{Pichl:2000ab}. At ${\cal O}(p^4)$
there are local scale-dependent amplitudes
\begin{eqnarray}
 \label{eq:e1e2p4kppee}
{\cal E}_{1 \mbox{\tiny ct}}^{(4)} & = & -i \, e \, \frac{G_8 M_K}{3
  F_0} \left[ -6M_K^2 z_2 ( N_{14} - N_{15} -N_{16}
-N_{17}) \right. \nonumber \\
&& \; \; \; \; \; \; \; \; \; \; \; \; \left. - 4 q^2(N_{14}^r -
N_{15}^r) \right] ,  \nonumber \\
 {\cal E}_{2 \mbox{\tiny ct}}^{(4)} & = & -i \, e \, \frac{G_8 M_K}{3
   F_0} \left[ 6M_K^2 z_2 ( N_{14} - N_{15} -N_{16}
-N_{17}) \right.  \nonumber \\
&& \;  \left. - 2 q^2(N_{14}^r + 2 N_{15}^r) + 6 q^2(N_{16}^r -
N_{17})  \right] ,
\end{eqnarray}
reducing to the scale-independent result
(\ref{eq:e4ctkp}) for
an on-shell photon. There is also
a divergent one-loop contribution at ${\cal O}(p^4)$. As in the
previous case, the magnetic amplitude is constant and coincides
with the case of a real photon in Eq.~(\ref{eq:m4ctkp}).
\end{enumerate}
The branching ratios of $K_L$ and $K_S$ decays were measured by NA48
\cite{Lai:2003ad,Batley:2011zza}:
\begin{eqnarray}
\mathrm{BR}(K_L \rightarrow \pi^+ \pi^- e^+ e^-) &=& (3.08 \pm 0.20) \times
10^{-7}, \nn
\mathrm{BR}(K_S \rightarrow \pi^+ \pi^- e^+ e^-) &=& (4.93  \pm 0.14)
\times 10^{-5}. \nn
\end{eqnarray}

For the $K_L$ decay \textcite{Pichl:2000ab} studied the impact of the weak
LECs appearing in the amplitudes. However, as
pointed out before, the main interest of this process is the
study of the interference between electric and magnetic amplitudes
provided by the measurement of the photon polarization
\cite{Heiliger:1993qt,Elwood:1995dj,Ecker:2000nj,Sehgal:2000gb,Sehgal:1999vg,Sehgal:1992wm}.
As we have seen before,
the electric amplitude of $K_L \rightarrow \pi^+ \pi^- e^+ e^-$ is
generated by the indirect CP-violating mixing of $K^0$ and
$\overline{K^0}$. Its interference with the CP-conserving magnetic amplitude
produces an asymmetry in the distribution of the angle
$\phi$ between the $e^+ e^-$ and $\pi^+ \pi^-$ planes in the kaon
center-of-mass system:
\begin{equation}
 \label{eq:acpphoton}
{\cal A}_{\mbox{\tiny CP}} \,  = \,
\frac{\int_{0}^{\pi/2} \frac{d \Gamma}{d \phi} \, d \phi -
  \int_{\pi/2}^{\pi} \frac{d \Gamma}{d \phi} \, d
  \phi}{\int_{0}^{\pi/2} \frac{d \Gamma}{d \phi} \, d \phi +
  \int_{\pi/2}^{\pi} \frac{d \Gamma}{d \phi} \, d \phi} .
\end{equation}
This observable was measured\footnote{The analogous asymmetry for
  the $K_S$ decay is compatible with zero \cite{Batley:2011zza}.}
by \textcite{Abouzaid:2005te} as ${\cal A}_{\mbox{\tiny CP}} = (14 \pm 2) \%$.
\textcite{Ecker:2000nj} observed
that the theoretical prediction of ${\cal A}_{\mbox{\tiny CP}}$ with
the leading electric and magnetic amplitudes in
Eqs.~(\ref{eq:m1kppgp4}) and (\ref{eq:e1kppee}) is short by almost a
factor two compared with experiment. With the inclusion of
an energy dependence in the magnetic amplitude along the model in Eq.~(\ref{eq:ga1a2kppg}), agreement with
the experimental determination is restored.

\subsection{Other decays}

\subsubsection{$K^0 \rightarrow \gamma \gamma \gamma$ }
\label{sec:kggg}
The CP-preserving amplitude for this process has
parity-violating and parity-conserving contributions:
\begin{equation} \label{eq:k3gamma}
 A(K^0 \rightarrow \gamma \gamma \gamma) = \varepsilon_{\alpha}^{*}(q_1) \varepsilon_{\beta}^{*}(q_2)
\varepsilon_{\gamma}^{*}(q_3) \left( M_{\mbox{\footnotesize PV}}^{\alpha
  \beta \gamma} + M_{\mbox{\footnotesize PC}}^{\alpha \beta \gamma}
\right) \, .
\end{equation}
The first contributes to $K_L \rightarrow \gamma \gamma \gamma$ while
the parity-conserving amplitude governs
$K_S \rightarrow \gamma \gamma \gamma$. The Lorentz structure of
both amplitudes is rather complex and can be
found in \textcite{Heiliger:1993ja} and \textcite{Ho:2010iy}. Both
amplitudes
are strongly suppressed
by angular momentum. This is due to the fact that any
two of the three photons
in the final state must have $J \geq 2$ as a result of gauge
invariance and Bose symmetry.
\par
Estimates of the branching ratios have recently been carried out
\cite{Ho:2010iy} by using naive dimensional analysis
to predict the order of magnitude of the LECs in the appropriate
chiral Lagrangian of ${\cal O}(p^{10})$ contributing at
leading order:
\begin{eqnarray}
&& 10^{-16} \leq \mathrm{BR}(K_L \rightarrow \gamma \gamma
\gamma) \leq 10^{-14}, \nonumber \\
&& 10^{-19} \leq \mathrm{BR}(K_S \rightarrow \gamma \gamma \gamma) \leq
10^{-17}.
\end{eqnarray}
There is a recent experimental upper bound on the
first process \cite{Tung:2010ep}:
\begin{equation}
\mathrm{BR}(K_L \rightarrow \gamma
\gamma \gamma) < 7.4 \times 10^{-8} \quad (\rm 90 \%  \, \, C.L.).
\end{equation}

\subsubsection{$K_L \rightarrow \gamma \gamma \ell^+ \ell^-$}
This decay is relevant in order to determine the background
subtraction for
$K_L \rightarrow \pi^0 \ell^+ \ell^-$
\cite{Greenlee:1990qy}.
Its theoretical description as a bremsstrahlung radiation off the
leptonic legs in  the decay $K_L \rightarrow  \gamma \ell^+ \ell^-$
appears to be in agreement with the experimental branching ratios
\cite{AlaviHarati:2000hr,AlaviHarati:2000tv}:
\begin{eqnarray}
\mathrm{BR}(K_L \rightarrow \gamma \gamma e^+ e^- \! , \, E_{\gamma}^*> 5~
\mbox{MeV})\;\;\; &\!\!\!\!\!\!\!\! =&\!\!\! (5.8  \pm 0.4) \times 10^{-7}, \nn
\mathrm{BR}(K_L \rightarrow \gamma \gamma \mu^+ \mu^- \! , \, m_{\gamma
  \gamma}> 1~ \mbox{MeV}) &=& (10^{+8}_{-6}) \times 10^{-9}. \nn
&&
\end{eqnarray}

\subsubsection{$K_L \rightarrow \gamma \nu \overline{\nu}$}
This decay exhibits an interesting interplay between long- and
short-distance contributions.  Both of them have been considered
\cite{Richardson:1995ye}: 1) A long-distance amplitude consisting of two
pieces, a weak transition in the $K_L$ leg followed by constituent
quark triangle diagrams, and a resonance-dominated amplitude; 2) A
short-distance contribution given by  $d \overline{s}
\rightarrow \gamma \nu
\overline{\nu}$. The results obtained within this approach give
\begin{eqnarray}
\mathrm{BR}(K_L \rightarrow \gamma \nu \overline{\nu})|_{\mathrm{ld}}
&\sim&
10^{-12}, \nonumber \\
\mathrm{BR}(K_L \rightarrow \gamma \nu \overline{\nu})|_{\mathrm{sd}}
&\sim& 10^{-11},
\end{eqnarray}
showing
that the process is short-distance dominated \cite{Ma:1978yf}.

The CP-conserving and CP-violating short-distance
contributions were estimated to be \cite{Geng:2000ny}
\begin{eqnarray}
\mathrm{BR}(K_L \rightarrow \gamma \nu \overline{\nu})|_{\mathrm{CPC}}
&=& 1.0 \times 10^{-13}, \nonumber \\
{\rm BR}(K_L \rightarrow \gamma \nu \overline{\nu})|_{\mathrm{CPV}}
&=& 1.5 \times 10^{-15},
\end{eqnarray}
somewhat at odds with the results mentioned
above. Given
the small size of these branching ratios, this process is a good
benchmark to look for effects beyond the SM \cite{Jiang:2003yp}.

\subsubsection{$K_S \rightarrow  \ell^+_1 \ell^-_1 \ell^+_2 \ell^-_2$}

If CP is conserved, these decays proceed via
\begin{equation}
K^0_1 \rightarrow \gamma^* \gamma^* \rightarrow  \ell^+_1 \ell^-_1
\ell^+_2 \ell^-_2 .
\end{equation}
To leading order, $\mathcal{O}(G_8 p^4)$, only one-loop diagrams with pions
contribute to the amplitudes.
The branching ratios were estimated by \textcite{birk}:
\begin{eqnarray}
\mathrm{BR}(K_S \rightarrow  e^+ e^- e^+ e^-) &=& 7 \times 10^{-11}, \nn
\mathrm{BR}(K_S \rightarrow  e^+ e^- \mu^+ \mu^-) &=&  8 \times 10^{-12},
\nn
\mathrm{BR}(K_S \rightarrow  \mu^+ \mu^- \mu^+ \mu^-) &=& 1 \times
10^{-14}.
\end{eqnarray}
No experimental limits are available.

\subsubsection{$K_L, K_S \rightarrow \pi^0 \pi^0 \gamma$}
The absence of charged particles in these decays implies that they do
not have an IB contribution. Moreover, in the CP limit the
amplitude for the $K_L$ decay is purely electric and the one of
$K_S$ is purely magnetic. In addition, Bose symmetry requires
both multipoles to be even. A consequence of all these features is
that the branching ratios are very small. They have still
not been observed experimentally
\cite{Abouzaid:2007qd}:
\begin{equation}
\mathrm{BR}(K_L \rightarrow
\pi^0 \pi^0 \gamma) < 2.43 \times 10^{-7}
\quad
({\rm 90 \% \, \, C.L.}).
\end{equation}
For the $K_L$ decay the leading multipole amplitude $E_2$ is at least
of ${\cal O}(p^6)$, because the  ${\cal O}(p^4)$ octet contributions, both
local and loop pieces, vanish \cite{Funck:1992wa}. The 27-plet contributions
to the ${\cal O}(p^4)$ one-loop amplitude give a tiny addition \cite{Mertens:2011ts}.
A naive chiral dimensional
estimate of the ${\cal O}(p^6)$ local contributions predicts
$\mathrm{BR}( K_L \rightarrow \pi^0 \pi^0 \gamma ) \simeq 10^{-10}$
\cite{Ecker:1993cq,D'Ambrosio:1994du,Mertens:2011ts}. However, other estimates
predict a larger branching ratio $\simeq 10^{-8}$ \cite{Heiliger:1993qi}.
The $K_S$ decay has not been
studied in detail but it is expected to give a much smaller branching
ratio \cite{Heiliger:1993qi}.

\subsubsection{$K_L,K_S \rightarrow \pi^0 \pi^0 \gamma \gamma$}
The decay $K_L \rightarrow \pi^0 \pi^0 \gamma \gamma$ was studied
at leading order in CHPT \cite{Dykstra:1991jd,Funck:1992wa}.
It was found to be dominated by the pion-pole contribution, showing a
strong dependence on the cut on the photon energies.
\par
The $K_S$ decay has no local contribution at ${\cal O}(p^4)$. The
amplitude at this order is therefore given by a finite one-loop piece
\cite{Funck:1992wa} that provides an unambiguous prediction:
\begin{equation}
{\rm BR}(K_S \rightarrow \pi^0 \pi^0  \gamma \gamma, \, m_{\gamma \gamma}
> 20 \,  \mbox{MeV}) = 4.7 \times 10^{-9}.
\end{equation}
There are no experimental bounds for these decays.

\subsubsection{$K \to 3 \pi \gamma$}

The experimental status of $K \to 3 \pi \gamma$ decays is still
rather meager. Only the two channels with a charged kaon have been
detected experimentally. None of the decay modes of a neutral kaon
have been seen.

Like in many other radiative $K$ decays, there is both an electric and a
magnetic amplitude. The magnetic contributions
are severely suppressed for all channels
\cite{D'Ambrosio:1996zy}, especially for the ones where the
non-radiative decay amplitudes are dominantly octet: $K^+ \to \pi^0
\pi^0 \pi^+ \gamma$, $K^+ \to \pi^+ \pi^+ \pi^- \gamma$ and $K_L \to
\pi^+ \pi^- \pi^0 \gamma$. For these channels, the rates can be
predicted with great accuracy by using the concept of ``generalized
bremsstrahlung'' \cite{D'Ambrosio:1996fy}. In this way, optimal use
can be made of available experimental information on the non-radiative
amplitudes. However, as long as the minimal photon energy
$E_\gamma^{\mrm{min}}$ is small ($E_\gamma^{\mrm{min}} \lsim 40$ MeV),
the difference between ``generalized bremsstrahlung'' and the
  leading Low contribution \cite{Low:1958sn} will not be accessible
experimentally in the foreseeable future. The
genuinely radiative contributions (direct emission) would
only matter for large $E_\gamma^{\mrm{min}}$, of course at the expense
of the number of events.

\begin{center}
\begin{table*}[floatfix]

\caption{\label{tab:k3pig} Comparison between theory
  \cite{D'Ambrosio:1996zy} and experiment \cite{Nakamura:2010zzb}
  for $K \to 3 \pi \gamma$ decays. The theoretical prediction for BR($K_S
  \to \pi^+ \pi^- \pi^0 \gamma$) depends on NLO weak LECs and is a rough
  estimate only. \\}

{ \begin{tabular}{|c|c|c|c|}
\hline
& & & \\[-.4cm]
 & $E_\gamma^{\mrm{min}}$(MeV) & BR (theory) & BR (PDG'10)
 \\[4pt]
\hline
 & & &    \\[-.3cm]
\hspace*{.2cm} $K^+ \to \pi^0 \pi^0 \pi^+ \gamma$ \hspace*{.2cm}&  10
& \hspace*{.3cm} $(3.78 \pm 0.05)\times 10^{-6}$ \hspace*{.3cm}
& \hspace*{.3cm} $(7.6 \,\mbox{}_{- 3.0}^{+6.0})\times 10^{-6}$
\hspace*{.3cm}
   \\[4pt]
\hspace*{.2cm} $K^+ \to \pi^+ \pi^+ \pi^- \gamma$ \hspace*{.2cm}&  5
& \hspace*{.3cm} $(1.26 \pm 0.01)\times 10^{-4}$ \hspace*{.3cm}
& \hspace*{.3cm} $(1.04 \pm 0.31) \times 10^{-4}$ \hspace*{.3cm}
   \\[4pt]
\hspace*{.2cm} $K_L \to \pi^+ \pi^- \pi^0 \gamma$ \hspace*{.2cm}&  10
& \hspace*{.3cm} $(1.65 \pm 0.03)\times 10^{-4}$ \hspace*{.3cm}
&    \\[4pt]
\hspace*{.2cm} $K_S \to \pi^+ \pi^- \pi^0 \gamma$ \hspace*{.2cm}&  10
& \hspace*{.3cm} $2 \times 10^{-10}$ \hspace*{.3cm}
&    \\[5pt]
\hline
\end{tabular}}
\end{table*}
\end{center}

The situation is different for the transition $K_S \to \pi^+ \pi^-
\pi^0 \gamma$. At lowest chiral order, the amplitude can only proceed
through a $\Delta I=3/2$ transition (via bremsstrahlung) and it is
therefore suppressed by the  $\Delta I=1/2$ rule. Similarly as for
$K^+ \to \pi^+ \pi^0 \gamma$, the NLO contributions generated by octet
operators become relatively more important. The electric amplitude is
dominated by NLO weak LECs, with a large theoretical uncertainty
\cite{D'Ambrosio:1996zy}.

The comparison between theory and experiment is displayed in Table
\ref{tab:k3pig}. For the two charged $K$ decays and for $K_L \to \pi^+
\pi^- \pi^0 \gamma$, the theoretical accuracy is only limited by the
precision with which the parameters of the non-radiative amplitudes
are known. The branching ratio for  $K_S \to \pi^+ \pi^- \pi^0 \gamma$
has a much larger theoretical uncertainty. Experimental information on
this mode would therefore be especially interesting, but the expected
rate is unfortunately very small.

\section{Conclusions and outlook}
\label{sec:conc}

Kaons have been at the center of many ground-breaking developments in
particle physics,
which are worth repeating here. From the introduction of internal ``flavor" quantum numbers (strangeness),
to parity violation, meson-antimeson mixing, quark mixing, CP violation, and the
suppression of flavor-changing neutral currents,
kaon  physics has played a key role in the construction of what
we now call the Standard Model.

In this review we have summarized the status of theoretical predictions for
all SM  allowed kaon decays with branching ratios greater than $10^{-11}$.
Theoretical predictions rely on the weak operator product expansion
to organize the short-distance dynamics,
the renormalization group to evolve the relevant operators down to the
hadronic scale,
and non-perturbative methods such as CHPT and lattice QCD
to deal with low-energy strong dynamics.
The accuracy of  theoretical predictions ranges  from excellent to
just fair,   depending on
whether the  decay amplitude is dominated by short- or long-distance effects.
In all cases  we have compared theoretical predictions to existing
data and we have discussed the impact of
future measurements.

It is remarkable that to date kaons still provide unique information
on physics at vastly separated energy scales:
\begin{itemize}
\item On one  hand, kaons are an excellent  probe of  low-energy
  strong interaction dynamics  ($E \sim \Lambda_{QCD}$).
While many kaon decays  are by themselves a valuable testing ground
for chiral effective theory techniques,
specific modes such as $K \to \pi \pi \ell \nu$ and $K \to  \pi \pi \pi$
allow one to  extract reliable  information on low-energy $\pi \pi$
  scattering, thus testing our understanding of
purely strong dynamics.

\item On the other  hand,  kaon decays
encode  information about flavor dynamics at the electroweak scale
$E \sim O(10^{2-3})$ GeV
through their sensitivity to  virtual exchange of  heavy SM particles
($W$, $Z$, $t$)
and possibly non-standard particles and their flavor structure.
Theoretical predictions for $K^0-\overline{K^0}$ mixing,
short-distance dominated rare decays,
and semileptonic branching ratios  have reached
a level of accuracy that already puts very stringent constraints on
extensions of the SM,
in ways often complementary to other  flavor observables and
electroweak precision tests.
Here we have not discussed in a systematic way  the impact of kaon
decays on extensions of the SM,
which can be found in reviews such as~\textcite{Buchalla:2008jp} and
\textcite{Antonelli:2009ws} in the broader context of
the low-energy/high-intensity frontier.
\end{itemize}

While by now kaon physics is a mature field,  there are a number of
challenges and exciting prospects
for the near future.
On the theoretical side the main challenge will involve improving our
control of strong interaction effects.
We anticipate that in the coming years lattice QCD (in synergy with
CHPT)  will play an increasing role in determining
the long-distance dynamics relevant for kaon decays.
This will not only  increase the accuracy of predictions for
semileptonic modes,  but will also put  the SM predictions  for
nonleptonic decays  on firmer ground,
in both cases  enhancing the constraining power of these decays  with
respect to models of New Physics.
On the experimental side, the world-wide program aimed at rare decays will face
the challenge of measuring branching ratios at the level of $10^{-11}$.
This program  will  produce high-statistics  results  for other
channels as a by-product,
and has a great potential to discover effects of New Physics in the
``golden modes" $K \to \pi \nu \bar{\nu}$.
Even in the event of agreement with the SM predictions, these
measurements will provide essential information on
the flavor structure of SM extensions at the TeV scale.
In both scenarios,   kaons  will play a central role at the low-energy
frontier of our search  for physics beyond the SM.

\section*{Acknowledgments}
As documented in the bibliography, many physicists have contributed to
the field of kaon decays. For their collaboration during the past 25
years, we thank especially J.~Bijnens, F.J.~Botella, A.G.~Cohen, G.~Colangelo,
G.~D'Ambrosio, C.A.~Dominguez,
J.F.~Donoghue, M.~Eidem{\"u}ller,  J.~Gasser, C.~Gatti, M.~Gianotti,
E.~Golowich,
D.~G\'omez Dumm, M.~Gonz\'alez-Alonso, B.~Guberina, G.~Isidori, M.~Jamin,
J.~Jenkins, R.~Kaiser,
J.~Kambor, A.~Kastner, M.~Knecht,
H.~Leutwyler, K.~Maltman, M.~Moulson, G.~M{\"u}ller, S.~Noguera, E.~Pallante,
M.~Palutan, J.A.~Pe\~narrocha, H.~Pichl, J.~Prades,
E.~de Rafael, I.~Rosell, H.~Rupertsberger, I.~Scimemi, P.~Talavera,
R.~Unterdorfer and D.~Wyler.
We also wish to thank B.~Bloch-Devaux, A.~Ceccucci,
E.~Cheu, M.~Contalbrigo, E.~Goudzovski, M.~Raggi and C. Smith for
correspondence.
This work has been supported in part by the EU MRTN-CT-2006-035482 (FLAVIAnet).
A.~P. and J.~P. are partially supported by MEC (Spain) under grant
FPA2007-60323, by the Spanish Consolider-Ingenio 2010 Programme CPAN
(CSD2007-00042) and by Generalitat Valenciana under grant
PROMETEO/2008/069. J.~P. is also partially supported by CSIC under grant
PII-200750I026. The work of V.C. is supported by the Nuclear Physics Office
of the U.S. Department of Energy under Contract No.~DE-AC52-06NA25396 and by
the LDRD program at Los Alamos National Laboratory.

\appendix

\section{One-loop functions}
The following functions occur in the one-loop amplitudes for rare and
radiative $K$ decays:
\begin{equation} \label{eq:1lfz}
 F(z) = \left\{ \begin{array}{lr}
                 1-\frac{4}{z}
                 \arcsin^2\left(\frac{\sqrt{z}}{2}\right) & z  \leq 4 \\
                 1+\frac{1}{z}\left( \ln
                 \frac{1-\sqrt{1-4/z}}{1+\sqrt{1-4/z}}  + i \pi
                 \right)^2  & z>4
                \end{array}
        \right.,
\end{equation}
\begin{equation} \label{eq:1lgz}
 G(z) = \left\{ \begin{array}{lr}
                 \sqrt{4/z-1} \arcsin \left( \frac{\sqrt{z}}{2}\right)
                 & z  \leq 4 \\
                  \frac{1}{2} \sqrt{1-4/z} \left( \ln
                 \frac{1+\sqrt{1-4/z}}{1-\sqrt{1-4/z}} - i \pi \right)
                  & z > 4
                \end{array}
        \right. ,
\end{equation}
\begin{eqnarray} \label{eq:1lhz}
 H(z) & = &  \frac{1}{2(1-z)^2} \left\{ z
 F\left(\frac{z}{r_{\pi}^2}\right)  - F\left(\frac{1}{r_{\pi}^2}\right)
\right.
\nonumber \\
&& \; \; \; \; \; \;\left. -2z \left[
 G\left(\frac{z}{r_{\pi}^2}\right)-  G\left(\frac{1}{r_{\pi}^2}\right)
 \right] \right\} \, ,
\end{eqnarray}
\begin{eqnarray} \label{eq:hijz}
\! \! \! \! \! \! \! \! \! \!  h_{ij}(z) & = & \frac{(4 \pi)^2}{z}
\left[ C_{20}(p^2, (p+q)^2, M_i^2, M_j^2) \right. \nonumber \\
&&   \left. \; \; \; \; \; \, \, \; \; \; \;   - \, C_{20}(p^2, p^2,
  M_i^2,M_j^2) \right] ,
\end{eqnarray}
with $p^2=M_{\pi}^2$ (for $h_{\pi K}$ and $h_{K\eta}$) or $p^2=M_K^2$
(for $h_{\pi \pi}$).
$C_{20}$ is defined by the three-point integrals for $q^2=0$:
\begin{eqnarray} \label{eq:hijc20}
 && \int \frac{d^4 \ell}{(2 \pi)^4} \frac{\ell^{\mu}
    \ell^{\nu}}{[\ell^2 - M_i^2][(\ell+q)^2 -
      M_i^2][(\ell-p)^2-M_j^2]} \nonumber \\
&& = i g^{\mu \nu} C_{20}(p^2, (p+q)^2, M_i^2, M_j^2) + \dots
\end{eqnarray}

\bibliographystyle{apsrmp}
\bibliography{kaons_archive_v4}

\end{document}